\documentclass[]{aa}

\usepackage{graphicx}

\newcommand{\Mdot}{\mbox{$\dot M$}}                 
\newcommand{\Msolyr}{~M$_{\odot}$\,yr$^{-1~}$}      

\begin{document}

\title{New observations of cool carbon stars in the halo
\thanks{Based on observations done at  Haute Provence Observatory
      operated by the Centre National de Recherche Scientifique (France)
}}

\author{N.~Mauron\inst{1}}

\offprints{N.~Mauron}

\institute{GRAAL, CNRS and Universit\'e  Montpellier II,  
 Place Bataillon, 34095 Montpellier, France\\ 
 \email{mauron@graal.univ-montp2.fr}
}

\date{}

\abstract{}
{We report new results of our search for rare, cool carbon 
stars located at large distances from the Galactic plane.} 
{Candidate stars were selected in the 2MASS point source catalogue with  
$JHK_{\rm s}$ colours typical of N-type carbon stars, with $K_{\rm s}$\,$\ga$6.0
 and with Galactic latitude $|b| > 20$\degr. Low resolution slit 
 spectroscopy was carried out on  58 candidates.}
{Eighteen new carbon stars were discovered. Six are remarkable by showing the 
two peculiarities of  a strong infrared excess
at 12 $\mu$m and a large height above the Galactic plane, from 1.7 to 6\,kpc. The number
of C stars with these properties has been increased to 16.   Mass-loss rates
were tentatively estimated by assuming that all these 16 stars are Miras and by 
using the correlation between $\Mdot$ and the
 $K$\,$-$\,[12] colour index. It is found that several stars have large mass loss,
with a median \Mdot\, of 4\,$\times$\,10$^{-6}$\,\Msolyr and a dispersion
of about a factor of 3 around this value.  It would be desirable to
detect their CO emission  to see whether, like one object already on the list, they display
a very low expansion velocity that could be the signature of AGB mass loss at low metallicity. 
The distances of our new carbon stars were determined by supposing them to be similar to those
of the Sagittarius dwarf galaxy.  These distances are relatively uncertain, but
they do indicate that eight stars might be more than 30\,kpc from 
the Sun, and  two at the unprecedented distance of 150\,kpc.
 }{}

\titlerunning{New carbon stars in the halo}
\authorrunning{N.~Mauron}
\maketitle


\section{Introduction} 

 Carbon (C) stars have been the subject of many studies, in particular because
 of their rich diversity. Carbon stars are also used for investigating the global 
 properties of galaxies such as their metallicity and star-forming history, 
 the enrichment of their interstellar medium in carbon and carbon-rich dust, or 
 the metal-poor stellar populations in the halos. 
 With considerable simplification, one can say that  one type of C star
 is composed of red, luminous, variable  stars on the asymptotic giant branch (AGB) for 
 which carbon is synthesized in the core and dredged up to the stellar outer layers. 
 These stars are the N-type  stars. Another family of C stars comprises binary
 stars where one of the components has already passed through the AGB phase and 
 polluted the  visible component  with carbon-rich material.  
 Among these objects are the CH-type giants and the carbon dwarfs, which are 
 generally older than N-type objects (for a review of carbon stars, see Wallerstein 
 \& Knapp \cite{walknapp98}). 
 
 This paper is centred on the search for new N-type stars that are located in the Galactic
 halo and  do not  belong to the Galactic thin disc like the large majority of known 
 cool C stars.    Bothun et al. (\cite{bothun91}) considered a sample of  32 faint C stars 
 in the halo and noted that a few of these stars in the Southern hemisphere were very red,
 indicating an intermediate-age population.
 Totten \& Irwin (\cite{ti98}) carried on the census and discovery of faint, high Galactic 
 latitude C stars by selecting very red objects on Schmidt plates, 
 which resulted in a list of 77 N-type or CH-type objects.
 Because the N-type stars were unlikely to have formed in the halo, Totten \& Irwin propose
 that they  come from the disruption of tidally captured dwarf satellite galaxies. 
 Subsequently, Ibata et al. (\cite{ibata01})  demonstrate that half of these 77 objects 
 originate in the Sagittarius dwarf galaxy, and are part of the tidal debris left by 
 this galaxy in its orbital path around the Milky Way.

 The goal of this work is to continue the exploration of the halo for new cool C stars and 
 to make a complete census of them, at least at Galactic latitude $|b|$\,$\ga$\,20\degr.
 The driving idea of this exploration is to exploit the 2MASS catalogue of point 
 sources (Cutri et al. \cite{cutri03}). A first reason is that cool C stars are very strong 
 near-infrared objects,  typically with $K_{\rm s}$\,$=$\,13 at 100\,kpc.
 Secondly, cool C stars occupy a restricted region in the colour-colour diagram,
 which helps in finding good C  candidates (see e.g. Fig.\,3 of MacConnell \cite{macconnell03}).
 Following our previous investigations (Mauron et al. \cite{mauron04}, \cite{mauron05}, 
 \cite{mauron07b}), two aspects
 appeared particularly interesting. One aspect concerns the cool C stars that one can detect 
 at a very large distance from the Galactic centre ($\ga$\,100\,kpc). These stars are the
 most luminous of a stellar system (e.g. a stream) that still comprises a small population 
 of stars of  intermediate age. Therefore, their presence might help for finding 
 where the more  numerous but fainter giants of this system would lie, 
 approximatively. Consequently, it seemed important to  us to discover
 N-type stars as far as possible within a large volume of the Galactic halo.

 The second aspect of our search for new N-type stars in the halo concerns stellar 
 evolution and, in particular, mass loss of AGB stars at low metallicity. According to 
 Zijlstra (\cite{zijlstra04}), there is indirect evidence that the mass-loss efficiency for 
 [Fe/H]$<$\,$-1$ is very low. But even for 0$<$[Fe/H]$<$\,$-1$,  the mechanism and the 
 strength of this mass loss  is more poorly documented than for AGB stars in the solar 
 vicinity. One of the reasons for this situation  is that  metal-poor AGB stars in the 
 Magellanic Clouds are too distant for the circumstellar gas (e.g. CO)
 to be detected with current millimetric instrumentation. However, 
 a small number of very red N-type C stars  are located far from the Galactic plane
 (Groenewegen et al. \cite{groe97}; Mauron et al. \cite{mauron07b}) which makes them 
 interesting to consider. 
 If they belong to a tidal stream or if they are members of the thick disc, there is 
 a significant probability that some of them might be metal-poor. At least two of them
 show signs of oxygen underabundance (Groenewegen et al. \cite{groe97}). Therefore, 
 these objects are key targets for investigating mass-loss at low metallicity. In 
 this paper, we discuss  our discovery  of 6 new, very red, mass-losing C stars with 
 distances from the Galactic plane between 1.7 and 6\,kpc.
 
 In the following sections, we report new observations of 58 candidate C stars 
 in the halo. This represents the last  observational part of our programme. 
 The general method of candidate selection was explained in Mauron et al. (\cite{mauron04}) 
 so is not repeated here, but some new considerations are given in Sect.\,3.1.
  In Sect.\,2, the observations and data reductions are described.
 In Sect.\,3, the results of our observations are analysed, especially the 
 spectra, the IRAS counterparts, and an estimation of mass-loss rates. 
 Section\,3.4 concerns the distances and  the location of our stars 
 in the halo. Final conclusions and suggestions are given in Sect.\,4.
 
 
 \section{Observations and reductions}

  The observations were made during the nights May 16 to 21 and 
  September 10 to 14, 2007 at the Haute-Provence Observatory.
 The 1.93-m telescope was equipped with the CARELEC spectrograph.
 This instrument was used with a 150 g mm$^{-1}$ grating. The detector is an 
 EEV 2048$\times$1024 CCD chip with 13.5$\times$13.5 $\mu$m pixels. 
 The resulting dispersion  is 3.6\,\AA\, per pixel. The slit width  is
 2.0 arcsec. The resolving power is $\lambda/\delta\lambda$\,=\,460, and 
 $\delta\lambda$\,=\,13\,\AA\,at 6000\,\AA\,, with a spectral coverage of
 4400 to 8400\,\AA. This low resolution was chosen to allow spectroscopy
 of some targets as faint as ($R$\,$\sim$ 17-18) for which the exposure time was
 $\sim$\,1 hour. In these spectra, even if the signal-to-noise ratio is low, 
 it is relatively easy to see the  molecular bands of C$_{2}$ and 
 CN typical of cool carbon stars. When the sky transparency was poor
 and/or the seeing was not good, spectroscopy was achieved for brighter targets with
 $R$\,$\sim$\,13-14.  We also  took spectra of 23 supplementary,  relatively bright, 
 candidate C stars detected  in the Byurakan objective-prism survey 
 (Gigoyan et al. \cite{gigoyan01}).  These objects needed   slit spectroscopy for 
 confirmation of their carbon-rich nature. All these supplementary spectra are 
 given in the Appendix.
 
 The reductions of the  CCD frames included 
 bias subtraction, flat-fielding, extraction of one-dimensional spectra 
 (object and sky), sky subtraction, cleaning of cosmic rays hits, 
 and wavelength calibration. The instrumental spectral 
 efficiency was  corrected with the spectrum of a standard photometric star
 observed during one night. No correction for atmospheric extinction was 
 attempted.  The obtained spectra are proportional to a signal expressed in
 erg s$^{-1}$ cm$^{-2}$ \AA$^{-1}$. However, no absolute calibration could be
 achieved because the sky was usually not photometric, and 
 because  strong slit losses occurred when the seeing was poor.
 Therefore, only  relative spectral distributions are drawn in our plots.

\begin{table*}[!ht]
	\caption[]{List of discovered  carbon stars and their characteristics}
	\label{table1}
	\begin{flushleft}
	\begin{tabular}{lcrrrrrrrrrl}
	\noalign{\smallskip}
	\hline
	\hline
	\noalign{\smallskip}
No.&  2MASS name  & $l$~~~~ & $b$~~ & $B$ & $R$~~ & $B$-$R$ & $J$~~ & $H$~~ & $K_{\rm s}$~~ & $J$-$K_{\rm s}$&Note\\
	\noalign{\smallskip}
	\hline
 84& \object{2MASS J003632.34$-$225451.0} &  83.435 & $-$84.602  & 19.4 &17.4 & 2.0 & 15.473 &  14.596 &  14.144 &  1.329& \\
 85& \object{2MASS J021926.96$+$355058.9} & 142.299 & $-$23.690  & 21.2 &14.4 & 6.8 & 11.719 &  9.914  &   8.518 &  3.201&1 \\
 86& \object{2MASS J022431.98$+$372933.0} & 142.664 & $-$21.786  & 21.9 &16.3 & 5.6 & 11.489 &   9.908 &   8.753 &  2.736& 1, 3 \\
 87& \object{2MASS J023904.88$+$345507.5} & 146.697 & $-$22.937  & 19.0 &16.6 & 2.4 & 12.264 &  10.151 &   8.386 &  3.878& 1\\
 88& \object{2MASS J035955.96$+$091904.4} & 181.024 & $-$31.584  & 19.0 &15.6 & 3.4 & 12.757 &  11.229 &  10.114 &  2.643& 1\\
 89& \object{2MASS J112705.59$+$101540.4} & 249.238 & $+$63.941  & 17.1 &14.3 & 2.8 & 13.144 &  12.260 &  11.673 &  1.471& \\
 90& \object{2MASS J130118.47$+$002950.7} & 308.423 & $+$63.264  & 21.2 &18.1 & 3.1 & 15.144 &  14.106 &  13.513 &  1.631& 1\\
 91& \object{2MASS J133557.07$+$062354.9} & 331.973 & $+$66.727  & 19.1 &15.0 & 4.1 & 12.270 &  11.195 &  10.575 &  1.695& \\
 92& \object{2MASS J144644.19$+$051238.0} & 359.497 & $+$54.876  & 18.7 &14.3 & 4.4 & 13.637 &  12.287 &  11.349 &  2.288& \\
 93& \object{2MASS J163631.69$-$032337.4} &  12.734 & $+$27.794  & 20.0 &13.6 & 6.4 & 10.004 &   7.897 &   6.098 &  3.906& 2\\
 94& \object{2MASS J172046.20$-$000423.5} &  22.082 & $+$20.023  & 15.6 &12.3 & 3.3 &  9.522 &   8.582 &   8.212 &  1.310&\\
 95& \object{2MASS J172554.39$+$030026.5} &  25.619 & $+$20.363  & 19.4 &16.1 & 3.3 & 14.794 &  13.889 &  13.409 &  1.385& \\
 96& \object{2MASS J181329.45$+$453117.5} &  73.053 & $+$25.349  & 18.2 &13.6 & 4.6 & 10.519 &   8.489 &   6.710 &  3.809& \\
 97& \object{2MASS J184950.90$+$621725.4} &  92.339 & $+$23.989  & 15.6 &12.2 & 3.4 &  8.804 &   7.972 &   7.495 &  1.309& \\
 98& \object{2MASS J195652.65$-$140520.5} &  27.282 & $-$20.741  & 17.7 &15.7 & 2.0 & 13.781 &  12.932 &  12.478 &  1.303& \\
 99& \object{2MASS J200303.83$-$194903.9} &  22.189 & $-$24.332  & 16.9 &13.1 & 3.8 & 10.531 &   9.633 &   9.110 &  1.421& \\
100& \object{2MASS J202000.43$-$053550.6} &  38.153 & $-$22.230  & 19.4 &16.0 & 3.4 & 12.044 &  10.190 &   8.613 &  3.431& 1\\
101& \object{2MASS J204817.91$+$102638.7} & 56.823  & $-$20.154  & 20.6 &15.2 & 5.4 & 10.059 &   8.786 &   7.830 &  2.229& \\
  
           \noalign{\smallskip}
   \hline
\end{tabular}
\end{flushleft}

{\small Notes: (1) the $B$ and $R$ magnitudes are from the USNO-B1.0 catalogue; 
(2) the $B$ mag is a lower limit (see text); (3) this star was also  identified to be a C star 
by Cruz et al.(\cite{cruz07})}
\end{table*}

\section{Results}

  Table\,\ref{table1} lists the main parameters  of the 18  objects that are under
analysis in this work. The columns of Table\,\ref{table1} give the running number 
following those of Mauron et al. (\cite{mauron07b}), the 2MASS name comprising the 
J2000 coordinates coded as JHHMMSS.ss$+$DDMMSS.s, the Galactic 
coordinates $l$ and $b$ in degrees, the $B$ and $R$ magnitudes and the $B-R$ colour 
index, the $J H K_{\rm s}$ magnitudes from the 2MASS point-source catalogue and the 
$J-K_{\rm s}$ colour index. 
	
	The $B$ and $R$ magnitudes are  from the USNO-A2.0 
catalogue of Monet et al. (\cite{monet98}).
When no such data was provided, we considered the data of the USNO-B1.0 catalogue 
(Monet et al. \cite{monet03}), which is indicated in the Notes of Table\,\ref{table1}. 
When two $B$ or $R$ magnitudes are given in USNO-B1.0, the average is adopted.
For star \#93, there is no available information for the $B$ magnitude in either catalogue.
Therefore, we adopted from USNO-A2.0 the $B$ magnitude of the faintest object within a radius 
of 2.5 arcmin, giving $B$ $\approx$ 20.0. The resulting $B$$-$$R$ (in fact a lower limit) 
is 6.4.  This is consistent with the strong $J-K_{\rm s}$ colour. An additional indication that 
$B$$-$$R$ must be very large is the fact that its spectrum  strongly rises to the red 
(see below). 

These $B$ and $R$ magnitudes are not very accurate, with an uncertainty of 
$\sim$0.4 mag., but they nevertheless give  useful information. For example,
it can be noted that out of 18 objects, 14 have $B-R \ge$3.0, which  agrees
with the typical colour of N-type stars.
 We also note that object \#84, the faintest in the $K_{\rm s}$-band and the
most distant in our list (see below), has  $B-R = 2.0$ and might have been eliminated
in the list of 2MASS candidates if $B-R$ had been required to be larger.

Two objects, \#84 and \#90, are very faint in the $K_{\rm s}$-band, with 
$K_{\rm s}$\,$=$\,$14.14$ and $K_{\rm s}$\,$=$\,$13.51$, respectively. In previous
studies, we had not attempted to detect objects as faint as that, and we had limited our
search for C stars among candidate objects having $K_{\rm s}$\,$<$\,13.5\,. For
the observations described here, we extended our selection of candidates
to $K_{\rm s}$\,$=$\,$14.5$.  This needed to eliminate a very large number ($\sim$2500)  of
contaminating galaxies that have $J$\,$H$\,$K_{\rm s}$ colours similar to C stars. 
These galaxies are present in the 2MASS point-source
catalogue despite their obviously appearing as diffuse objects in the 
digitized POSS plates. This search for faint, good quality candidates  resulted in finding 
24 best cases  with $K_{\rm s}$ between 13.5 and 14.5, $\delta$\,$\ge$\,$-25$\degr\, 
and $R$\,$\la$\,$18.0$. The $R$-band constraint
was applied to allow spectroscopy to be carried out with our instrumentation.
Of these 24 objects,  ten were observed: two were carbon stars, six M-type stars, 
and the remaining  two  were difficult to classify, but certainly not carbon-rich. This
suggests that several very distant C stars remain to be discovered 
with $R$\,$\la$\,$18.0$ in the northern hemisphere, but probably not a large number.  
A more sensitive instrumentation than 
the CARELEC spectrograph  is needed to search for C stars beyond $R$\,$=$\,18. 

One source has already been classified as a carbon star: object \#93 is \object{CGCS 3716} in the
catalogue of Galactic carbon stars of Alksnis et al. (\cite{alksnis01}). This object  is 
\object{IRAS 16339-0317}. Little-Marenin et al. (\cite{little87}) detected the SiC 11.2~$\mu$m feature in
its IRAS low resolution spectra. Epchtein et al. (\cite{epchtein90}) also classified it as a carbon 
star by considering its location in combined IRAS and $JHKL$ colour-colour diagrams
that separate oxygen-rich and carbon-rich giants. Our spectrum firmly establishes
the star to be carbon-rich by showing a very clear CN molecular band.
Another source (\#97) is known as a variable star of the Mira Ceti type
(Dahlmark \cite{dahlmark98}, Kazarovets et al. \cite{kazarovets00}). 
Its name is \object{IZ Dra} and its period  312 days. Our observations show that it is carbon-rich.

Finally, it can be noted that two objects (\#95 and \#98) that are quite faint in $K_{\rm s}$
($\sim$ 13), were found near the limit of our survey in Galactic latitude, 
at $b$\,$=$$+20.3$\degr and at 
 $b$\,$=$\,$-20.7$\degr. We shall see that these objects have distances of about 100 and 60 kpc, 
 respectively, with \#95  remarkable because of its large distance to
 the main plane of the Sgr Stream. Traditionally, faint high-latitude carbon stars were 
 searched for at $|b|$\,$\ga$\,30\degr\,, but our experiment shows that  searching for 
 distant  C stars closer to the Galactic plane is feasible.

\subsection{Spectra}

All  obtained spectra   (except the three in Fig.\,\ref{fig1}) are presented in the Appendix. 
In Fig.~\ref{fig1}, we show three representative spectra. Object \#89  is
relatively bright ($R$\,$=$\,$14.3$) so that the signal to noise ratio
is high; this star is moderately red in the near-infrared ($J-K_{\rm s}$\,$=$\,$1.47$).
Object \#84 is the faintest in $K_{\rm s}$ with $K_{\rm s}$\,$=$\,14.14 and 
$J$\,$-$\,$K_{\rm s}$\,$=$\,$1.33$. Its spectrum has  the lowest  S/N in our sample
but  clearly shows the C$_{2}$ bands at $\lambda$\,$<$\,$5600$\AA\,. Finally, 
object \#100  is one of the reddest sources ($J$\,$-$\,$K_{\rm s}$\,$=$\,$3.43$), displaying
a strongly rising continuum, with the CN-band at 7800\,\AA\, deep enough
to prove its carbon-rich nature.\\

Many objects display H$\alpha$ in emission. For several
objects, this emission is faint (\#84, 85, 87, 94), while for others the emission
is strong (\#86, 88, 90, 91, 96, 97, 98). The fraction of stars with this emission line
is therefore $\sim$ 50\%,  in fair agreement with our previous studies. 
There is one object, \#97, for which H$\beta$ and  H$\gamma$ in emission are also 
visible. Finally, object \#96 shows H$\alpha$, H$\beta$, and [O\,III] $\lambda$ 5007\,\AA\, 
in emission and might be a symbiotic star with a hot ionizing companion.\\

\begin{figure}[!ht]
\resizebox{8cm}{!}{\rotatebox{-90}{\includegraphics{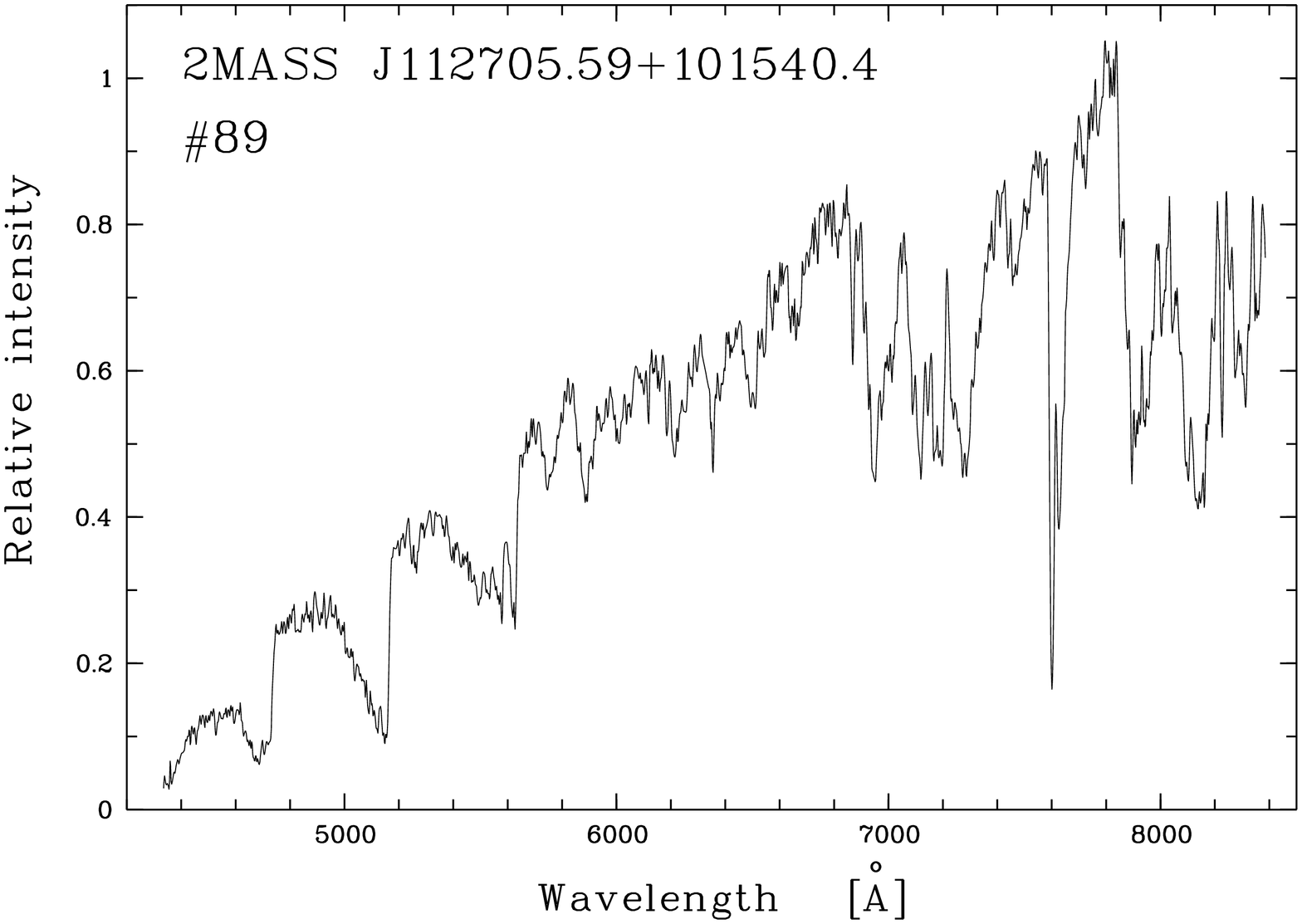}}}
\resizebox{8cm}{!}{\rotatebox{-90}{\includegraphics{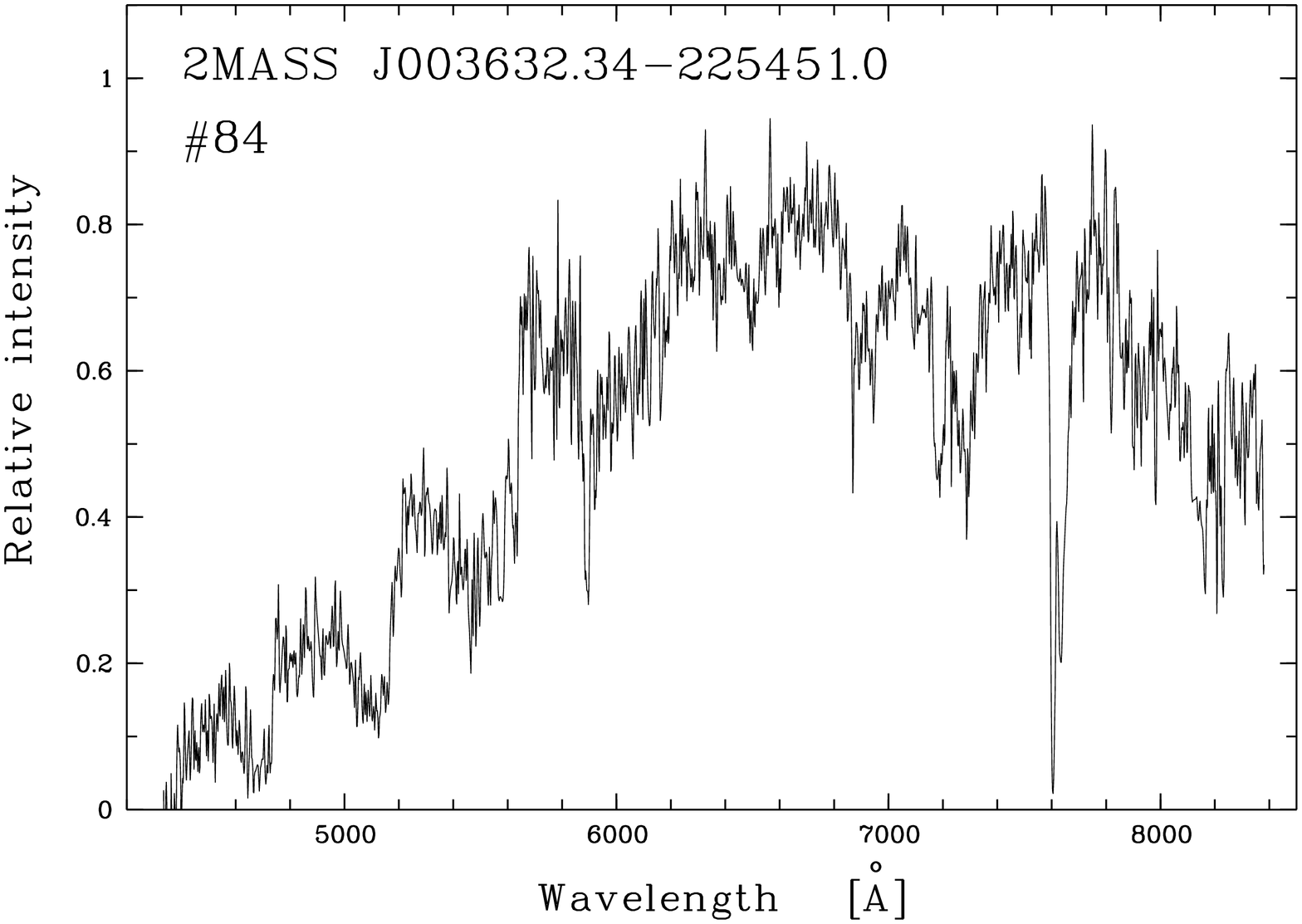}}}
\resizebox{8cm}{!}{\rotatebox{-90}{\includegraphics{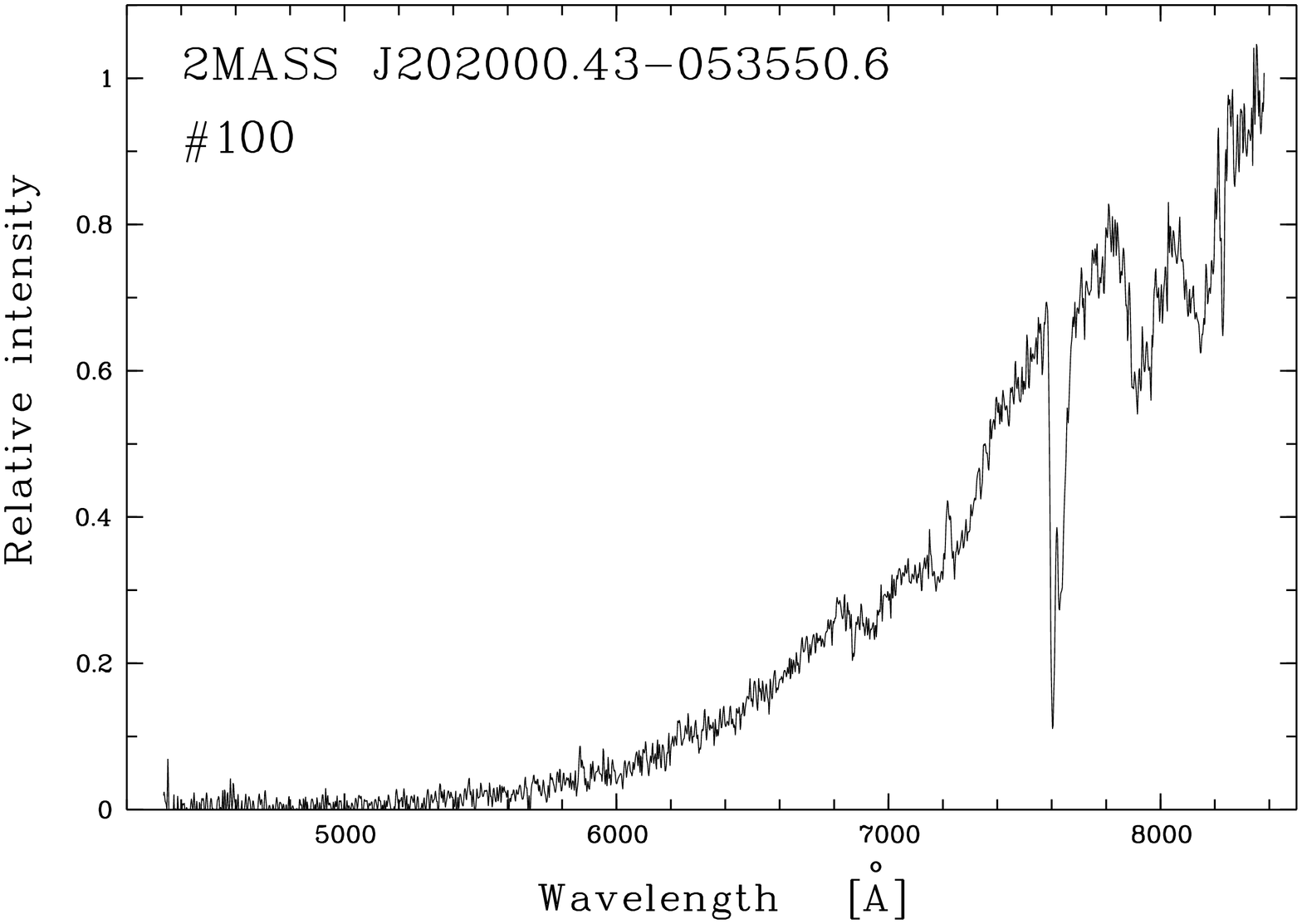}}}

\caption[]{ Three typical spectra of our sample. {\it Top panel\,:} Object \#89,
a relatively bright source, $R$\,$=$\,$14.3$, with $J$\,$-$\,$K_{\rm s}$\,$=$\,$1.47$.
{\it Middle panel\,:} Object \#84, the faintest source in $K_{\rm s}$ with 
$K_{\rm s}$\,$=$\,$14.14$,
$R$\,$=$\,$17.4$, $J$\,$-$\,$K_{\rm s}$\,$=$\,$1.33$.
{\it Bottom panel\,:} object \#100, a very red source in $J$\,$-$\,$K_{\rm s}$, 
with $R$\,$=$\,$16.0$, $K_{\rm s}$\,$=$\,$8.61$, $J$\,$-$\,$K_{\rm s}$\,$=$\,$3.4$\,.
 }
\label{fig1}
\end{figure}

\subsection{Objects with IRAS counterparts}

 For all objects in Table\,\ref{table1}, we searched for IRAS counterparts and found eight cases,
 with six sources from the IRAS point-source catalogue (Beichman et al. \cite{beichman88})
 and two additional sources from the IRAS faint source catalogue (Moshir et al. \cite{moshir89}).
 The results are in Table\,\ref{table2}. In this table, one finds the names and the galactic 
 coordinates, the IRAS 12\,$\mu$m flux, $K_{\rm s}$ and $J-K_{\rm s}$ from 2MASS, the 
 distances to the Sun $d$ and the heights $Z$ above the Galactic plane
(see below for details on the derivations of $d$ and $Z$).
 There are five objects with $J-K_{\rm s} > 3$ that have counterparts with 12-$\mu$m  
 fluxes ($f_{\rm 12}$) between $\sim$~0.4 and 14~Jy. The three
 other objects with smaller $J-K_{\rm s}$  generally have lower $f_{\rm 12}$.
 No star with $K_{\rm s}$\,$\ga$\,8.8 was found to be detected by IRAS.

In Table~\ref{table2}, one can see that the distances from the Galactic plane are between 1.7 
to  6\,kpc. This indicates that these
sources do not belong to the Galactic thin disc, since the scale height
of disc carbon stars is only 200\,pc (Claussen et al. \cite{claussen87}, 
Groenewegen et al. \cite{groe92}; Bergeat et al. \cite{bergeat02}). 
It would be important to measure the kinematics of these stars. This would allow us 
to see whether they might belong to the thick disc (although this disc is rather 
old for dusty AGB C stars to be present) or are members of a stream in the 
Galactic halo, especially the Sgr Stream. 
 
These dusty C stars deserve attention because their $Z$ are 
larger than most of the AGB C stars known. Compared to C stars of the Galactic disc,
 their circumstellar envelopes will have a different environment, probably with a  
 lower interstellar density, and  weaker ultraviolet ambient radiation from hot 
stars of the Galactic disc. Therefore, it is quite possible that the extension
of the CO emitting envelope will be unusually large. For a typical mass-loss rate
of 4\,$\times$\,10$^{-6}$ \Msolyr (the median mass loss rate in Table\,\ref{table3}, see below), 
the CO photodissociation radius for AGB C stars in the disc is about 0.05 pc 
(from equation 7.9 of Olofsson \cite{olofsson04}). If it is twice larger for our 
high $Z$ stars ($\sim$ 0.1\,pc), this corresponds, at a median distance 
of $\sim$\,10~kpc,  to an angular diameter of 4$''$, which should be easily 
detected and resolved by future millimeter interferometers like ALMA. 

\subsection{Mass-loss rates}

For stars with available IRAS 12-$\mu$m fluxes, it is interesting to see what could be
their mass-loss rates, even if it is only estimated approximatively. In 
a study of Galactic, Mira-type carbon stars, Whitelock et al.\,(\cite{whitelock06}) 
derived a tight correlation between the $K$-[12] colour and
the total mass-loss rate $\Mdot$. In this correlation, $K$ is  the mean  magnitude
(averaged over time), and [12] is the IRAS magnitude defined as 
$-$2.5\,log$_{\rm 10}$($f_{\rm 12}/28.3$) 
where $f_{\rm 12}$ is in Jy. For a given $K$-[12], $\Mdot$ is given by:

\smallskip

\noindent ${\rm log}$\,$\Mdot = -7.668  +0.7305$\,$(K-{\rm [12]})$

$-5.398 \times 10^{-2}$\,$(K-{\rm [12]})^{2} +1.343 \times 10^{-3}$\,$(K-{\rm [12]})^{3}$\,\,\,.

\smallskip

 For establishing this correlation, Whitelock et al. calculated $\Mdot$  by using
the relation of Jura\,(\cite{jura87}) where $\Mdot$ depends on
the  IRAS 60-$\mu$m flux, on the distance of the object, on the wind velocity, and 
on several other parameters of the circumstellar shells (see 
Eq.\,(11) in Whitelock et al. 2006). In particular, the Jura relation assumes 
a gas-to-dust mass ratio of 220 (Jura \cite{jura86}).  

Whitelock et al.\,(\cite{whitelock06}) noted that the correlation between
$\Mdot$ and $K$-[12] established for Galactic C stars is also verified for the 
carbon Miras of the Large Magellanic Cloud. In this case, the LMC C Miras were 
taken from the work
by Van Loon et al. (\cite{vanloon99}) who adopt  a gas-to-dust ratio of 500.  Consequently, 
the $K$-[12] versus
$\Mdot$ relation holds, provided one supposes that there is less dust produced 
in the winds of C stars at lower metallicity, which is also favoured by recent
Spitzer observations of C stars in external galaxies (e.g. Matsuura et al. 
\cite{matsuura07}).
 
Although we do not know the metallicity of our sources, we  suppose that 
the Whitelock et al. relation can be applied to them. 
We must  also assume that our stars are Mira variables and that the 2MASS 
$K_{\rm s}$ magnitude is not very different from  $K$ averaged over time (if the
difference is $\sim$\,0.5 mag, it will change $\Mdot$ by only 40 percent).
With these assumptions, we obtain the results of Table~\ref{table3}, in which
$\Mdot$ is estimated  from the $K$-[12] colour  and  is uncertain by about a 
factor of 2. The first six objects are from this work. The following ten objects, 
designated by their IRAS names, are objects with $J$\,$-$\,$K_{\rm s}$\,$>$\,3 
extracted from Table\,2 of Mauron et al.\,(\cite{mauron07b}).
 In  Table~\ref{table3}, we have
not included Objects  \#97  and \#101 which are relatively blue with 
small $J$\,$-$\,$K_{\rm s}$ indices and $K$-[12] in the range of 2.0-2.3, making the
assumption of Mira-type variability more questionable. 
The distances and height above the Galactic plane are also mentioned in  Table~\ref{table3}, but
it must be emphasized that the $\Mdot$ values that we obtained above  do not depend 
on  $d$ or $Z$ since they are directly derived from $K$\,$-$\,[12].

Table\,\ref{table3} shows that, presently, we have identified 16 C stars out of 
the Galactic plane, with 6 coming from this work. Several objects have $\Mdot$ 
of the order of 4\,$\times$\,10$^{-6}$ \Msolyr or even larger. 
It can be noted that both the distances and the mass-loss rates 
have been derived homogeneously for all of them, so that useful comparisons can be 
made between objects. For example,  one may consider \object{IRAS 12560+1656} as a reference object 
since it was detected in CO with such a low and unusual expansion velocity of 
3\,km\,s$^{-1}$ probably due to its low metallicity. Then, a look at Table\,\ref{table3} shows that
six objects have larger  $\Mdot$, together with smaller distances $d$. These six
cases should be easier to detect in CO than IRAS\,12560+1656.

  \begin{table*}[!ht]
        \caption[]{Objects of Table\,1 with IRAS counterparts}
	\label{table2}
        \begin{center}
        \begin{tabular}{rrcrrrrrrrl}
        \noalign{\smallskip}
        \hline
        \hline
        \noalign{\smallskip}
  \#& IRAS name & 2MASS  name & $l$\hspace{5mm}  & $b$\hspace{5mm} & $f_{12}$ & $K$\,\,\,   & $J-K$  & 
  $d$\hspace{2mm} &
  $Z$\,\,\,     & Note\\
          &  &             &  (deg)\,\,\,\, & (deg)\,\,\,\, &  (Jy)    & (mag) &  (mag) & (kpc)& (kpc) &\\
        \noalign{\smallskip}
        \hline
        \noalign{\smallskip}

 \#85&  02164$+$3537  &  J021926.96$+$355058.9  & 142.299 &$-$23.690 & 0.41  &  8.52 & 3.20 &  13.3 & $-$5.3 &\\
 \#86& F02214$+$3716  &  J022431.98$+$372933.0  & 142.664 &$-$21.786 & 0.16  &  8.75 & 2.74 &  16.2 & $-$6.0 & 1\\
 \#87&  02360$+$3442  &  J023904.88$+$345507.5  & 146.697 &$-$22.937 & 1.16  &  8.39 & 3.89 &  11.0 & $-$4.3 &\\
 \#93&  16339$-$0317  &  J163631.69$-$032337.4  &  12.734 &$+$27.794 &14.57  &  6.10 & 3.91 &   3.7 & $+$1.7 &\\
 \#96&  18120$+$4530  &  J181329.45$+$453117.5  &  73.053 &$+$25.349 & 7.86  &  6.71 & 3.81 &   5.1 & $+$2.2 &\\
 \#97&  18493$+$6213  &  J184950.90$+$621725.4  &  92.339 &$+$23.989 & 0.23  &  7.49 & 1.31 &   7.4 & $+$3.0 &\\
\#100&  20173$-$0545  &  J202000.43$-$053550.6  &  38.153 &$-$22.230 & 0.45  &  8.61 & 3.43 &  13.3 & $-$5.0 &\\
\#101& F20458$+$1015  &  J204817.91$+$102638.7  &  56.823 &$-$20.154 & 0.17  &  7.83 & 2.23 &  11.7 & $-$4.0 & 1\\
       
   \noalign{\smallskip}
   \hline
\end{tabular}
\end{center}
{\small Notes: (1) the IRAS name and flux come from the IRAS Faint Source Catalogue} 
\end{table*}

  \begin{table}[!ht]
        \caption[]{Halo carbon stars with predicted high mass-loss 
	 rates \Mdot\,\, in units of 10$^{-6}$\,\Msolyr.} 
	 \label{table3}
	
        \begin{center}	 
        \begin{tabular}{crrrr}
	\hline
	\hline
        \noalign{\smallskip}
   Name & $K$-[12] & $d$\,\,\,   & $Z$\,\,\,   & \Mdot \\
        &  (mag)   & (kpc) & (kpc)  & \\
        \noalign{\smallskip}
        \hline
        \noalign{\smallskip}

\#85   & 3.92 & 13 & $-$5.3 & 3.0\\
\#86   & 3.13 & 16 & $-$6.0 & 1.5\\
\#87  & 4.92 & 11 & $-$4.3 & 6.0\\
\#93  & 5.38 &  3.7 & $+$1.7 & 8.0\\ 
\#96  & 5.32 &  5.0 & $+$2.2 & 8.0\\
\#100 & 4.11 & 13 & $-$5.0 & 3.5\\

\noalign{\smallskip}
\noalign{\smallskip}

01582$+$0931   & 4.43 &  7 &  $-$5.5   & 4.0\\
03242$+$1429   & 4.01 & 13 &  $-$7.5   & 3.0\\
03582$+$1819   & 5.68 & 17 &  $-$7.5   & 10.0\\
04188$+$0122   & 4.11 &  6 &  $-$3.5   & 3.5\\
08427$+$0338   & 4.65 &  6 &  $+$2.5   & 5.0\\
08546$+$1732   & 6.47 & 36 &  $+$21.0  &14.0\\
12560$+$1656   & 3.91 & 11 &  $+$11.0  & 3.0\\
20005$+$7737   & 4.31 & 14 &  $+$5.5   & 4.0\\
20176$-$1458   & 4.21 & 14 &  $-$6.0   & 3.5\\
21064$+$7749   & 4.41 &  4 &  $+$1.3   & 4.0\\

   \noalign{\smallskip}
   \hline
\end{tabular}
\end{center}
\end{table}

\subsection{Distances and positions in the halo}

\begin{figure}[!ht]
\resizebox{8cm}{!}{\rotatebox{-90}{\includegraphics{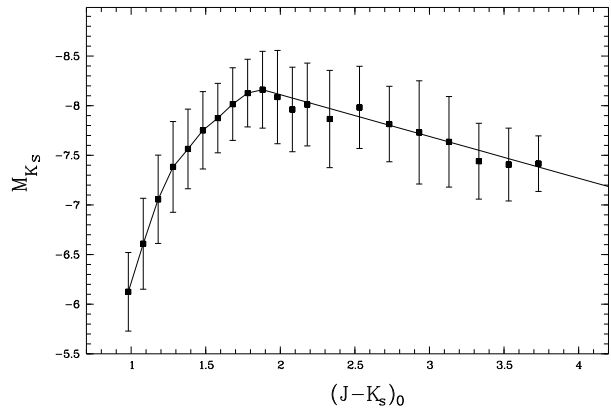}}}

\caption[]{ Absolute magnitudes of the LMC carbon stars in the $K_{\rm s}$ band
 as a function of $(J-K)_{\rm 0}$ (solid line). The filled squares indicate averaged 
 $M_{\rm K}$ from a given bin in colour, and the error bars indicate 
 the dispersion in each bin ($\pm$ $\sigma$ is drawn).}
\label{fig2}
\end{figure}

\begin{figure}[!ht]
\resizebox{8cm}{!}{\rotatebox{-90}{\includegraphics{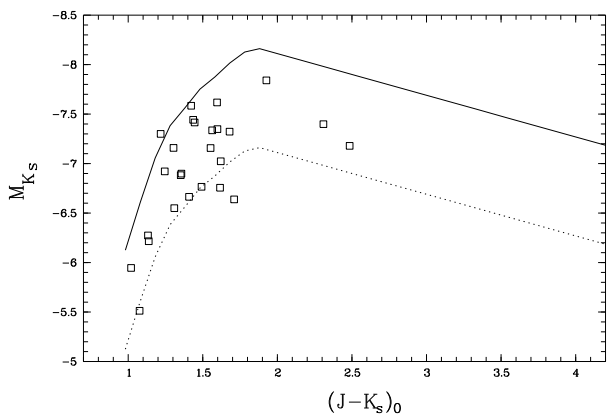}}}

\caption[]{Comparison of LMC and Sgr absolute magnitudes as a function of colour. 
The LMC is represented by the solid line as in Fig.\,2. The open squares
are 26 stars of Sgr. The dotted line is the LMC line shifted down by 1.0 mag. On
average, the Sgr stars appear to be less luminous than LMC ones 
by $\sim$\,0.5\,magnitude (see text). }
\label{fig3}
\end{figure}

 The distances of our new C stars  are a priori difficult to estimate because 
 they are field stars. We follow the method described by Mauron et al. (\cite{mauron04}) 
 and adopted in our previous works on halo C stars, but we provide here more
detailed information. Only near infrared data from 2MASS ($J$ and $K_{\rm s}$) 
are used because they are less affected than optical data by interstellar or 
circumstellar extinction,  or by variability. The $B$ or $R$ magnitudes 
are not used.
  
 The first step in the method consists in
obtaining $K_{\rm s}$-band absolute magnitudes for C stars of the LMC. The catalogue 
of LMC optically-selected C stars of Kontizas et al. (\cite{kontizas01}) provides a first sample
of C stars to consider.  In particular, this sample contains $\sim 5000$ 
C stars with colours between $J-K_{\rm s}$\,$=$\,1.0 and $J-K_{\rm s}$\,$=$\,2.0\,.  
 A second sample of C stars is obtained by considering
the regions labelled J and K in the 2MASS colour-magnitude diagrams, as
defined by Nikolaev \& Weinberg (\cite{niko00}, hereafter NW00; see their Fig.~3).
Region J contains $\sim$\,7000 objects with  1.4\,$<$\,$J$$-$$K_{\rm s}$\,$<$\,2.0, and 
region K contains 1600 objects (often called obscured AGB stars)  
with 2.0\,$<$\,$J$$-$$K_{\rm s}$$<$\,4.0. According to NW00 (their Fig.~8),
the objects located in these two regions are in a large majority, but not exclusively,
carbon stars.  Several works in the past have found OH/IR stars
in region K. For example, Wood et al. (\cite{wood92}) searched for OH emission on  54 very red objects
and found it for 6 objects.  Five of them  have near infrared photometry, and two are in region K.
Whitelock et al. (\cite{whitelock03}) also considered obscured C-rich and O-rich  AGB stars 
in the LMC: of the 19 O-rich objects with available $JHK$ photometry, 7 are in region K.
But more recent work based on $JHK$, 8$\mu$m photometry, and $Spitzer$ infrared spectroscopy  
suggest that the overwhelming majority of AGB stars are carbon 
stars (Buchanan et al. \cite{buchanan06}).  Also, among the 250 brightest and compact 
8\,$\mu m$ sources in the LMC, Kastner et al. (\cite{kastner07}) find
that there are 7 times more C-rich AGB stars than O-rich AGB stars. By considering the
2MASS photometry of these 250 sources (from their Table~1), we find that region K contains
41  C AGB stars, 5 O-rich AGB stars, 3 HII regions, and no red supergiants. Although
these numbers are small compared to the total population of region K, 
they constitute an indication  that region K is very probably dominated by carbon stars. 
Consequently, we assume that this region K is suitable for
determining the $K_{\rm s}$-band absolute magnitude of  very red LMC C stars.

We adopt a mean LMC interstellar  extinction $E(B-V)$\,$=$\,$0.13$ 
(Van den Bergh \cite{vandenbergh00}), and corresponding extinction for the 
2MASS photometry are  $A_{\rm Ks}$\,$=$\,0.367\,$E(B-V)$ and 
$E(J-K_{\rm s})$\,=\,0.535\,$E(B-V)$. A distance modulus $m-M$\,$=$\,$18.50$ is assumed 
(Van den Bergh \cite{vandenbergh00}; see also Fig.~8 of  
Clementini et al. \cite{clementini03}). 

The resulting absolute magnitude
for LMC C stars $M_{\rm K}$  is calculated for bins in the $(J-K_{\rm s})$$_{0}$ 
colour and  is shown in Fig.\,\ref{fig2}. The number of stars in the bins is in the range  50 to
 1500. In the colour  interval where the two samples
mentioned above overlap, the agreement is very good and their average has been adopted.
The error bars represent the dispersion of $M_{\rm K}$ in each bin ($\pm$ $\sigma$ is drawn). 
These dispersions are about 0.40\,$\pm 0.05$\,mag. 

The second point that we consider is how C stars of the Sgr dwarf galaxy 
compare in $M_{\rm K}$ with those of the LMC. No complete catalogue of C stars of Sgr
is available, but Whitelock et al. (\cite{whitelock99}) have published a list of 26 C stars
which are radial velocity members of Sgr. The 2MASS photometry of these stars is
available. The interstellar reddening was found for each star from the maps of
Schlegel et al.\,(\cite{schlegel98}), with $E(B-V)$ found to be in the range 0.09 to 0.18 mag. 
A distance modulus of Sgr $m-M$\,$=$\,17.0 was adopted, which is between the
16.9 taken by Majewski et al. (\cite{majewski03}) and 17.18 adopted by 
Whitelock et al. (\cite{whitelock99}). This allows us to plot in Fig.\,\ref{fig3} 
the $M_{\rm Ks}$ of these stars versus their $(J-K_{\rm s})_{0}$, and
to compare them to the LMC C stars. This figure shows that  the Sgr stars are
on average 0.5\, magnitude less  luminous in $K_{\rm s}$ than the LMC stars.
 Beyond  $J$\,$-$\,$K_{\rm s}$\,$\sim$\,2.5, there are no Sgr points, and we shall assume that
the shift of 0.5\,mag between LMC and Sgr is still valid in this region.
[One can note that this luminosity shift of 0.5 magnitude between LMC and Sgr C stars is 
 significantly reduced (to $\approx$\,0.16) if one adopts (1) a closer distance for the LMC,
 e.g. $m-M$\,$=$18.39 from van Leeuwen et al. (\cite{vanleeuwen07}), or  $m-M$\,$=$18.34 
 from An et al. (\cite{an07}), and (2)
 a greater distance for Sgr, e.g. $m-M$\,$=$17.18 (mentioned above). In this case,
 the distances of our objects derived below would have to be increased by $\sim$ 10\%].  

A remark by the referee was that for stars as red as $J-K_{\rm s}$\,$\sim$\,3, circumstellar
extinction will be significant and should be taken into account when determining
distances. To our knowledge, there is no way to measure the circumstellar  extinction
of a dusty C star, especially when only photometry is available. But its effect 
is included in the absolute magnitude when we have determined $M_{\rm K}$  
for LMC stars: in Fig.~\ref{fig2}, the  $K_{\rm s}$-band luminosity decreases beyond 
$J-K_{\rm s}$\,$\sim$\,2 and this is due to circumstellar dust extinction 
(NW00).

Thus, we assume that our stars are identical  to those of the Sgr dwarf 
galaxy and have similar absolute magnitude  in the $K_{\rm s}$-band for a
given $J-K_{\rm s}$ colour. The distances of our objects are determined by 
using the LMC $M_{\rm Ks}$ values $plus$ 0.5\, magnitude.  Interstellar absorption 
was taken into account although this effect is small, since the largest $E(B-V)$ is 
0.38\,mag. The results are in Table\,\ref{table4}. This table lists the running number,
the galactic coordinates $l$ and $b$ in degrees, the interstellar colour excess 
measured in the direction of the object, $E_{\rm B-V}$ from Schlegel et al. maps, 
the dereddened $J-K_{\rm s}$ colour, the adopted $K_{\rm s}$-band absolute magnitude, 
the distance $d$ to the Sun in kpc, the galactocentric $XYZ$ coordinates in kpc, and 
the distance $D$ to the Sgr Stream average plane, in kpc.
Given the above considerations, the relative uncertainty on these distances is 
at least of the order of $\pm$20\% ($\pm$\,1\,$\sigma$). In particular, stars having
blue colours  with $(J-K_{\rm s})$$_{0}$\,$\sim$\,$1.0-1.2$ like \#94 and \#98 have
very uncertain distances because there are few points at this colour in Fig.~\ref{fig3}, 
and the slope of  $M_{\rm K}$  with colour is very steep.

 There are four objects with their distances  to the Sun 
below 10~kpc, 6 with 10\,$<$\,$d$\,$<$\,20\,kpc,  5 with 20\,$<$\,$d$\,$<$\,100\,kpc, 
and 3 with $d$\,$>$\,100\,kpc. The two objects with the greatest distances are
\#84 ($d$\,$\approx$\,165\,kpc) and \#90 ($d$\,$\approx$\,150\,kpc).
 We checked that  these distant stars are not carbon dwarfs, especially for the
most distant ones (\#84, \#90, \#94).  Although the USNO-B1.0 catalogue gives 
a non-zero proper motion for object \#84 ($\mu(\alpha)$\,$=$\,$-$12\,mas\,yr$^{-1}$, 
$\mu(\delta)$\,$=$\,$-8$\,mas\,yr$^{-1}$), 
this motion is too small 
to be considered with certainty, as was shown for several similar cases in Mauron
et al.\,(\cite{mauron05}). The USNO-B1.0 gives zero proper motion  for the
other two distant objects \#90 and \#95. An additional suggestion that these objects 
are AGB stars comes from their $JHK_{\rm s}$ colours, which put them outside the
locus of carbon dwarfs in  the colour-colour diagrams shown by Totten 
et al.\,(\cite{tiw00}) or by Downes et al. (\cite{downes04}).

A check on the distances adopted in this work is possible for a few objects.  
Object \#97 (IZ Dra)  is a Mira variable with a known period (312\,days). Then,
one can apply the period luminosity $P$\,-$K$ of Feast et al. (\cite{feast89}) if one
assumes that the 2MASS $K_{\rm s}$ magnitude is not too different from the 
time-averaged magnitude. The Feast et al. relation is $K$\,$=$\,$-3.3$~log\,$P$\,+18.98
for Miras of the LMC, for which we adopt a distance modulus of 18.50. Therefore,
the absolute magnitude $M_{\rm K}$ is given by 
$M_{\rm K}$\,$=$\,$-3.3$\,log\,$P$\,$+$\,$0.48$. Then, one obtains 
$d$\,$=$\,11\,kpc for IZ Dra, which is more than our estimate (7.4\,kpc), but
the discrepancy is nearly acceptable.

Our estimates of distances can also be evaluated by considering
two objects in Table~\ref{table3}, \object{IRAS 08546+1732} and \object{IRAS 12560+1656}. Their periods
are 390\,days (see Groenewegen et al. \cite{groe97}). For the first object, 
$K_{\rm s}$\,$=$\,10.71
and the distance derived from the $P$-$K$ relation is 57\,kpc,
whereas our method gives 36 kpc.
For the second, $K_{\rm s}$\,$=$\,7.82 and 15\,kpc  is obtained, 
whereas our method gives 11\,kpc. Therefore, the distances adopted
in this work (in Tables 2, 3, and 4)  appear a little too small but still 
reasonably  correct given the uncertainties mentioned above. 
At least for the reddest objects, a
$K$-band monitoring would help to establish their variability and, if they are Miras,
to determine  these distances more precisely.

\begin{figure*}[!ht]
      \resizebox{\hsize}{!}{
      {\rotatebox{-90}{\includegraphics{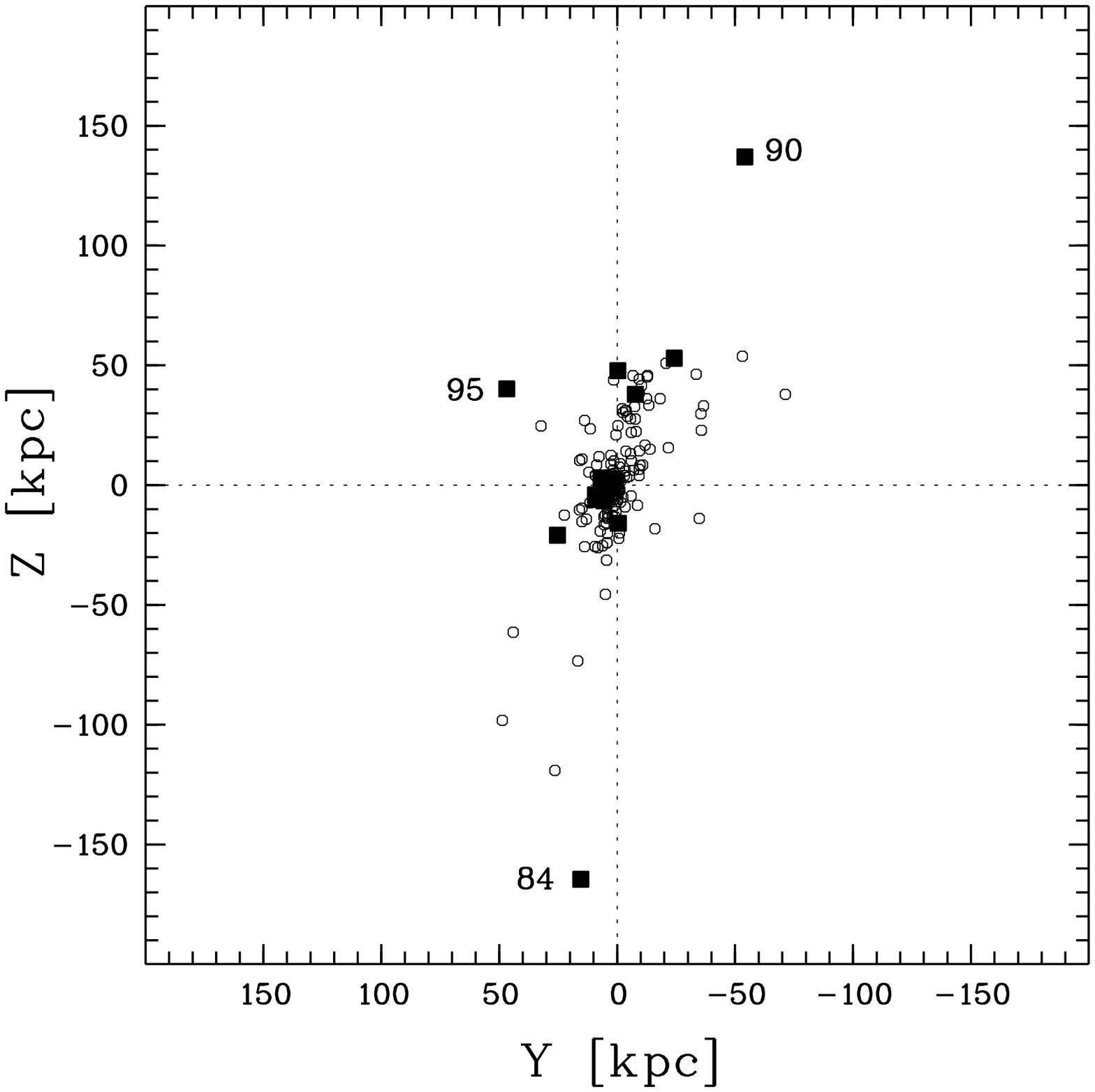}}}
      {\rotatebox{-90}{\includegraphics{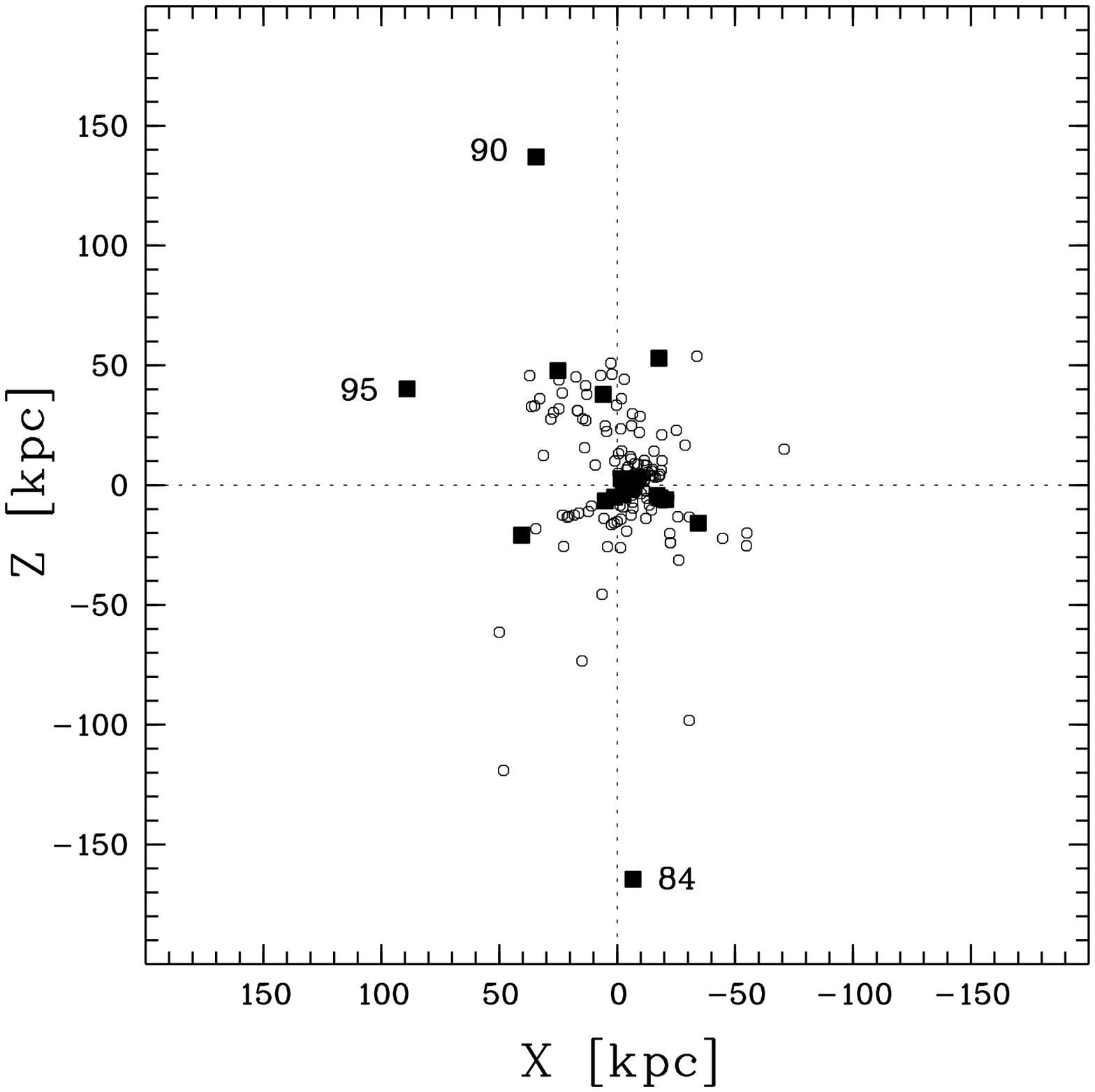}}} }

      \caption[]{ Plot in galactocentric $XYZ$ coordinates
       of the observed halo C stars, indicated by filled squares.
      The XYZ system is such that the Sun is at $X$\,$=$\,$-8.5$\,kpc, $Y$\,$=$\,$0$, $Z$\,$=$\,$0$; 
      the $Y$-axis is positive towards  $l$\,$=$\,$+90$$\degr$ and  the $Z$ axis is positive to 
      $b$\,$=$\,$+90$$\degr$. In the left panel, the Sgr Stream is seen close 
      to edge-on, slightly inclined
      clockwise with respect to the $Z$ axis. It is seen face-on in the right panel.}
 
\label{fig4}
\end{figure*}

We also calculated for Table\,\ref{table4} the galactocentric $XYZ$ coordinates of each object,
as well as their distance $D$ to the main plane of the Sgr Stream as given by 
Newberg et al. (\cite{newberg03}). Plots of the $YZ$ and $XZ$ are given in Fig.~\ref{fig4}. 
Note that their sizes are 400\,$\times$\,400~kpc in order to emphasize
the position of very distant objects. In the $YZ$ panel, the Sgr stream is seen 
nearly edge-on, slightly inclined clockwise from the $Z$ axis. 

The $YZ$ plot shows that objects  \#84 and \#90 are not very far
from the Stream plane, only  by 22 kpc (see Table\,\ref{table4}), which represents an angle 
of  about 8 degrees as seen from the Galactic centre.  Object \#84 suggests that the 
Sgr stream  extends to $\sim$\,150\,kpc toward the South pole if our
distances are correct, and the qualitative agreement of these observations 
with the model of Law et al. (\cite{law05}) concerning the existence of a 
southern warp is reinforced (see Fig.~4 of Mauron et al. \cite{mauron05}).
Object \#90 is, in contrast, far toward the North Galactic pole where no warp is
predicted. Finally, object \#95 is located by $\sim$\,50\,kpc from the Sgr plane and is isolated in 
this region of the halo, but this may be due in part to its galactic coordinates, $l = 25.6$\degr, 
$b = +20.3$\degr. At this low latitude, it becomes more difficult to find faint 
($K_{\rm s} \sim 13 $) carbon stars, because numerous interlopers, such as young stars, come
into the candidate list and this decreases the probability of discovering a new faint C star.

\begin{table*}[ht]
        \caption[]{Properties of the halo C stars.}
	\label{table4}
        \begin{flushleft}
        \begin{tabular}{rrrrcrrrrrrr}
        \noalign{\smallskip}
        \hline
        \hline
        \noalign{\smallskip}
No.& $l$\,\,\,\, & $b$\,\,\,\, & $E_{\rm B-V}$ &  $(J-K_{\rm s})_0$ & $K_{\rm s}$ &  $M_{\rm Ks}$ & $d$\,\,\,\, & $X$ & $Y$ & $Z$ & $D$\\
        \noalign{\smallskip}
        \hline
        \noalign{\smallskip}
	
84  &  83.435 & $-$84.602 & 0.018  & 1.319 &  14.144 & $-$6.95 &  165 &   $-$6.7 &   15.4  &$-$164.5 &$-$22.7\\
85  & 142.299 & $-$23.690 & 0.070  & 3.164 &   8.518 & $-$7.12 &   13 &  $-$18.1 &    7.4  &  $-$5.3 &    7.4\\
86  & 142.664 & $-$21.786 & 0.038  & 2.716 &   8.753 & $-$7.31 &   16 &  $-$20.5 &    9.1  &  $-$6.0 &    9.0\\
87  & 146.697 & $-$22.937 & 0.055  & 3.849 &   8.386 & $-$6.83 &   11 &  $-$16.9 &    5.5  &  $-$4.3 &    5.7\\
88  & 181.024 & $-$31.584 & 0.276  & 2.495 &  10.114 & $-$7.40 &   30 &  $-$34.4 &  $-$0.5 & $-$15.9 & $-$1.7\\
89  & 249.238 & $+$63.941 & 0.037  & 1.451 &  11.673 & $-$7.20 &   59 &  $-$17.7 & $-$24.2 &    53.0 & $-$9.8\\
90  & 308.423 & $+$63.264 & 0.026  & 1.617 &  13.513 & $-$7.43 &  153 &     34.4 & $-$54.1 &   137.0 &$-$22.5\\
91  & 331.973 & $+$66.727 & 0.031  & 1.678 &  10.575 & $-$7.51 &   41 &      5.9 &  $-$7.7 &    37.9 &    1.3\\
92  & 359.497 & $+$54.876 & 0.039  & 2.267 &  11.349 & $-$7.50 &   58 &     25.1 &  $-$0.3 &    47.8 &    9.5\\
93  &  12.734 & $+$27.794 & 0.388  & 3.698 &  6.098  & $-$6.90 &  3.7 &   $-$5.3 &     0.7 &     1.7 &    1.7\\
94  &  22.082 & $+$20.023 & 0.315  & 1.141 &  8.212  & $-$6.38 &  7.9 &   $-$1.7 &     2.8 &     2.7 &    3.7\\
95  &  25.619 & $+$20.363 & 0.126  & 1.318 & 13.409  & $-$6.95 &  115 &     89.1 &    46.8 &    40.2 &   49.3\\
96  &  73.053 & $+$25.349 & 0.033  & 3.791 &   6.710 & $-$6.86 &  5.1 &   $-$7.1 &     4.4 &     2.2 &    5.5\\
97  &  92.339 & $+$23.989 & 0.055  & 1.280 &   7.495 & $-$6.88 &  7.4 &   $-$8.8 &     6.8 &     3.0 &    8.1\\
98  &  27.282 & $-$20.741 & 0.264  & 1.162 &  12.478 & $-$6.47 &   59 &     40.5 &    25.3 & $-$20.9 &   17.3\\
99  &  22.189 & $-$24.332 & 0.164  & 1.333 &   9.110 & $-$6.98 &   16 &      5.0 &     5.5 &  $-$6.6 &    3.7\\
100 &  38.153 & $-$22.230 & 0.074  & 3.391 &   8.613 & $-$7.02 &   13 &      1.1 &     7.6 &  $-$5.0 &    6.3\\
101 &  56.823 & $-$20.154 & 0.078  & 2.187 &   7.830 & $-$7.53 &   12 &   $-$2.5 &     9.2 &  $-$4.0 &    8.3\\

   \noalign{\smallskip}
   \hline
\end{tabular}
\end{flushleft}
\end{table*}

\section{Conclusions and suggestions} 

\hspace{3mm} In order to find new  cool carbon stars in the halo, we have carried out
 spectroscopy of 58 candidates  selected in the 2MASS catalogue through 
  their near-infrared colours and $|b|$\,$>$20\degr\,. Eighteen new cool carbon stars
were discovered. The spectra of these rare stars generally show strong C$_{2}$ and CN 
bands typical of AGB carbon stars, with H$\alpha$ often in emission, indicating 
pulsating atmospheres.  Of these 18 new C stars, 10 have IRAS counterparts with detected 
fluxes  at 12\,$\mu$m. For 6 of them, the 12$\mu m$ infrared excess is large, and
their heights above  the Galactic plane are  between 1.7 to 6 kpc. After adding 
these 6 dusty C stars to previously known  cases, a list of 16 similar cases is 
established (Table\,3). This list contains  one object (IRAS 12560+1656) for which  
CO emission was detected (Groenewegen et al. \cite{groe97}). Its CO line shows a unique 
expansion velocity of only 3 km\,s$^{-1}$.
 
 \vspace{2mm}

\,\, Mass-loss rates \Mdot\,\,  have been homogeneously estimated 
by assuming that the relation between
$K$\,$-$\,[12]\, and \Mdot \,\, established by Whitelock et al. for Galactic and LMC 
Mira-type C stars can be applied to the 16 objects. It is found that the \Mdot\,\, estimates 
are of the order of 4\,$\times$\,10$^{-6}$\,\Msolyr , with a range of a factor of\,\,3.
 These strong mass-loss rates suggest that
further investigation of these halo AGB stars should be carried out. Several of them
might be metal-poor and similar to IRAS\,12560+1656, helping us to understand 
 mass loss of AGB C stars with low metallicity better.

\vspace{2mm}

\,\, Although distances are uncertain by at least $\pm 20$\%  in relative error 
($\pm$\,1\,$\sigma$), taking them at face value shows that three objects might be as far as 
 100~kpc from the Sun, with two reaching unprecedented distances of  $\sim$\,150\,kpc. 
One object (\#84) is located  at 165\,kpc toward the South Galactic pole 
and is relatively close, given uncertainties, to a region of the halo where 4 C stars have 
been  discovered previously, and where a loop of the Sgr Stream is predicted.
In the future, it would be interesting to measure the radial velocity of these 5 
stars and to see if this southern warp can be detected with fainter but much more numerous
M giants in similar regions in the sky and  at similar velocities.

\begin{acknowledgements}

The author thanks the anonymous referee for useful remarks, and 
Pauline Mc\,Nish and Joli Adams for correcting the English of the manuscript.
The author also thanks the staff of Observatoire de Haute-Provence, which is supported
by the French Centre National de Recherche Scientifique.
The use of  the Two Micron All Sky Survey is acknowledged.
2MASS is a joint project of the University of Massachusetts and  the Infrared
Processing and Analysis Center / California Institute of Technology, funded
by  the National Aeronautics and Space Administration (NASA) and the
 National Science Foundation (NSF).  Finally, this
work benefitted from using the CDS database in Strasbourg (France).

\end{acknowledgements}


 \appendix
 \section{Spectra of halo C stars}
 
 This appendix presents  in a first part the spectra of 15 stars  discussed in this paper. 
 The three spectra for objects \#84, \#89, and \#100 are in Fig.~\ref{fig1}.  The 23 spectra that 
 follow are low-resolution spectra that were needed for confirming candidate carbon stars 
 found in the First Byurakan Survey (FBS) and listed in Gigoyan et al. (\cite{gigoyan01}) 
 or in Mauron et al. (\cite{mauron07a}). The names of the sources and the J2000 coordinates 
 are given in the plots. There are 10 FBS objects, originally  carbon star candidates, that 
 were found to be M-type dwarfs. They are listed in the following table.

  \begin{table}[!ht]
        \caption[]{ Carbon star candidates from the FBS found to be M dwarfs}
	\label{tablea1}
	 \begin{center}	 
        \begin{tabular}{lr}
	\hline
	\hline
        \noalign{\smallskip}
   Name & Coordinates J2000\\
        \noalign{\smallskip}
        \hline
        \noalign{\smallskip}

\object{FBS 0300$-$030}  & 03 03 03.2 $-$02 51 12\\
\object{FBS 0310$+$016}  & 03 12 39.9 $+$01 51 42\\
\object{FBS 0310$+$043}  & 03 13 19.5 $+$04 30 25\\
\object{FBS 0318$+$048A} & 03 21 03.7 $+$05 03 29\\
\object{FBS 0318$+$048B} & 03 21 07.2 $+$05 02 33\\
\object{FBS 0322$+$041}  & 03 25 27.2 $+$04 16 01\\
\object{FBS 1058$+$083}  & 11 01 16.5 $+$08 02 31\\
\object{FBS 1435$-$092}  & 14 37 46.4 $-$09 26 59\\
\object{FBS 2143$-$081}  & 21 46 38.0 $-$07 53 11\\
\object{FBS 2144$-$089}  & 21 46 40.6 $-$08 41 05\\

   \noalign{\smallskip}
   \hline
\end{tabular}
\end{center}
\end{table}
 
 \smallskip
 \smallskip
 \smallskip
  \begin{figure*}
  \caption[]{Spectra of new carbon stars in the halo}
  \resizebox{8.5cm}{!}{\rotatebox{-90}{\includegraphics{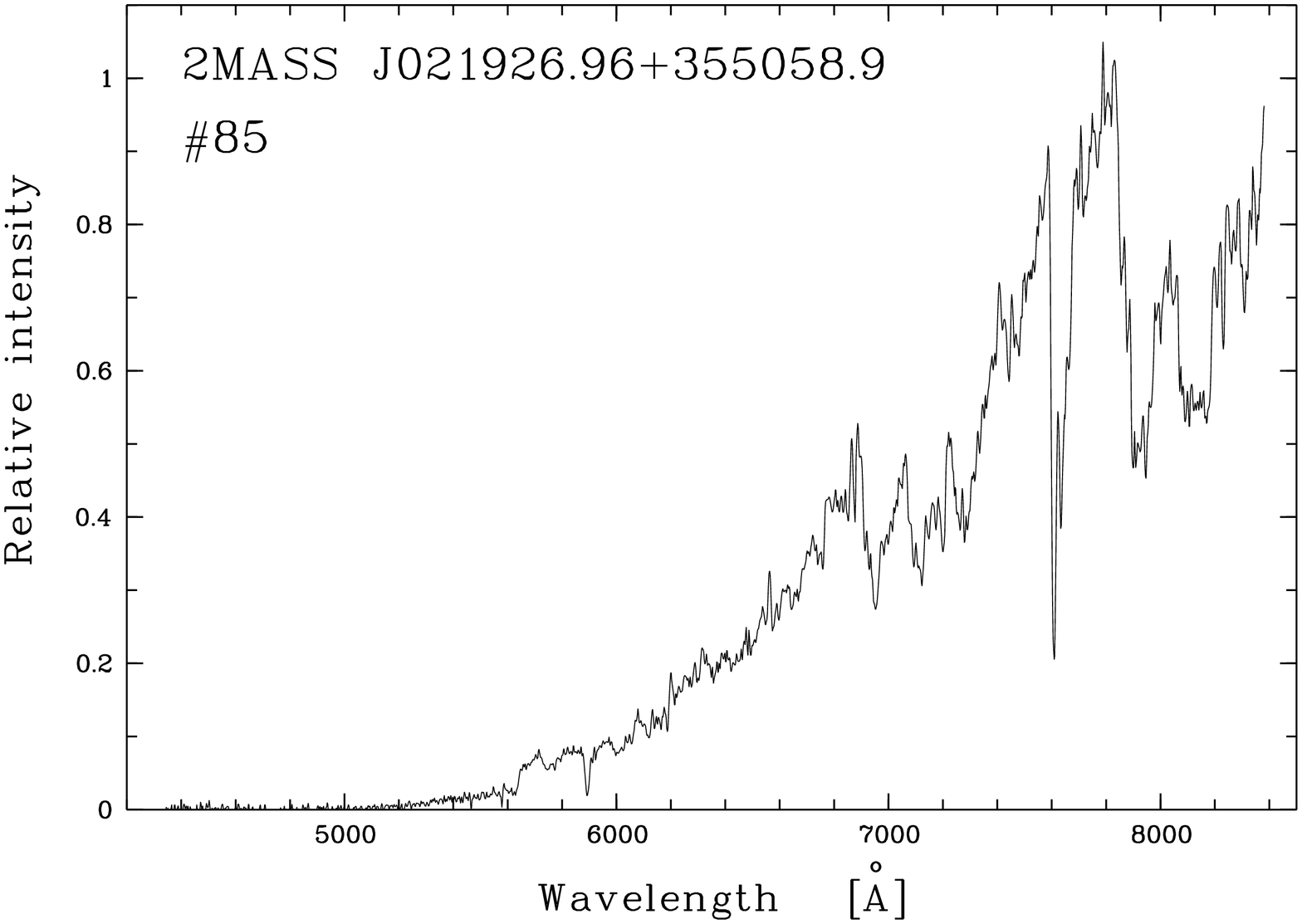}}}
  \resizebox{8.5cm}{!}{\rotatebox{-90}{\includegraphics{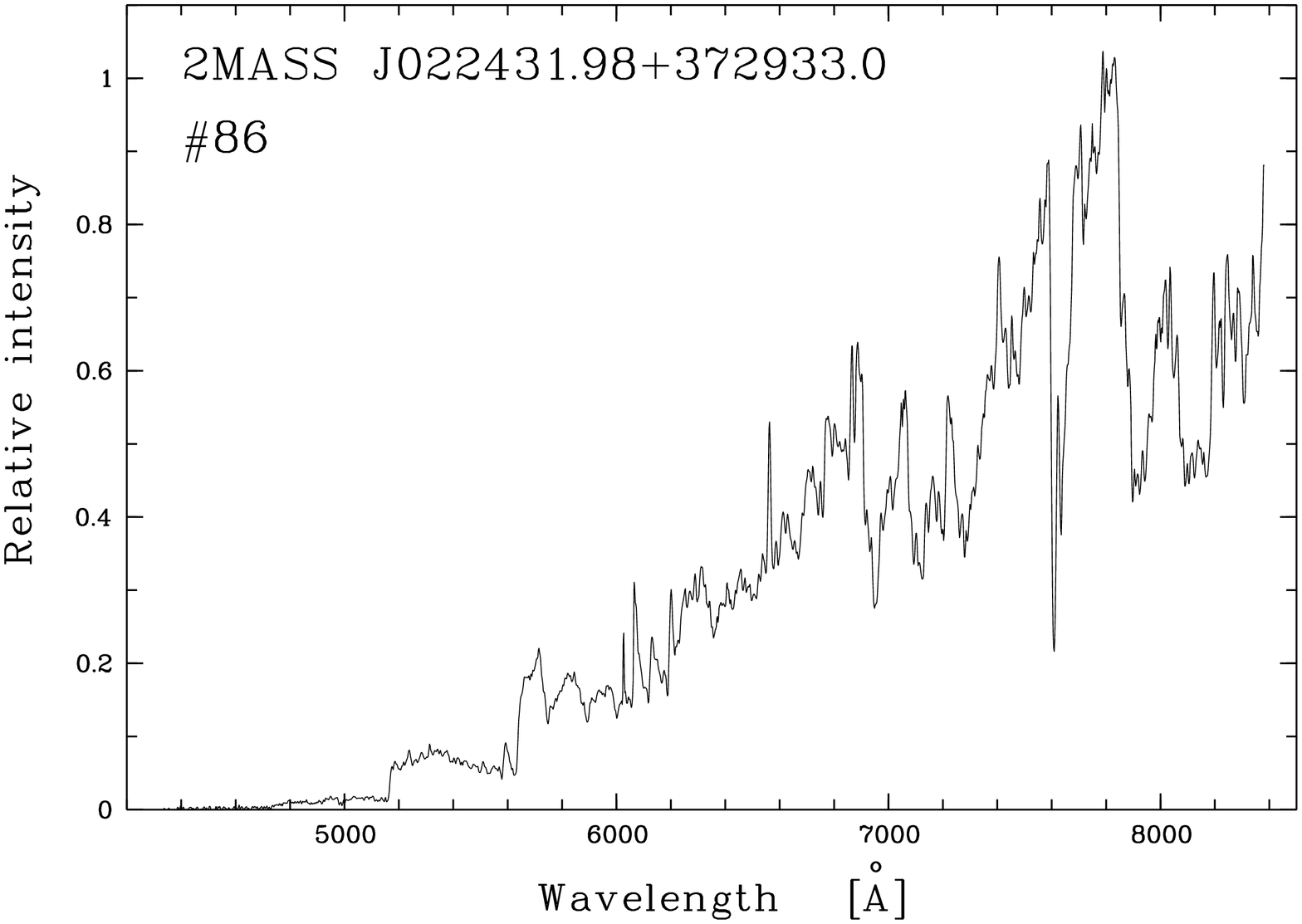}}}
  \resizebox{8.5cm}{!}{\rotatebox{-90}{\includegraphics{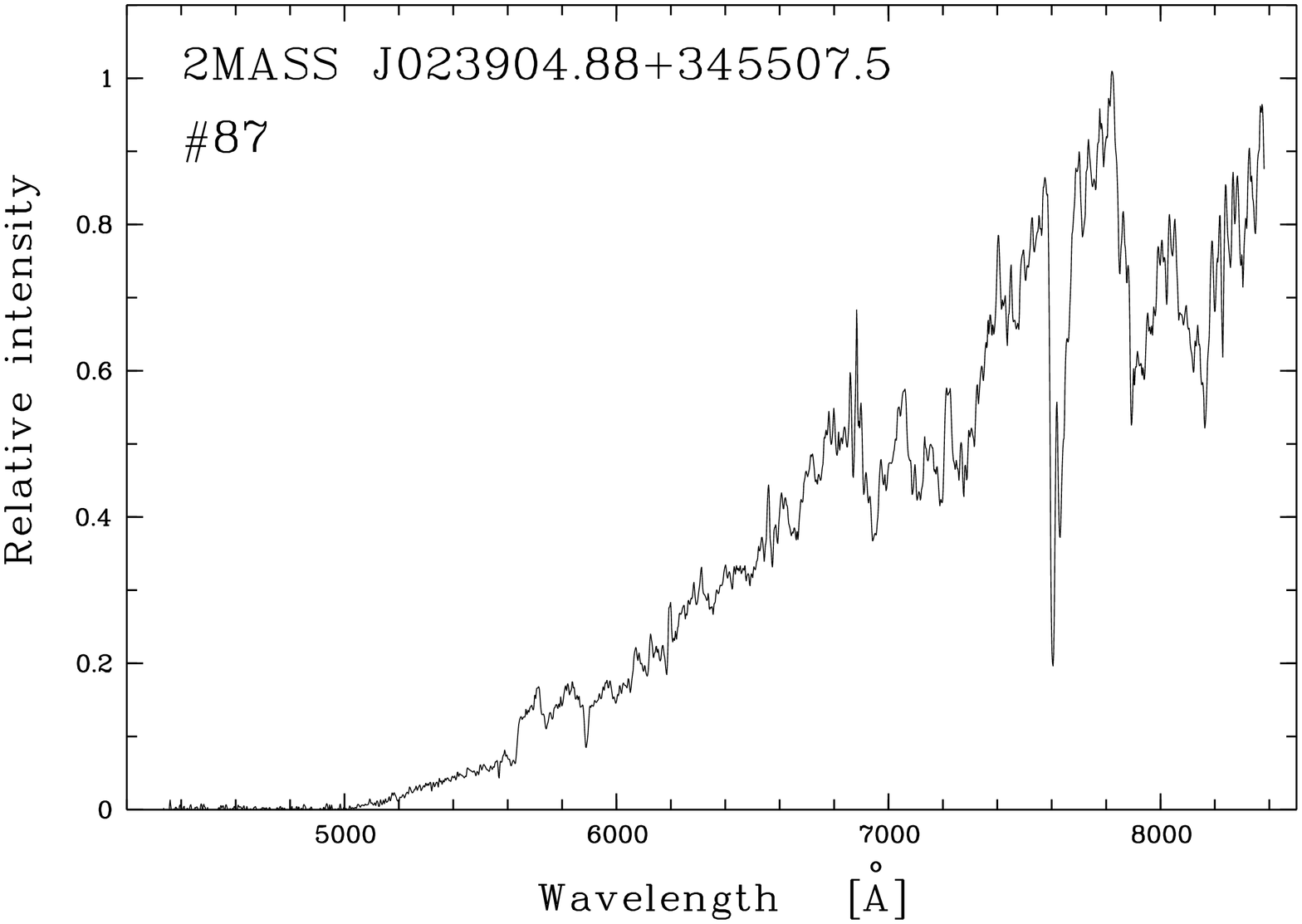}}}
  \resizebox{8.5cm}{!}{\rotatebox{-90}{\includegraphics{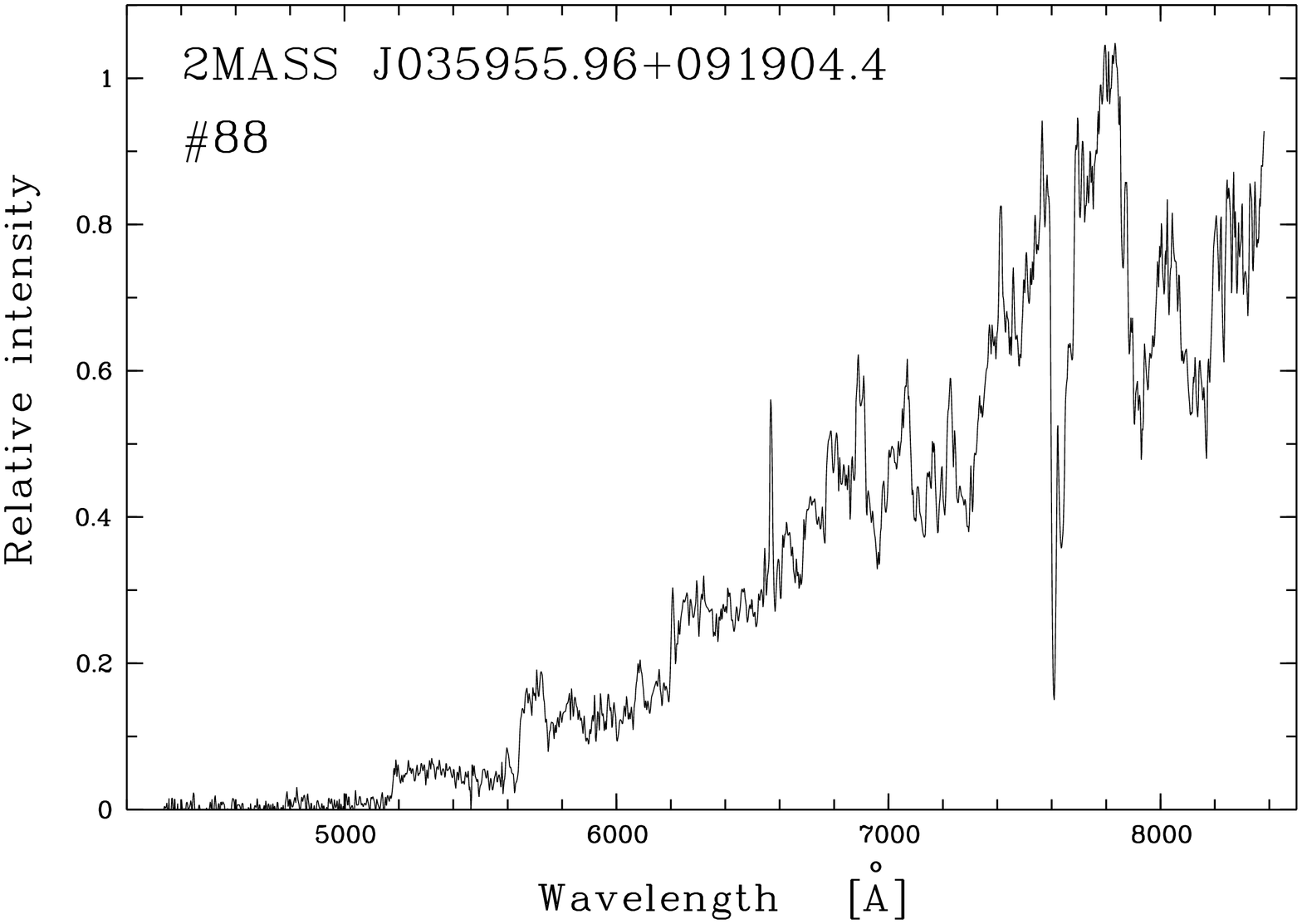}}}
  \resizebox{8.5cm}{!}{\rotatebox{-90}{\includegraphics{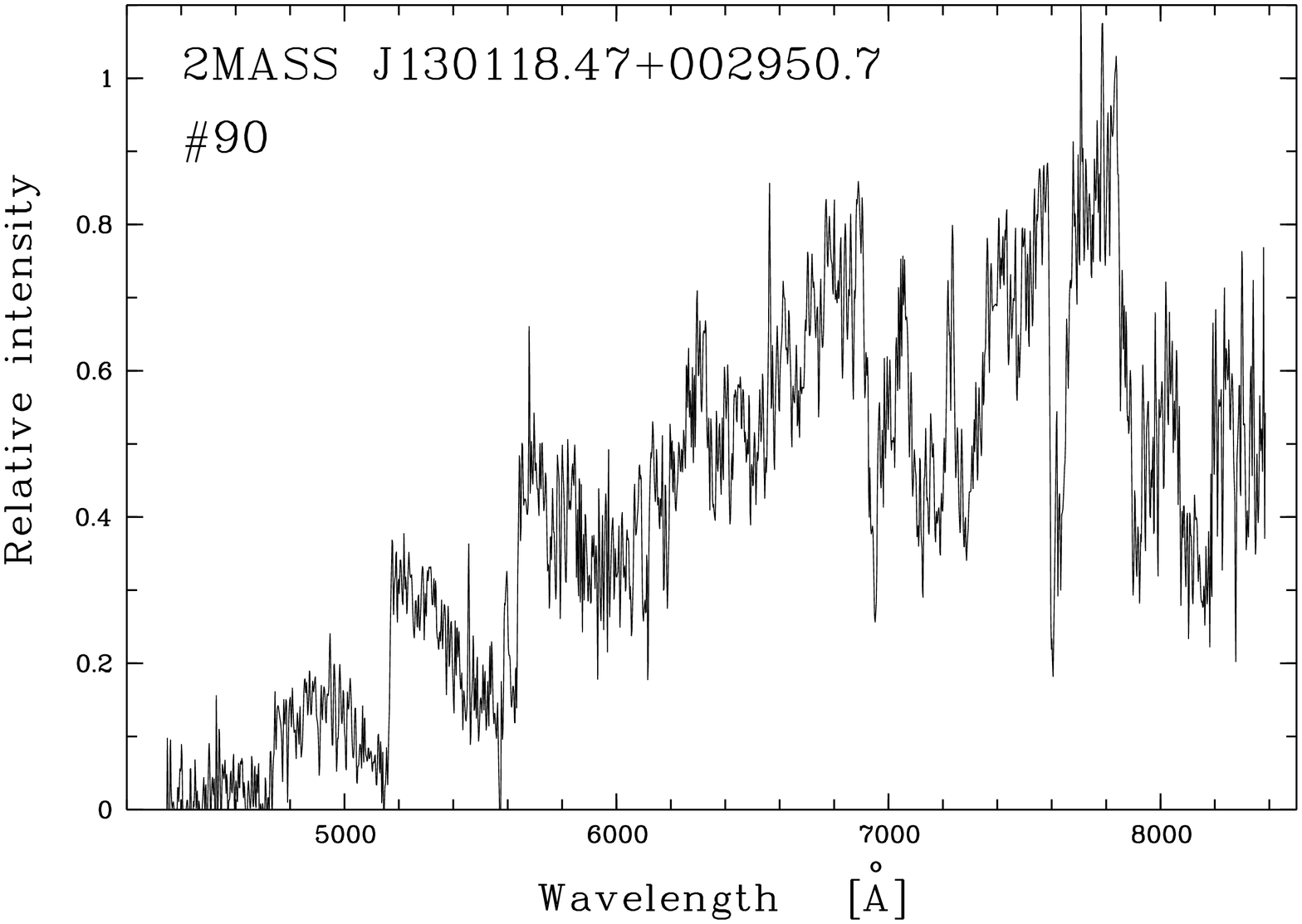}}}
  \resizebox{8.5cm}{!}{\rotatebox{-90}{\includegraphics{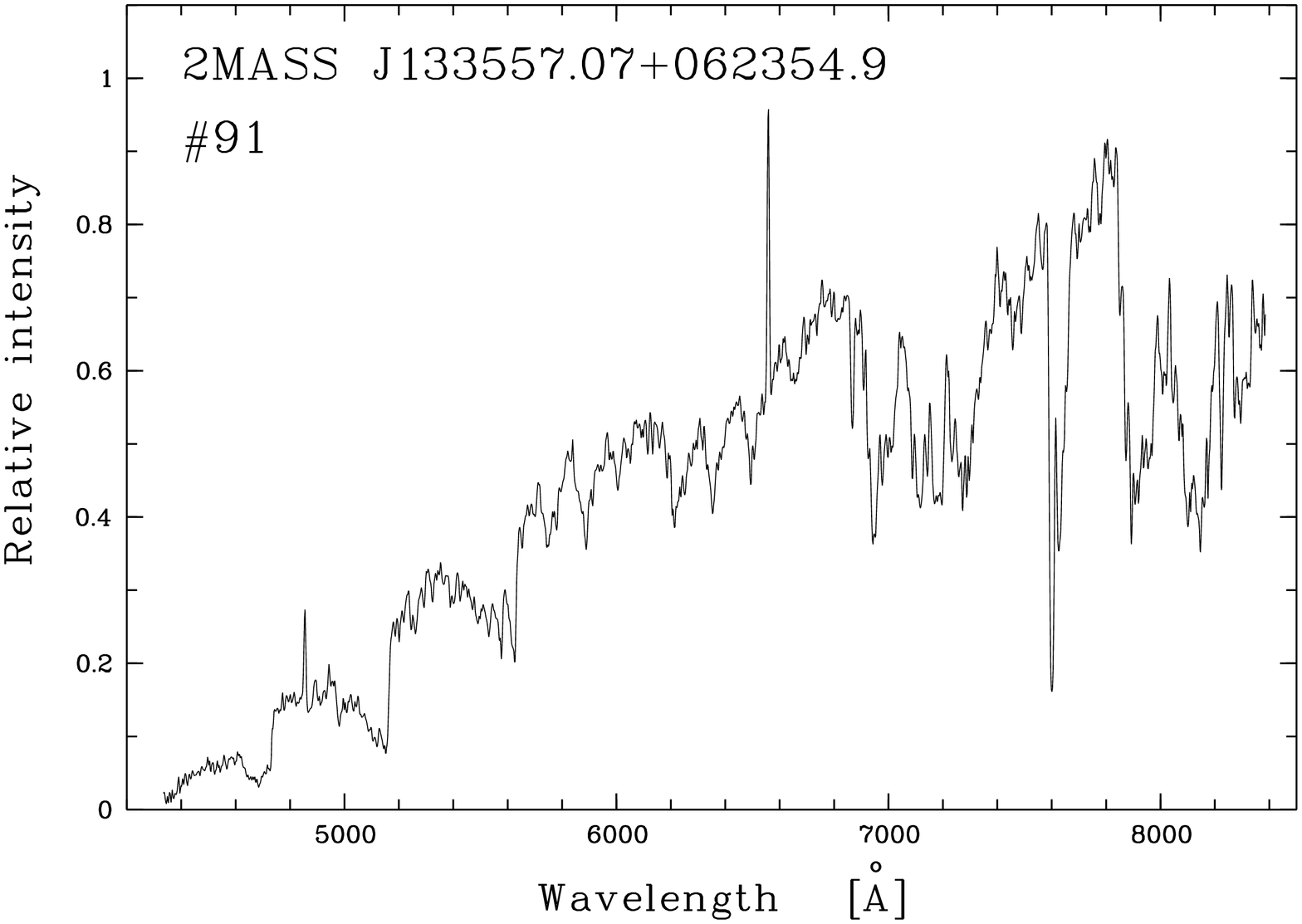}}}
  \resizebox{8.5cm}{!}{\rotatebox{-90}{\includegraphics{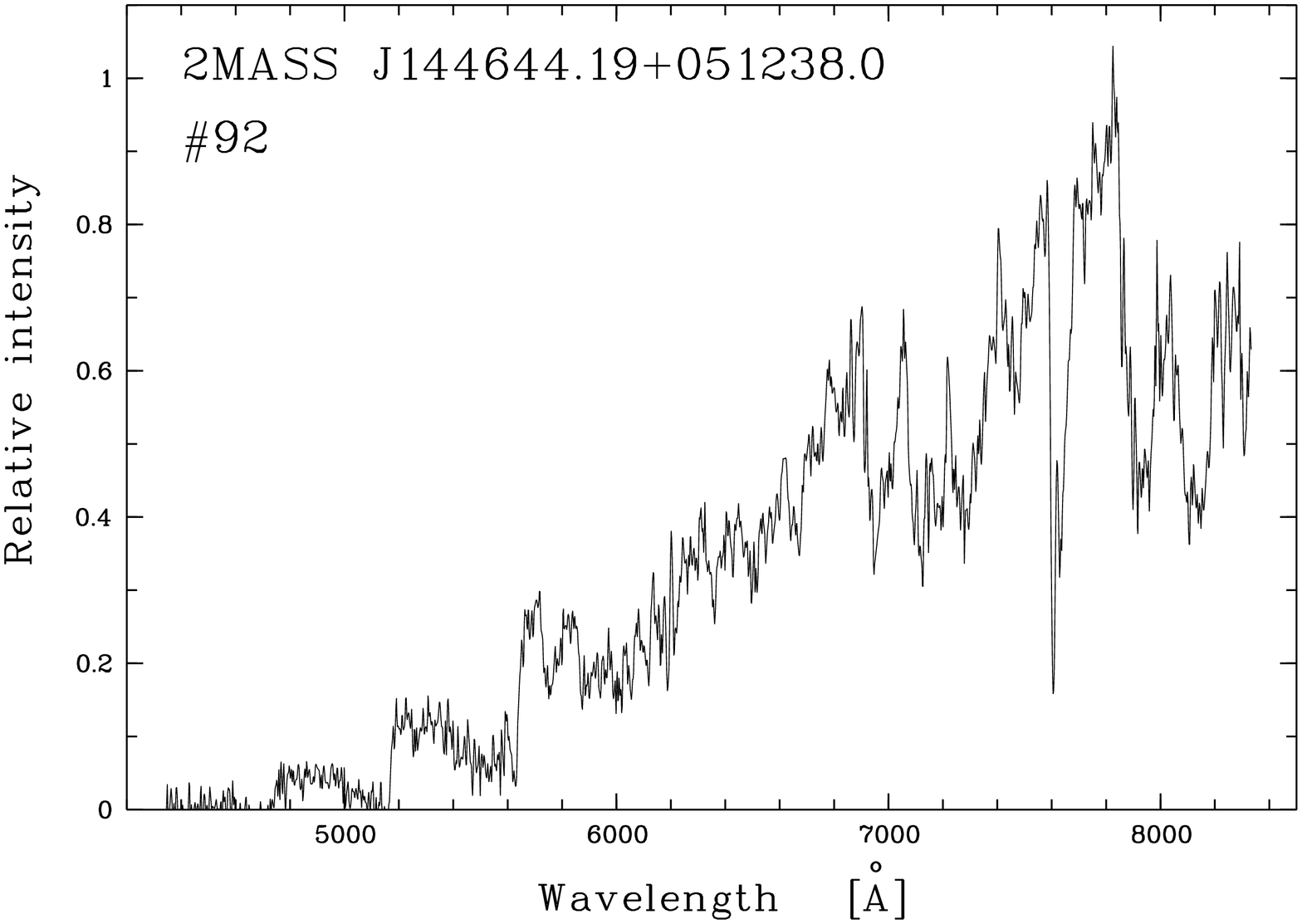}}}
  \resizebox{8.5cm}{!}{\rotatebox{-90}{\includegraphics{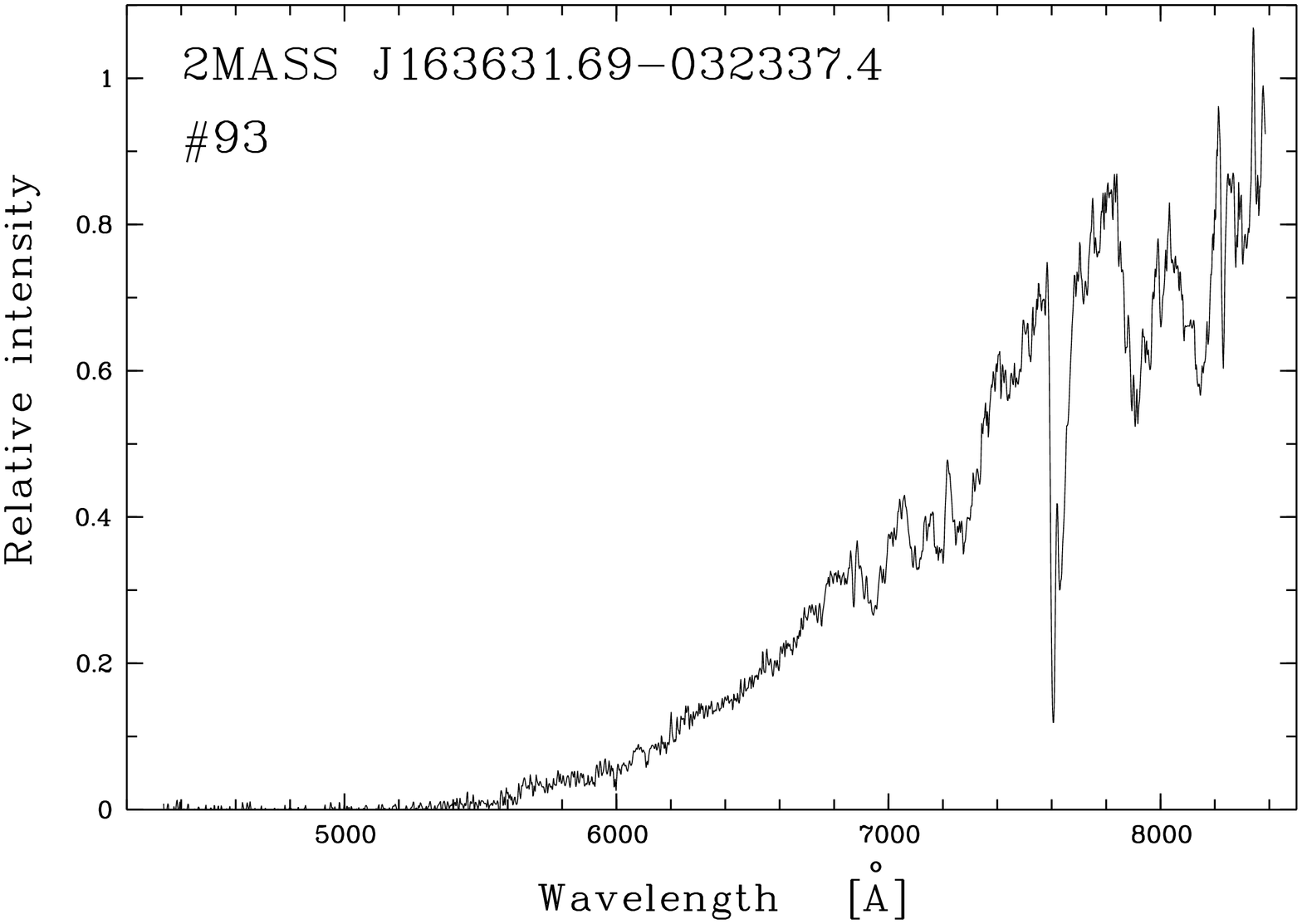}}}
  \end{figure*}

\addtocounter{figure}{-1}

  \begin{figure*}
  \caption[]{{\it Continued}}
  \resizebox{8.5cm}{!}{\rotatebox{-90}{\includegraphics{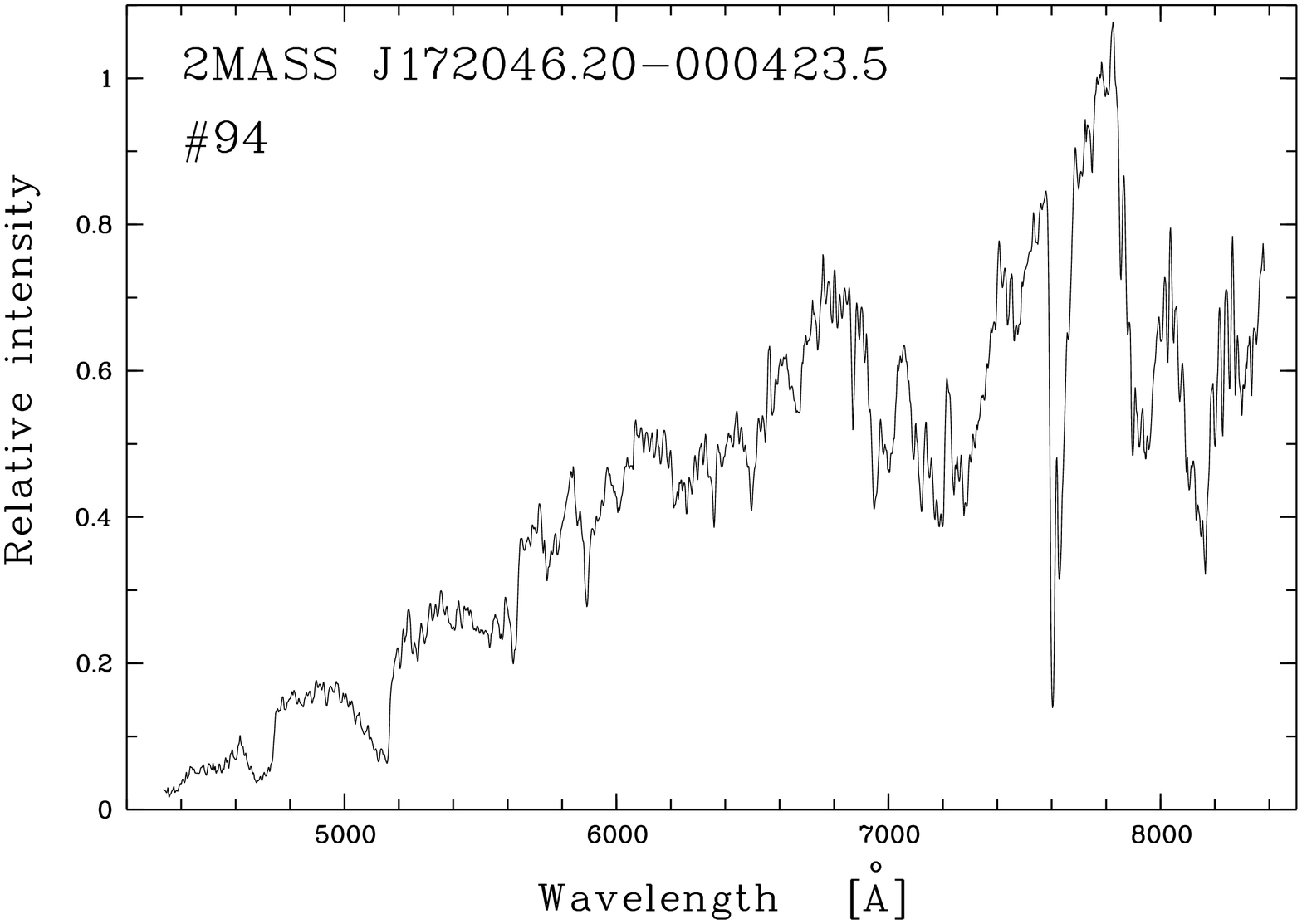}}}
  \resizebox{8.5cm}{!}{\rotatebox{-90}{\includegraphics{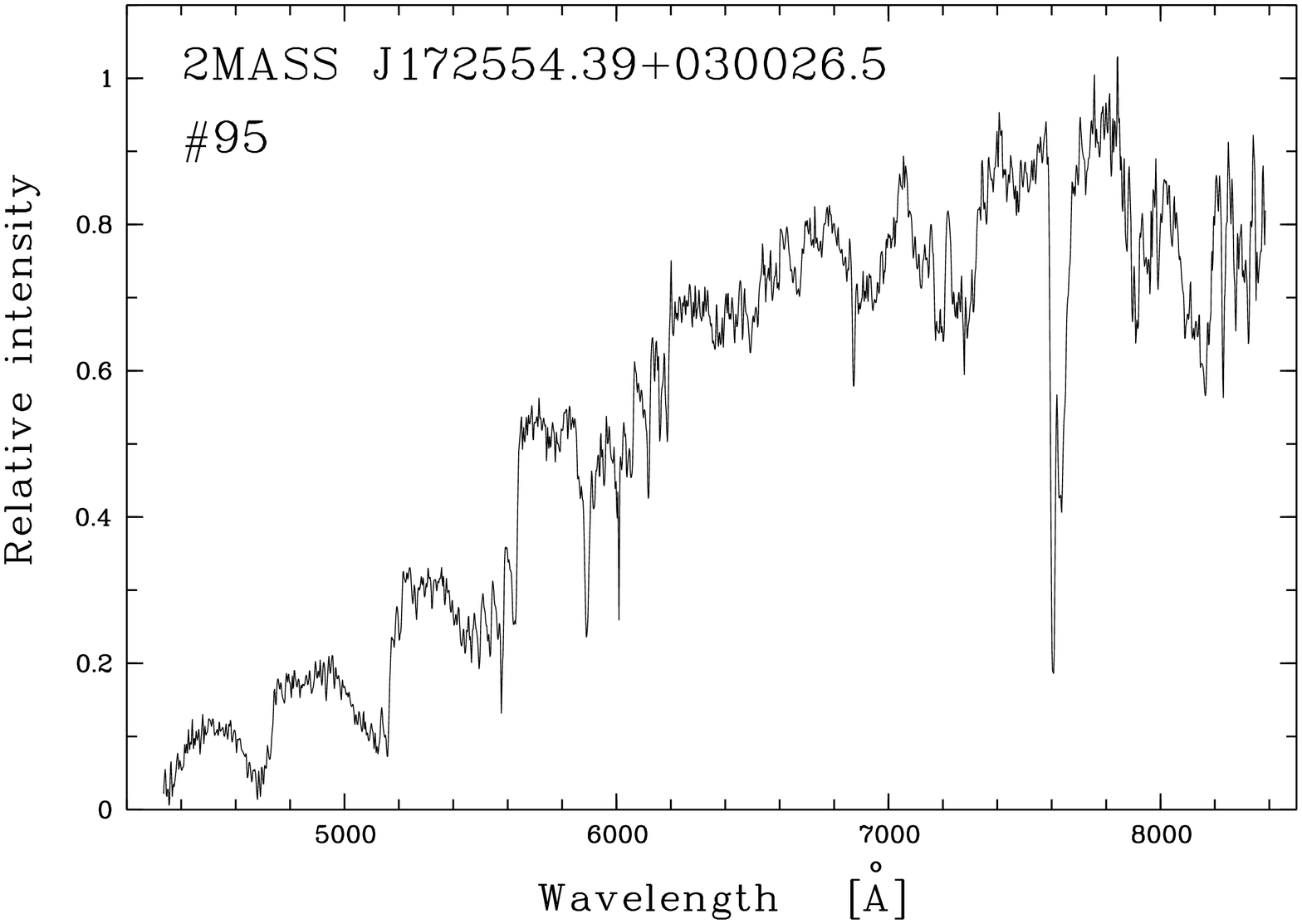}}}
  \resizebox{8.5cm}{!}{\rotatebox{-90}{\includegraphics{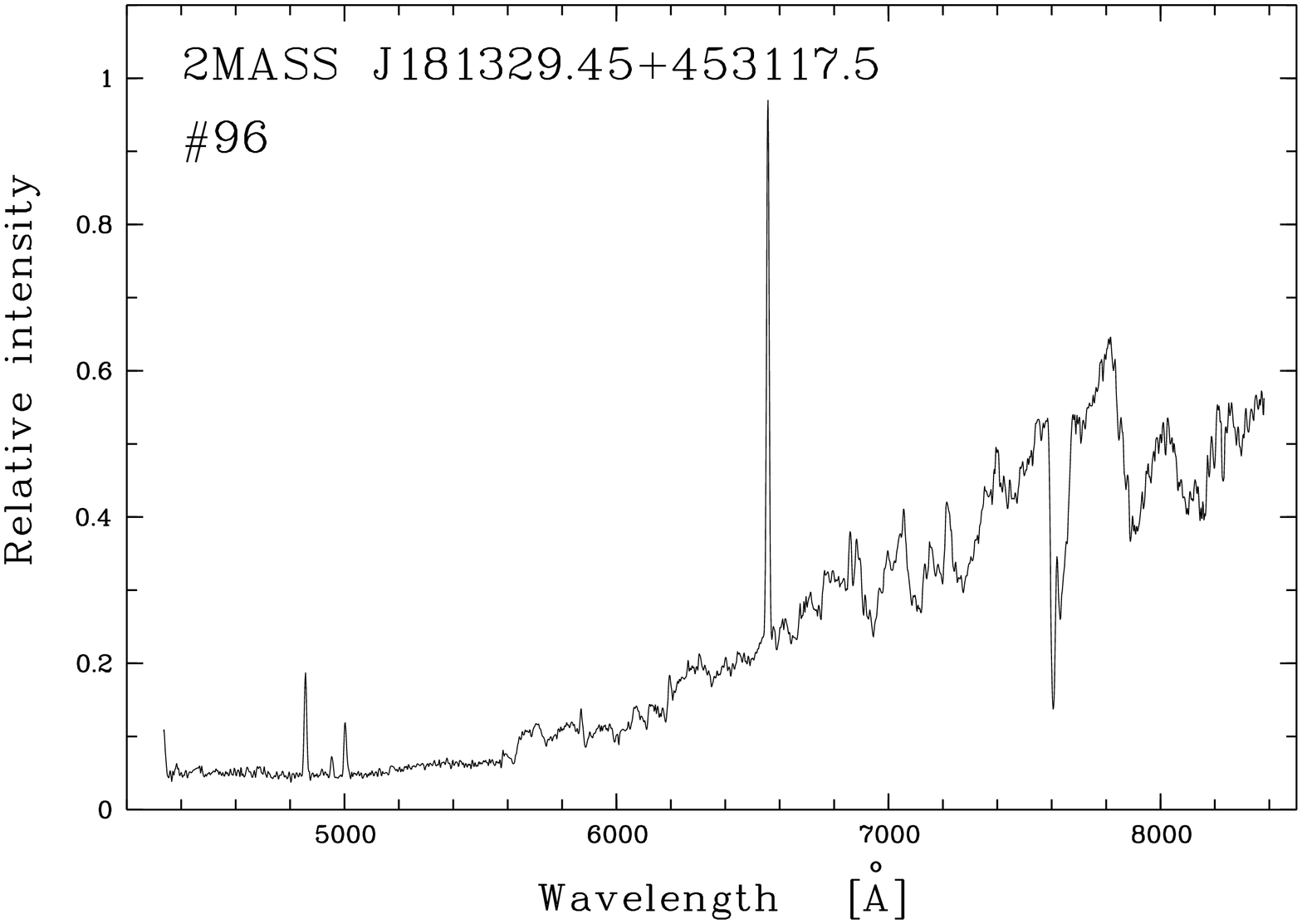}}}
  \resizebox{8.5cm}{!}{\rotatebox{-90}{\includegraphics{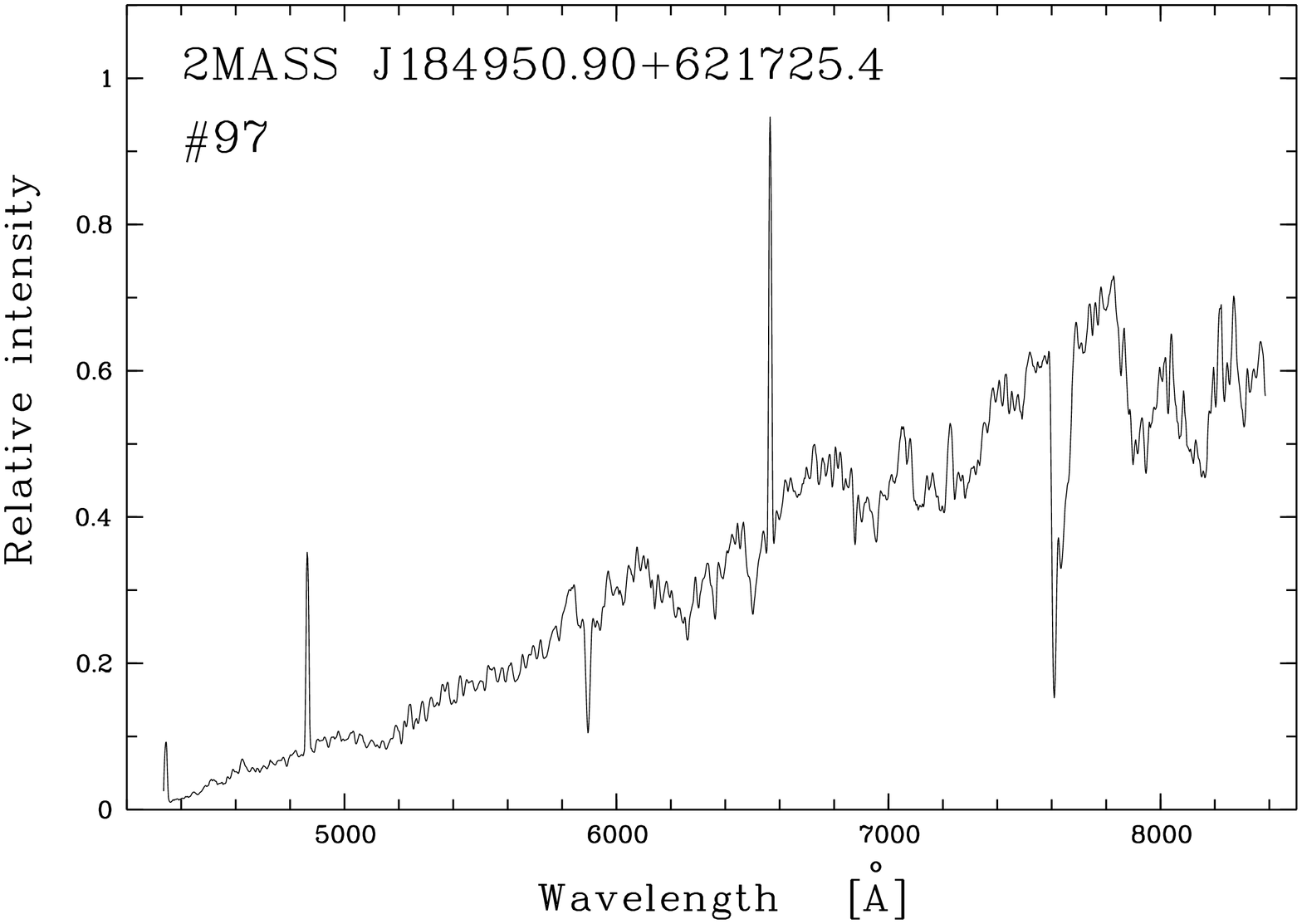}}}
  \resizebox{8.5cm}{!}{\rotatebox{-90}{\includegraphics{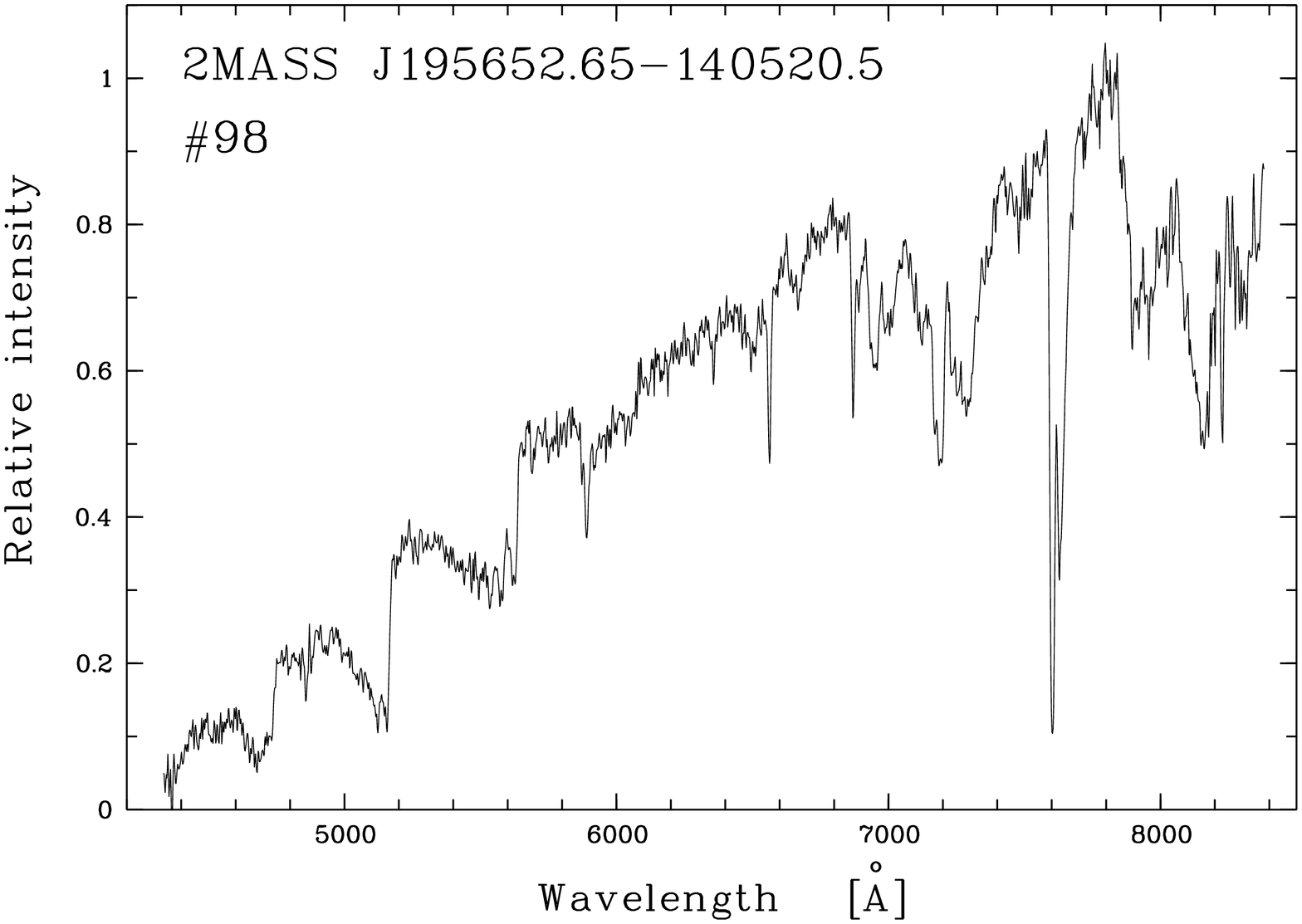}}}
  \resizebox{8.5cm}{!}{\rotatebox{-90}{\includegraphics{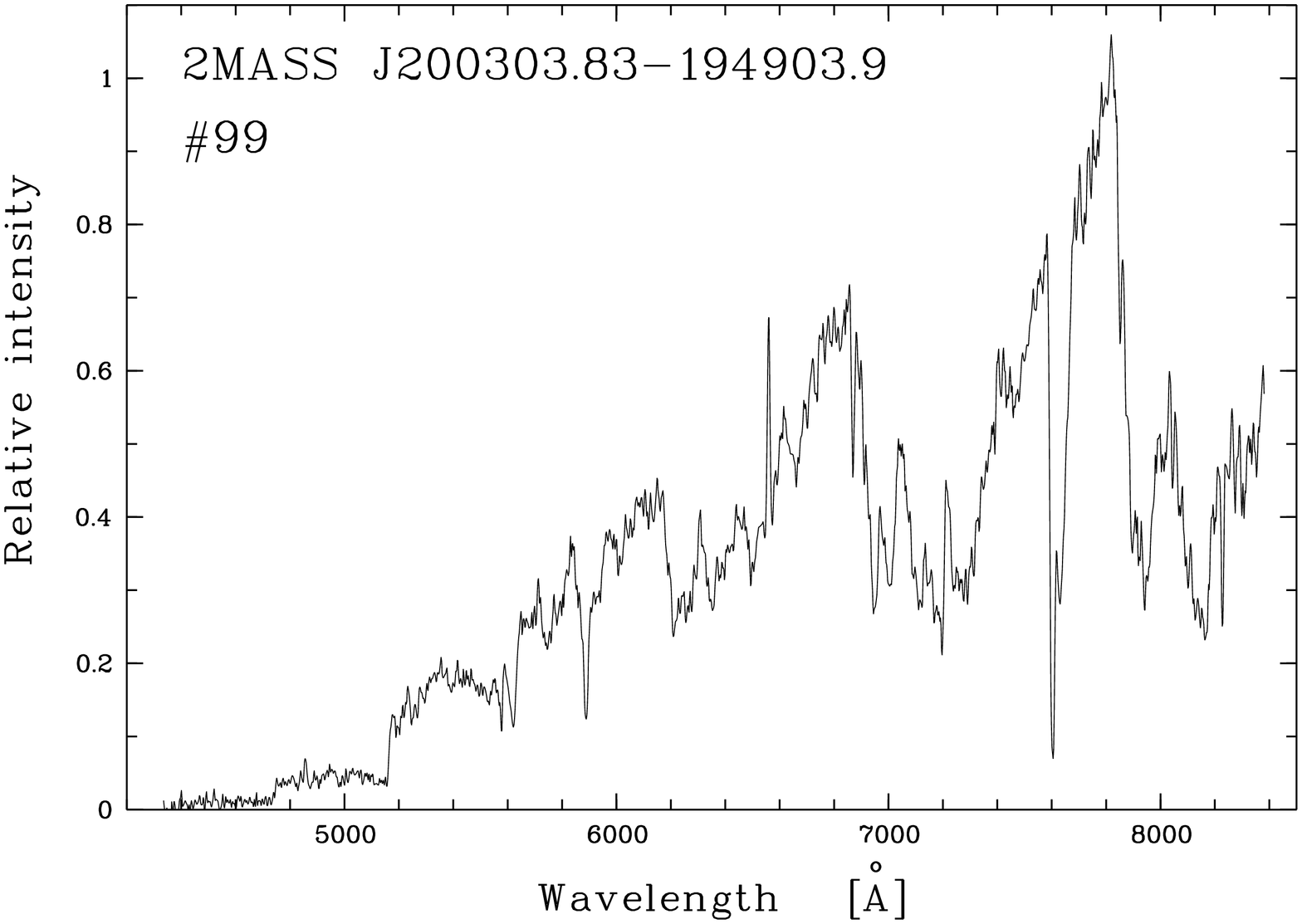}}} 
  \resizebox{8.5cm}{!}{\rotatebox{-90}{\includegraphics{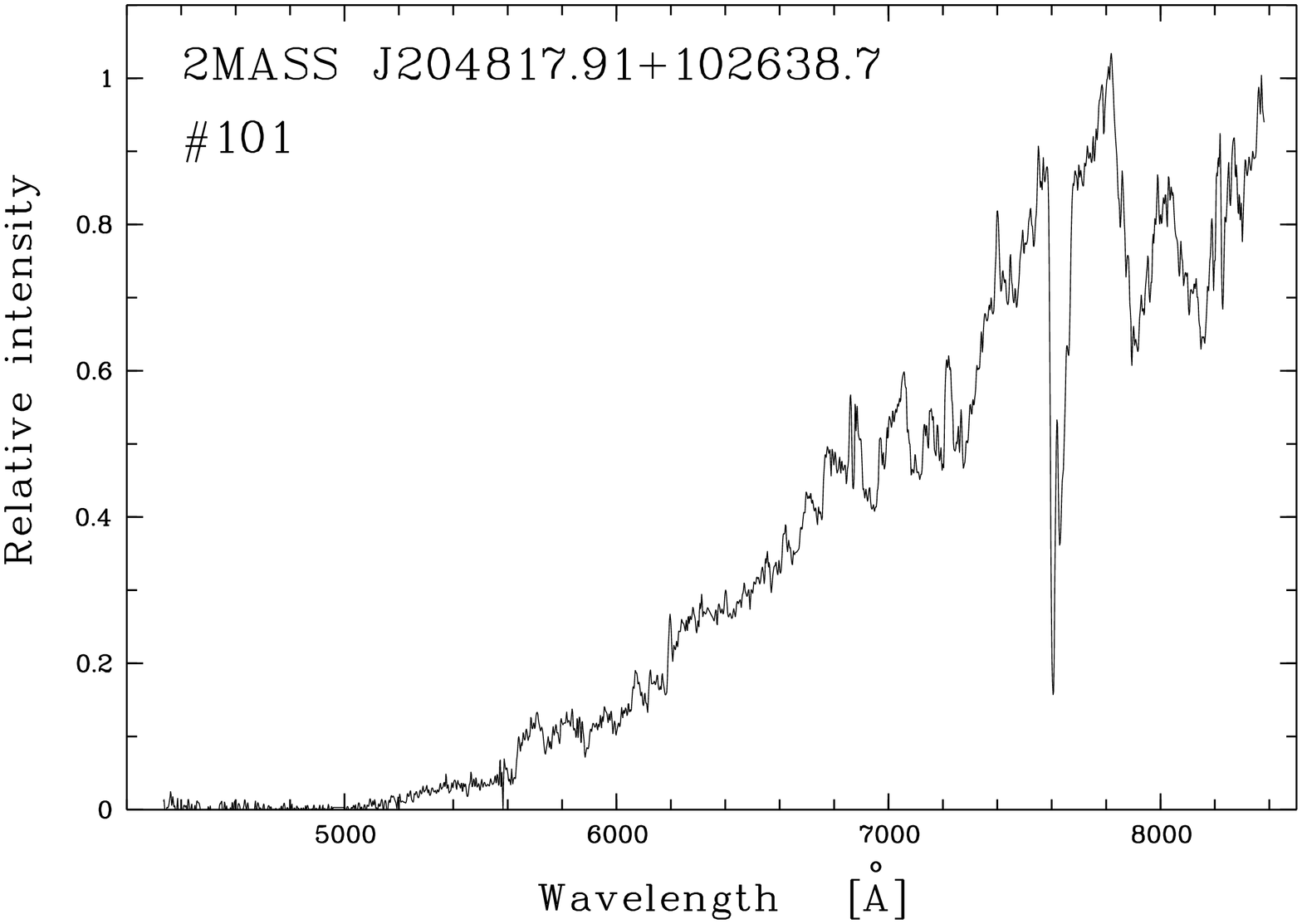}}}
  \end{figure*}
  

   \begin{figure*}
   \caption[]{Spectra of FBS candidate carbon stars}
   \resizebox{8.5cm}{!}{\rotatebox{-90}{\includegraphics{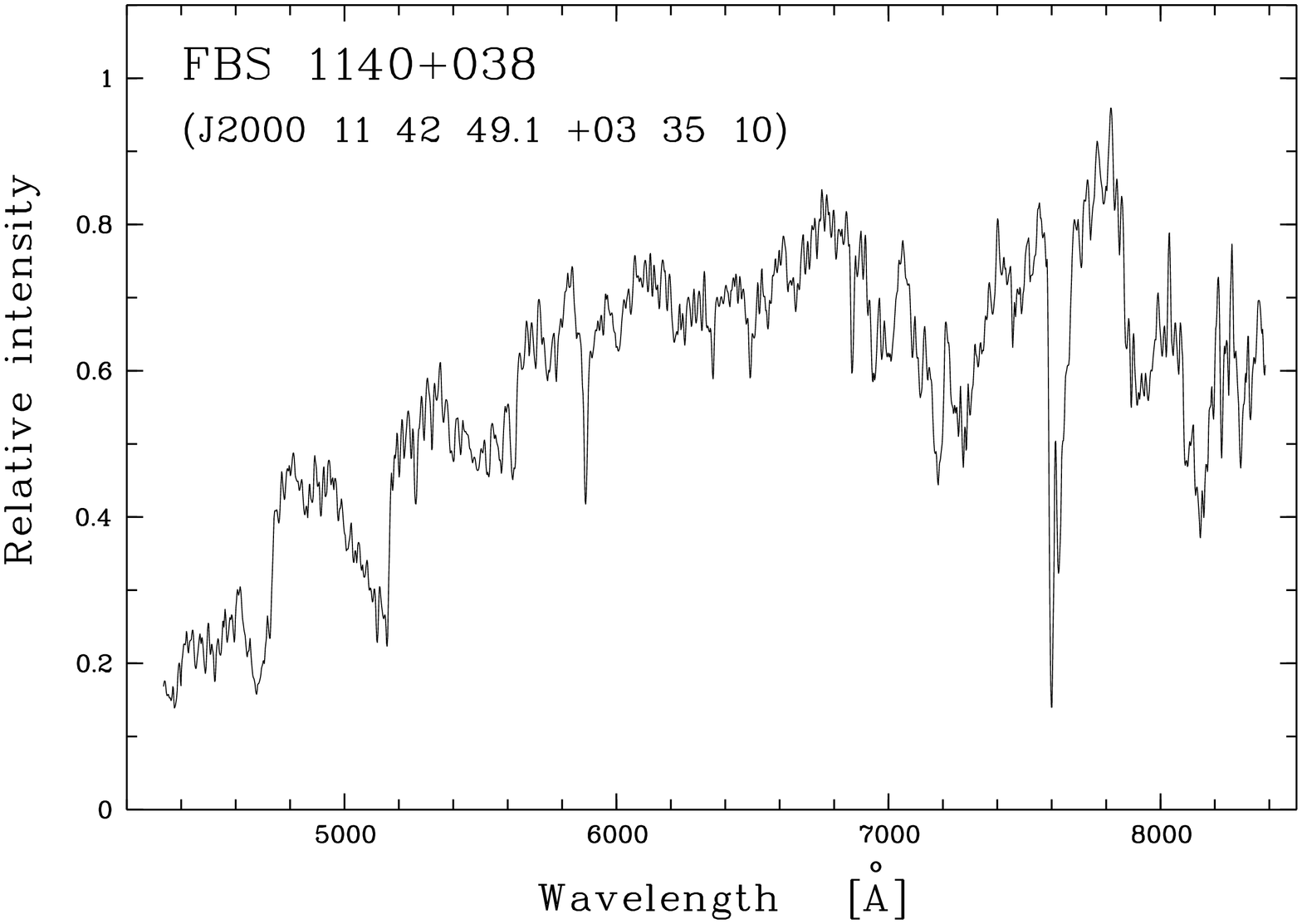}}}
   \resizebox{8.5cm}{!}{\rotatebox{-90}{\includegraphics{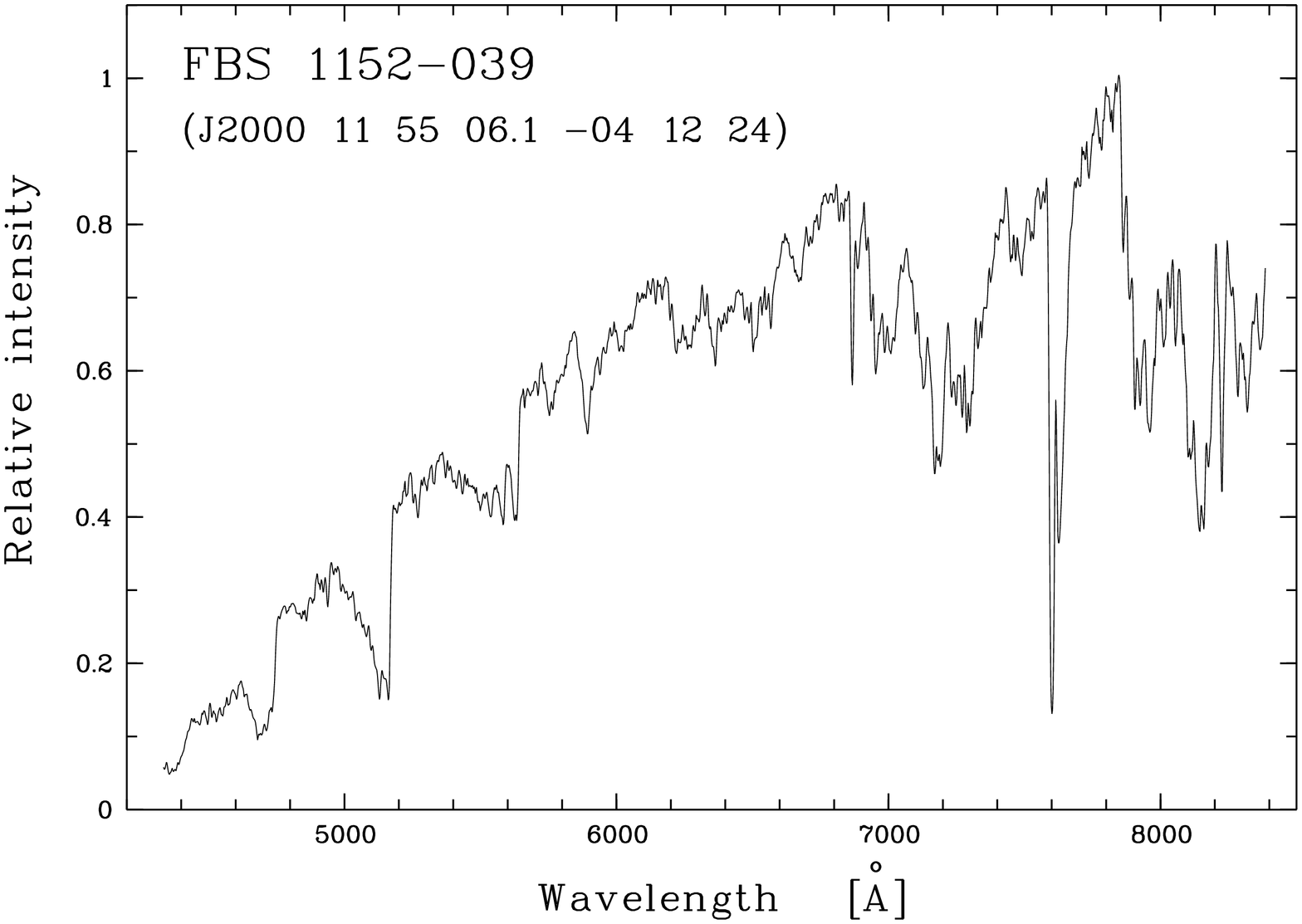}}}
   \resizebox{8.5cm}{!}{\rotatebox{-90}{\includegraphics{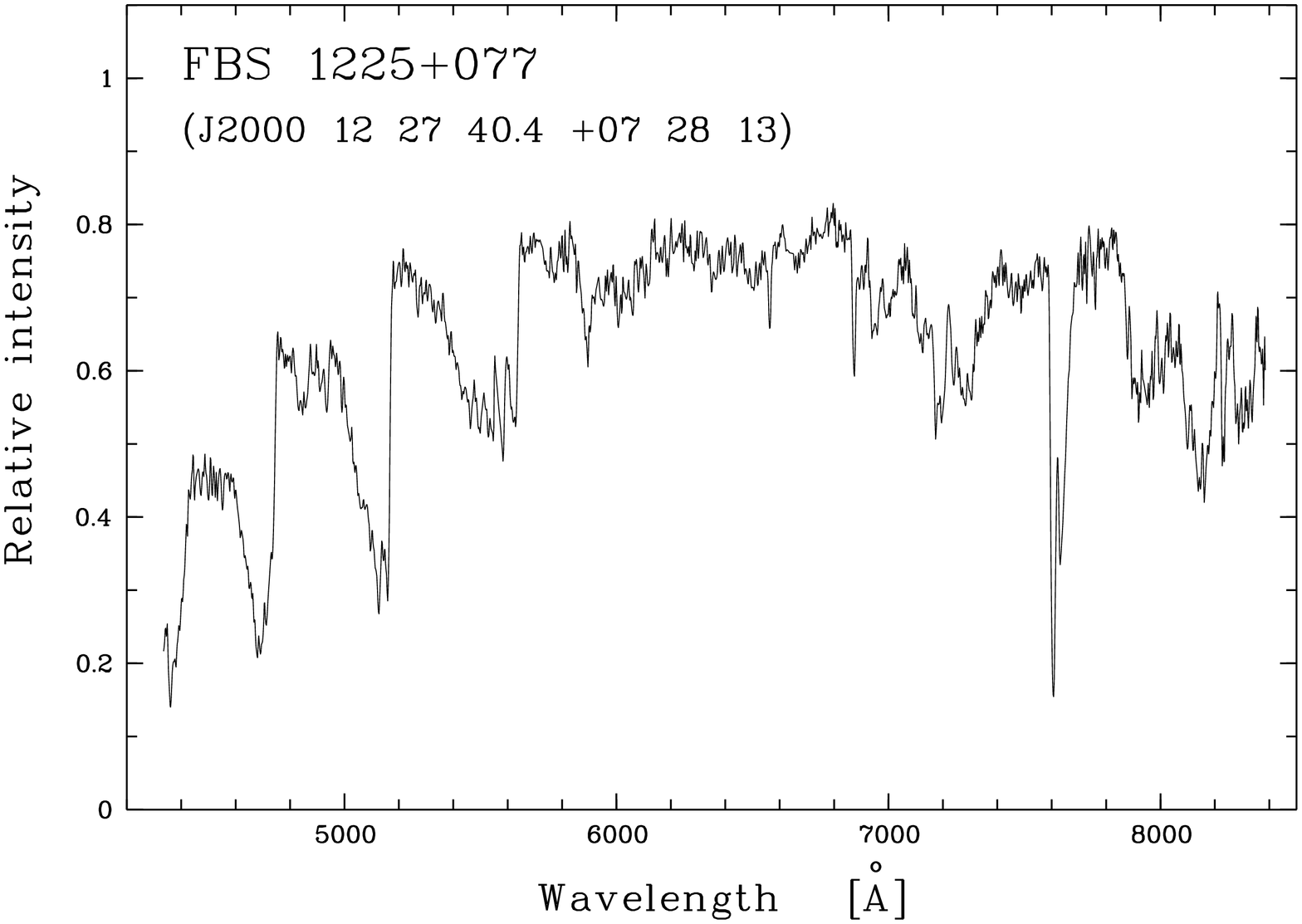}}}
   \resizebox{8.5cm}{!}{\rotatebox{-90}{\includegraphics{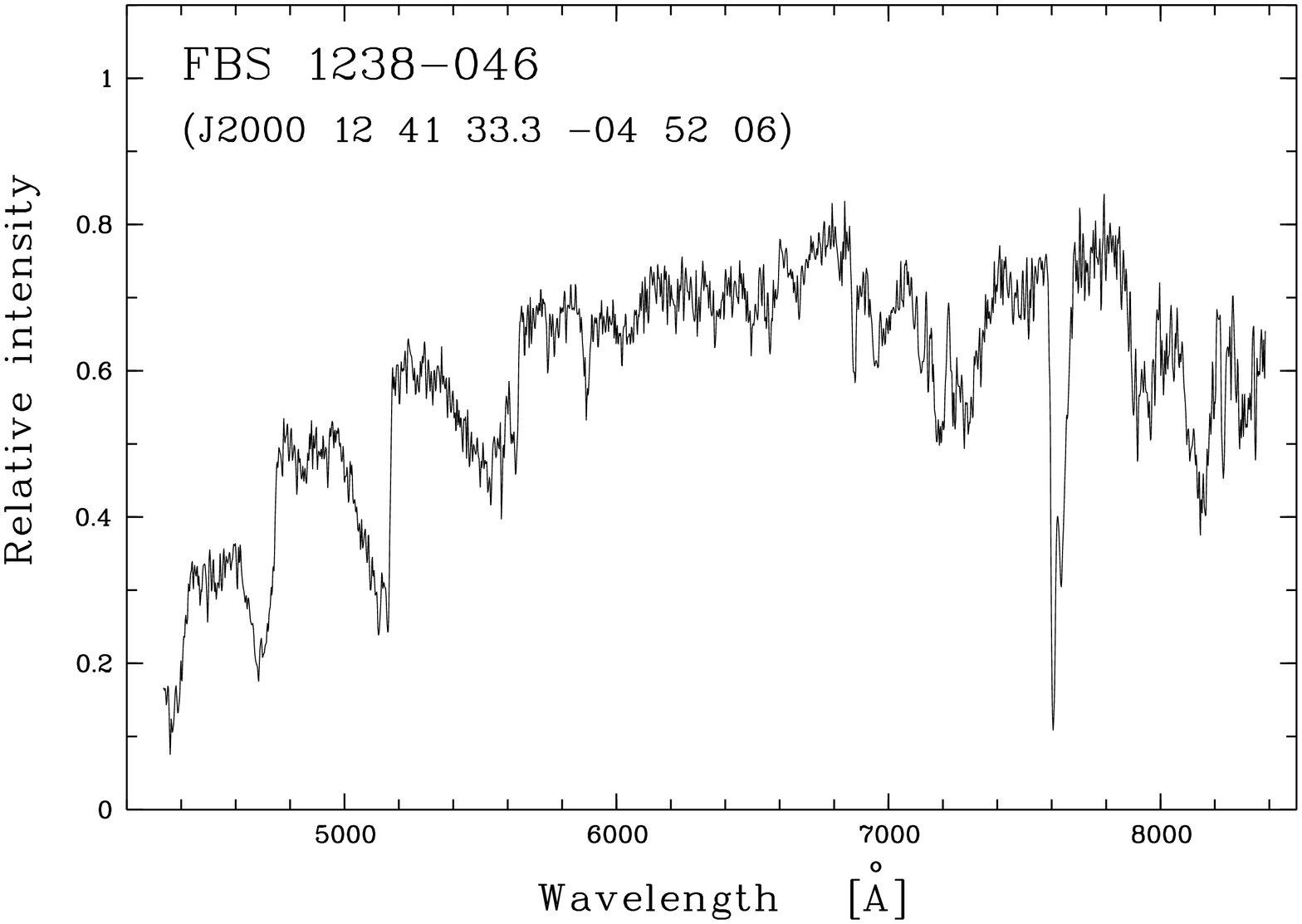}}}
   \resizebox{8.5cm}{!}{\rotatebox{-90}{\includegraphics{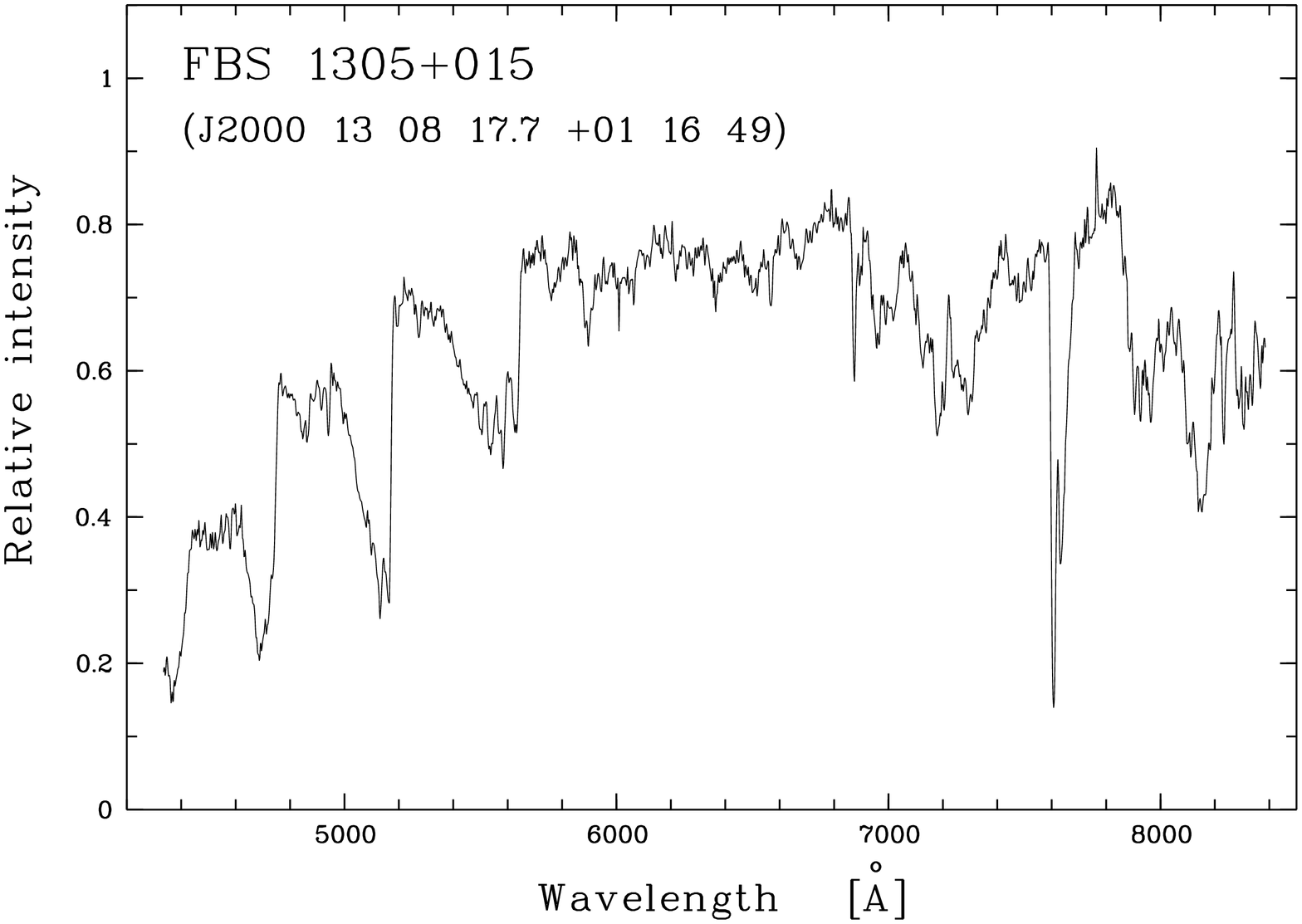}}}
   \resizebox{8.5cm}{!}{\rotatebox{-90}{\includegraphics{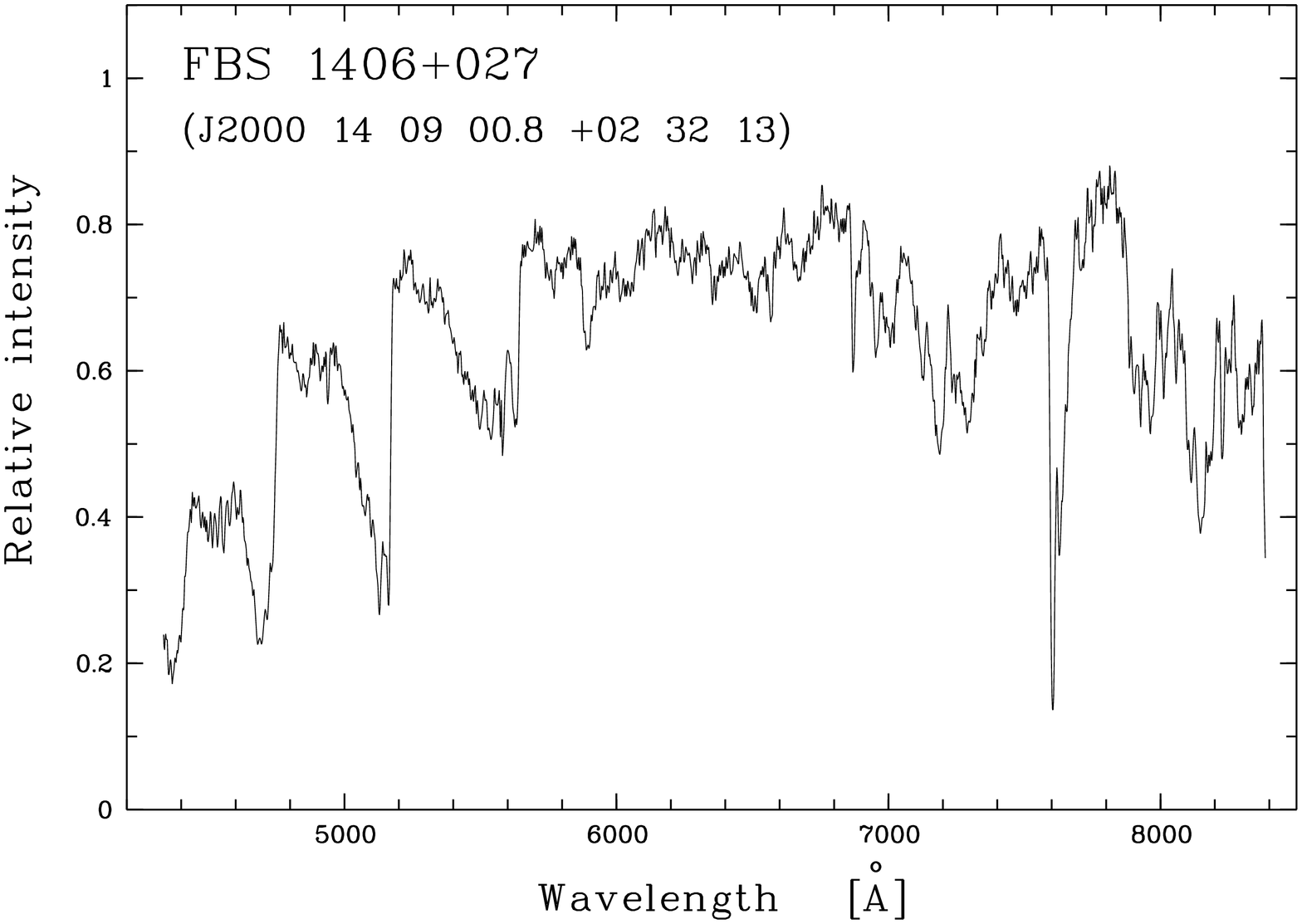}}}
   \resizebox{8.5cm}{!}{\rotatebox{-90}{\includegraphics{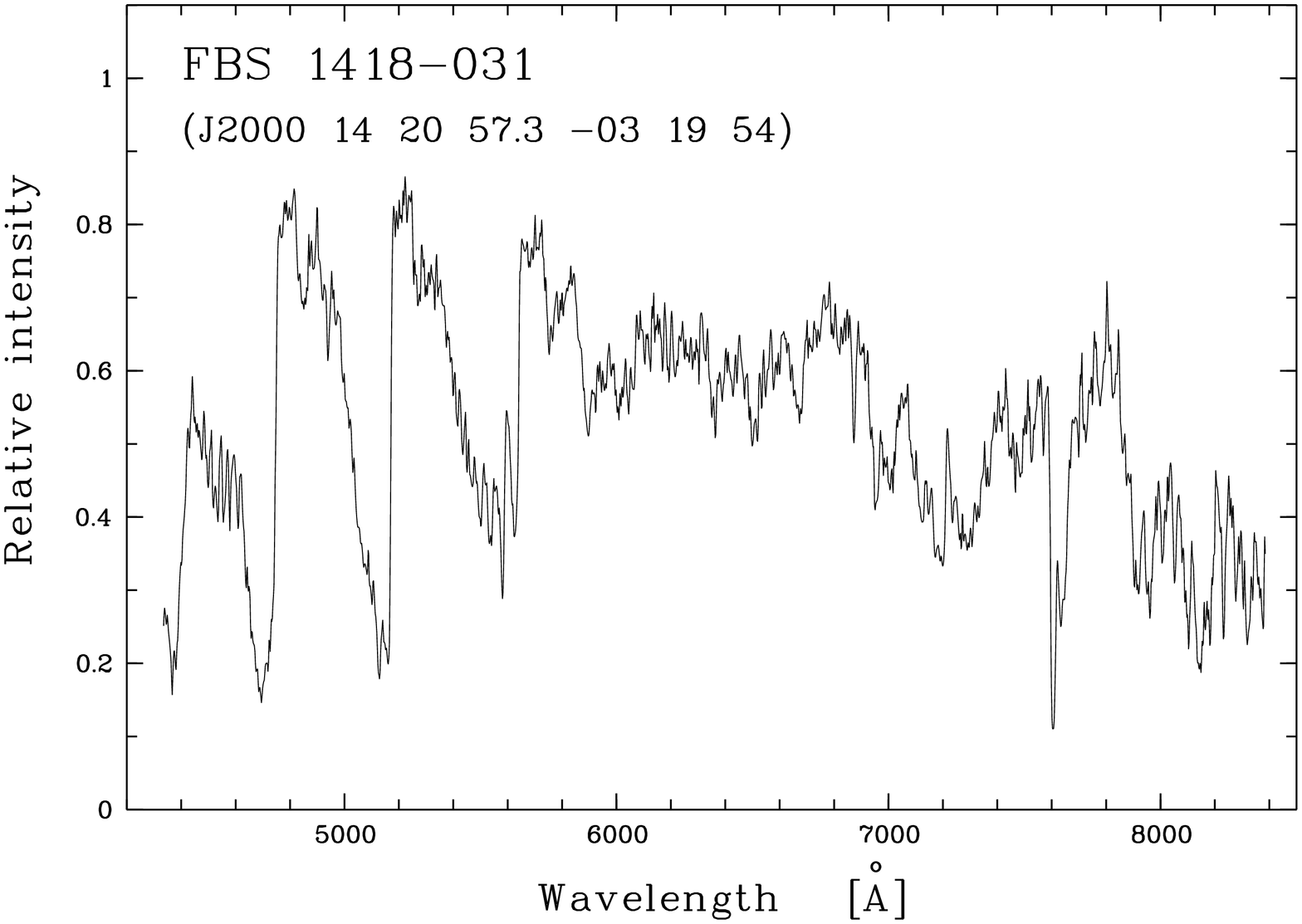}}}
   \resizebox{8.5cm}{!}{\rotatebox{-90}{\includegraphics{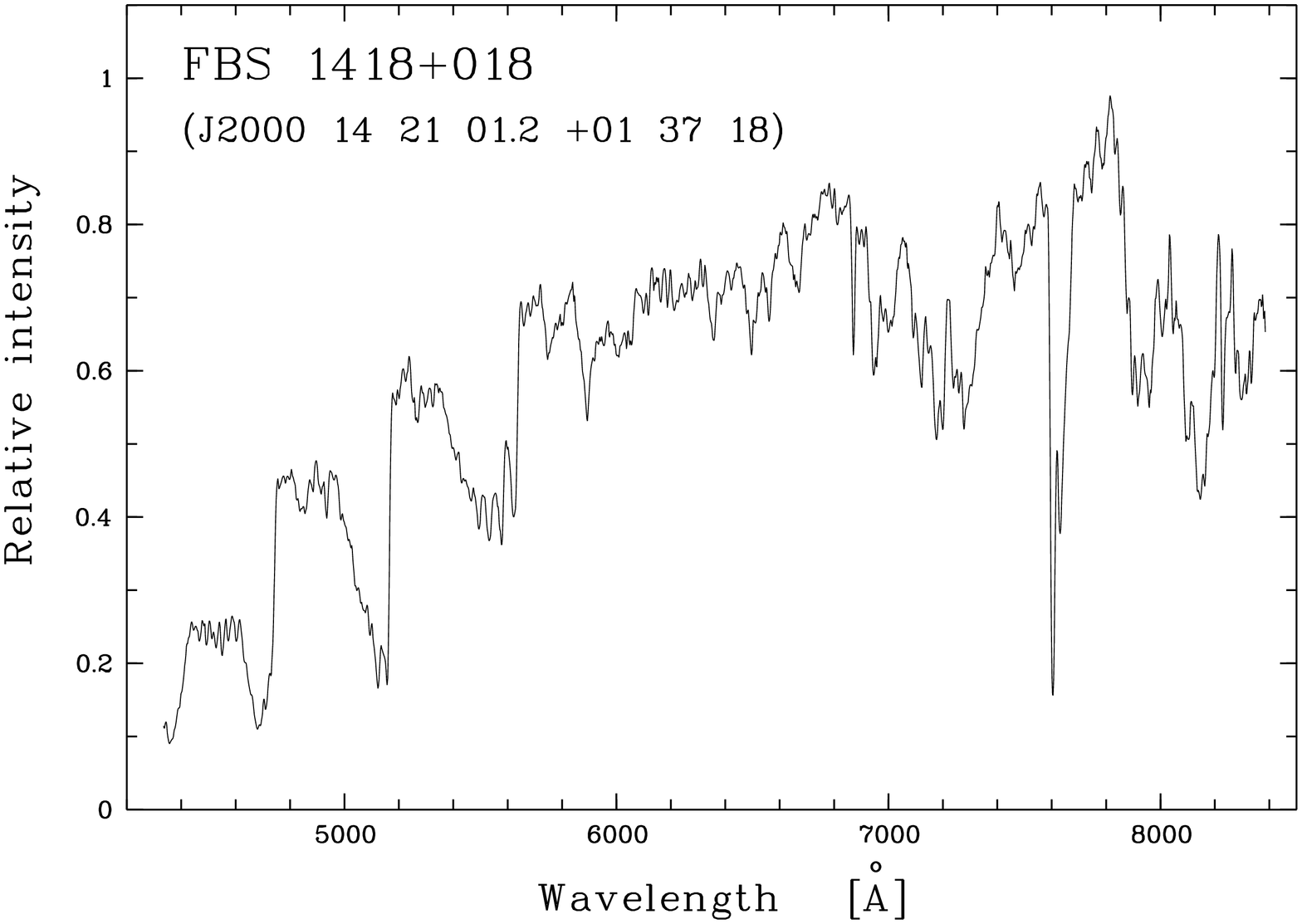}}}
   \end{figure*}

\addtocounter{figure}{-1}

   \begin{figure*}
   \caption[]{{\it Continued}}
   \resizebox{8.5cm}{!}{\rotatebox{-90}{\includegraphics{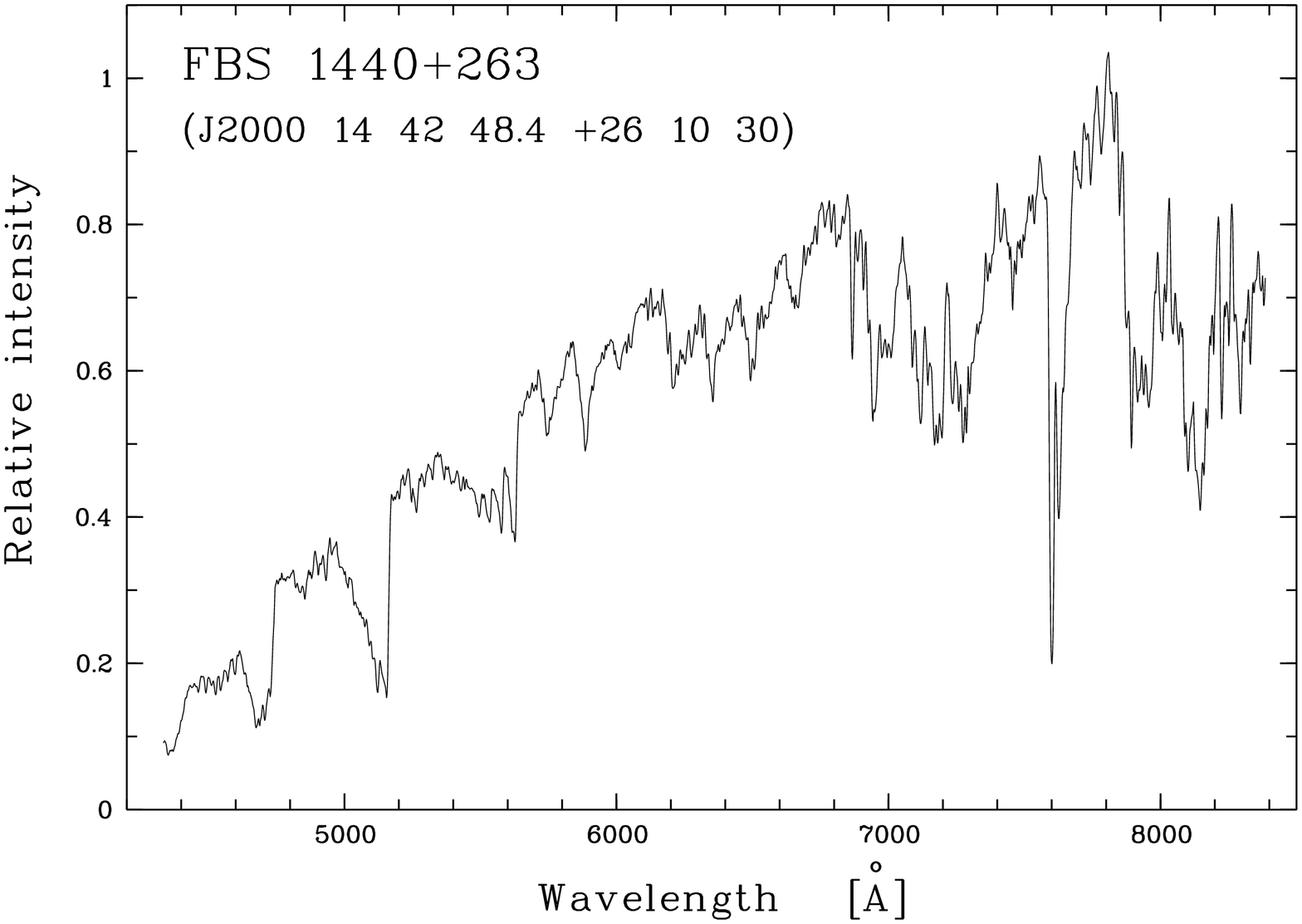}}}
   \resizebox{8.5cm}{!}{\rotatebox{-90}{\includegraphics{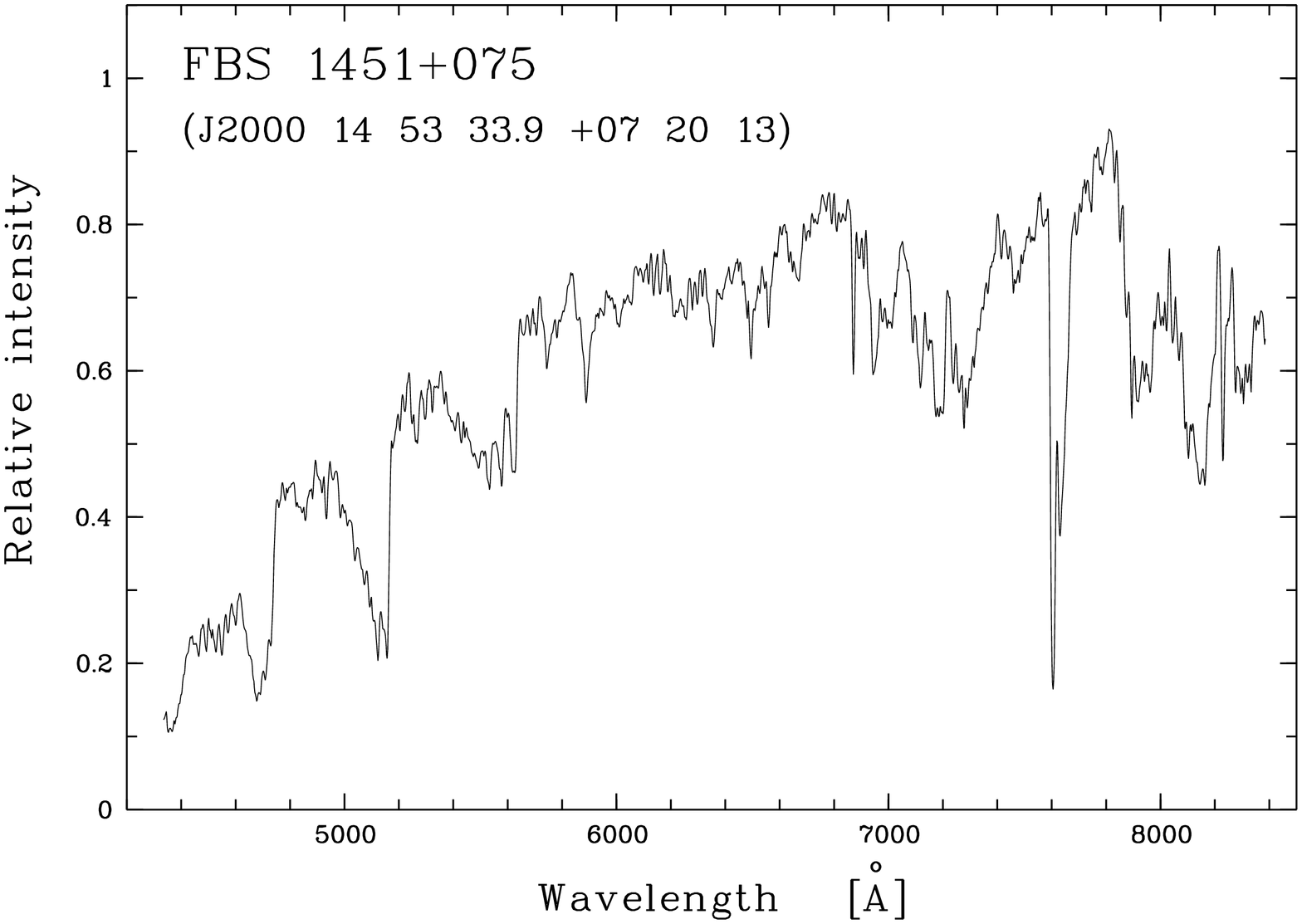}}}
   \resizebox{8.5cm}{!}{\rotatebox{-90}{\includegraphics{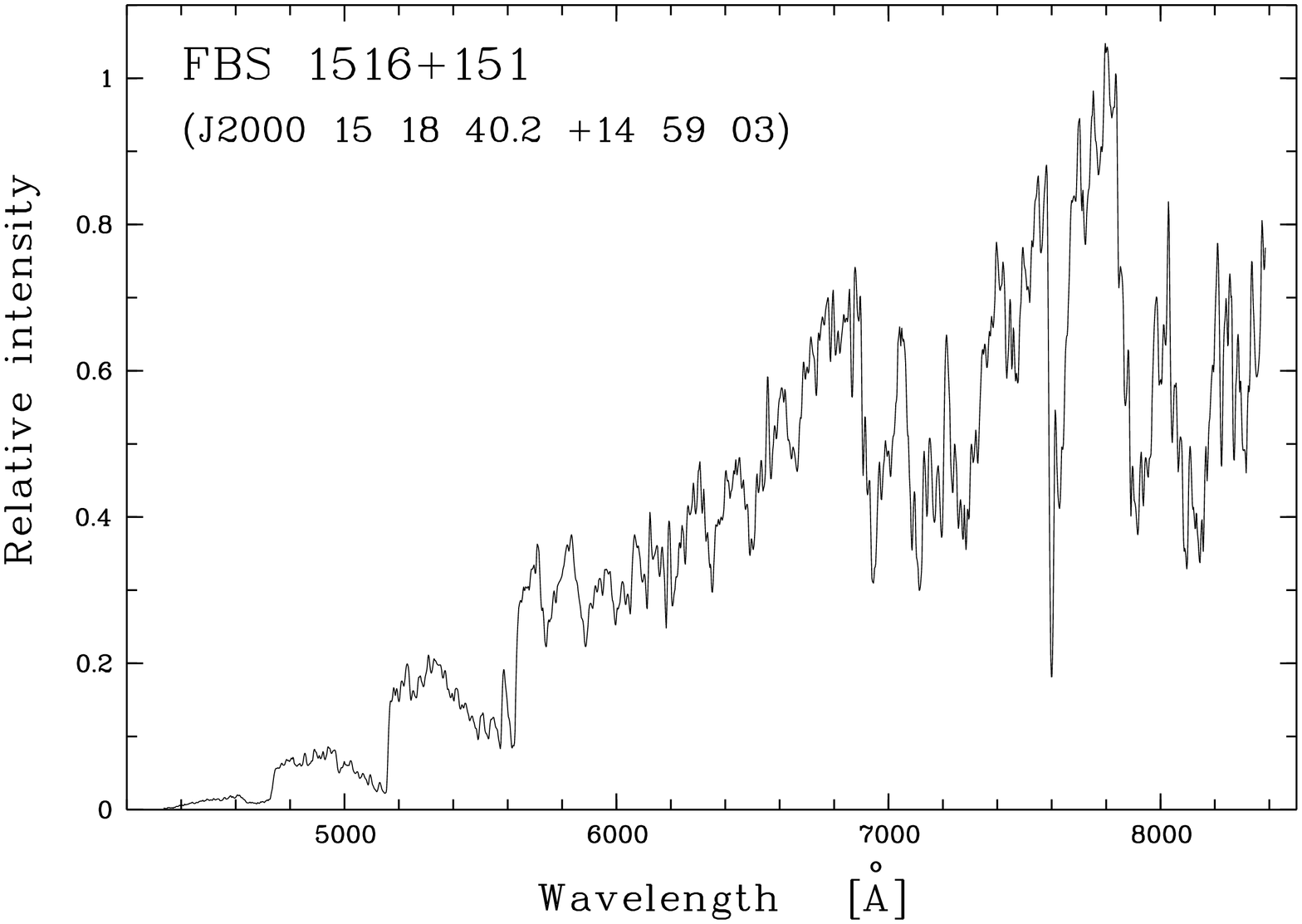}}}
   \resizebox{8.5cm}{!}{\rotatebox{-90}{\includegraphics{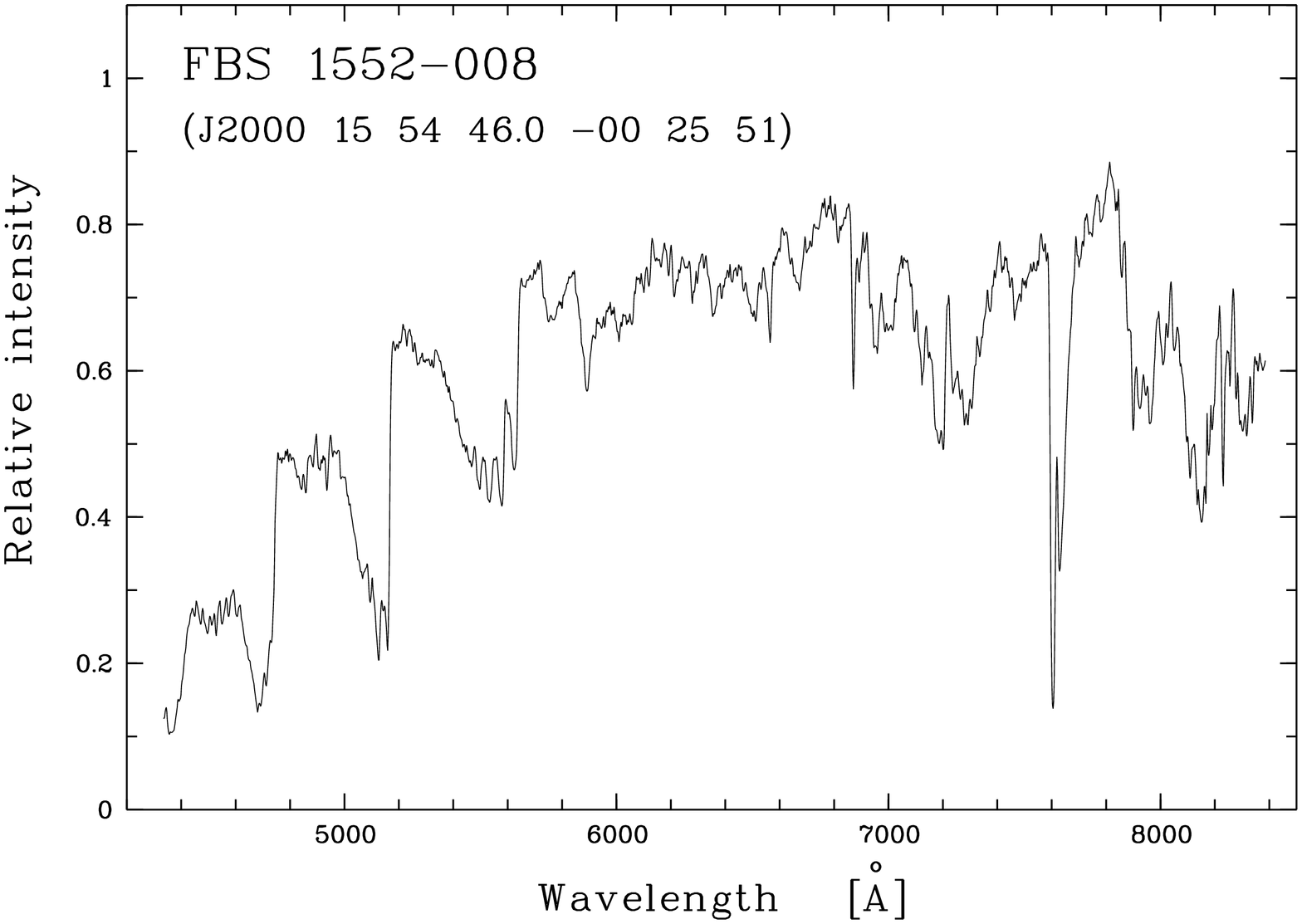}}}
   \resizebox{8.5cm}{!}{\rotatebox{-90}{\includegraphics{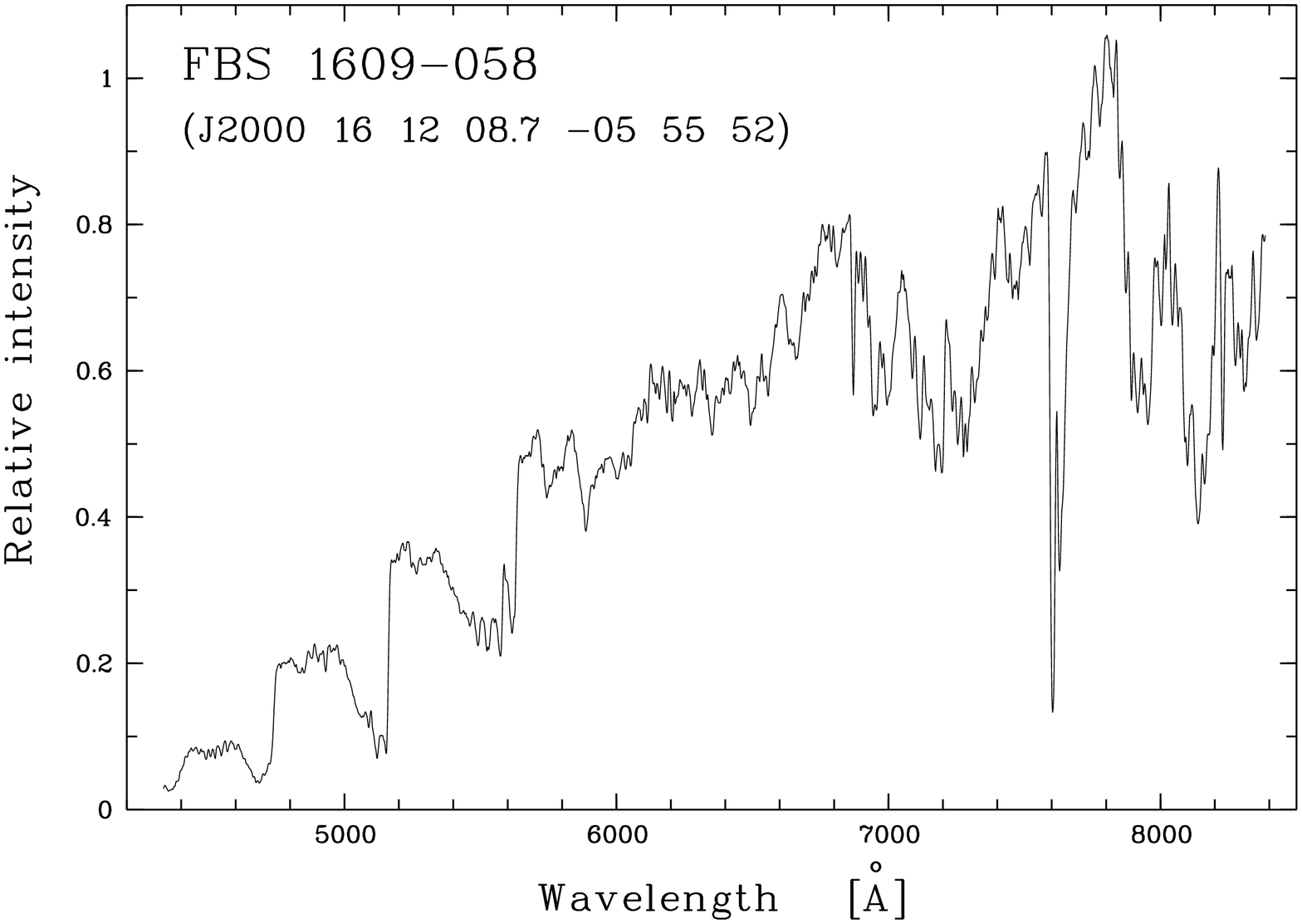}}}
   \resizebox{8.5cm}{!}{\rotatebox{-90}{\includegraphics{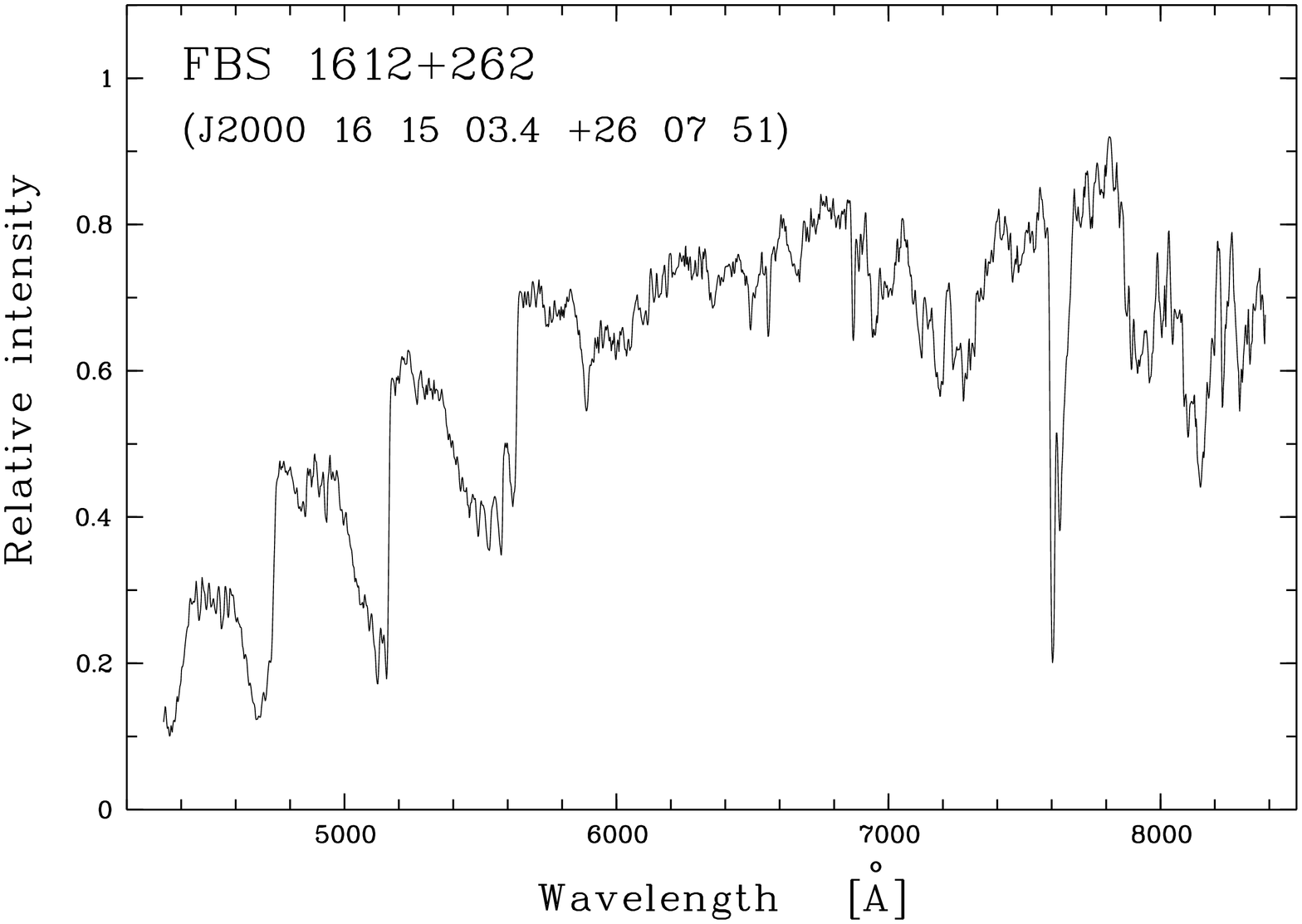}}}
   \resizebox{8.5cm}{!}{\rotatebox{-90}{\includegraphics{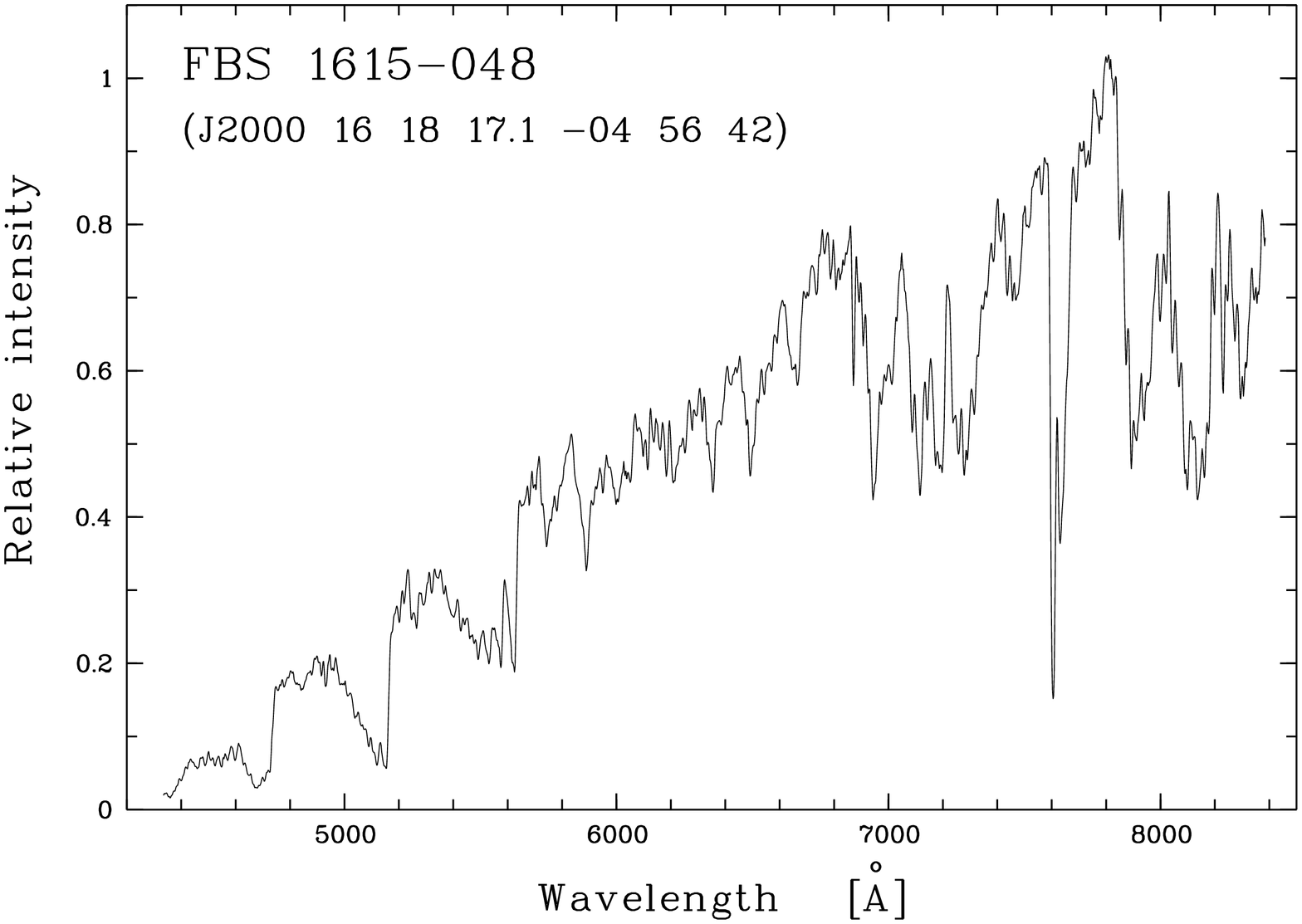}}}
   \resizebox{8.5cm}{!}{\rotatebox{-90}{\includegraphics{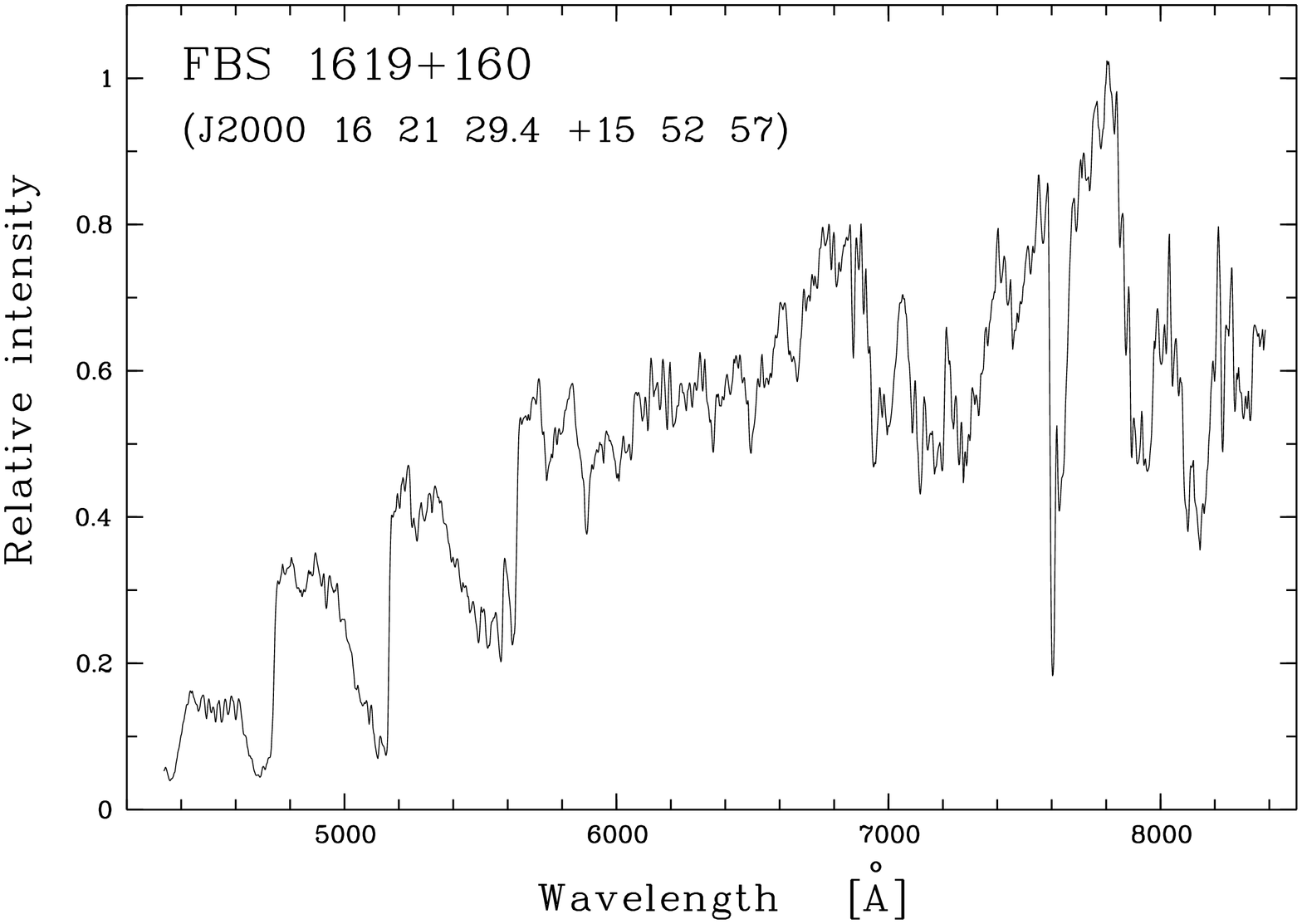}}}
   \end{figure*}
 
 \addtocounter{figure}{-1}

   \begin{figure*}
   \caption[]{{\it Continued}}
   \resizebox{8.5cm}{!}{\rotatebox{-90}{\includegraphics{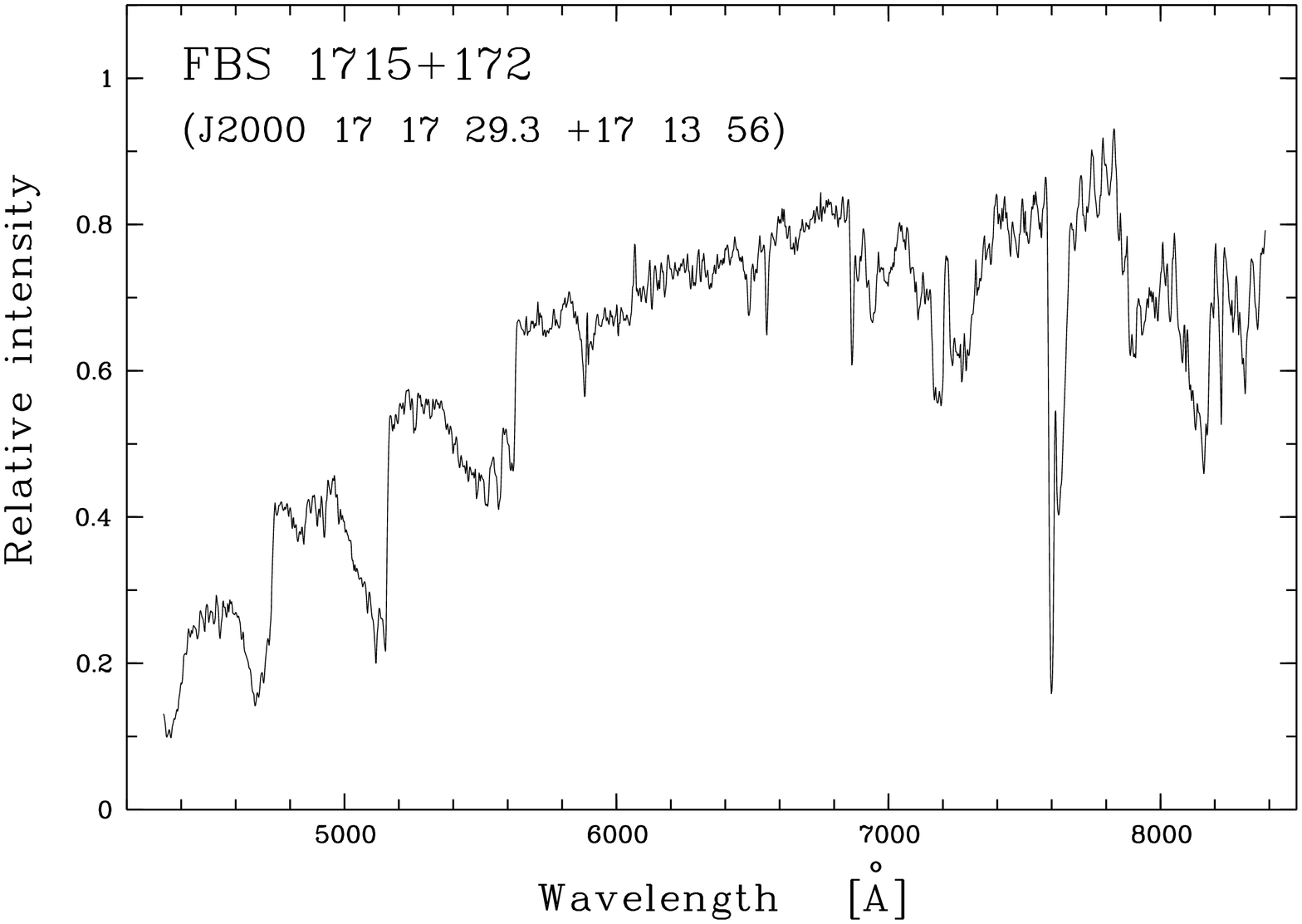}}}
   \resizebox{8.5cm}{!}{\rotatebox{-90}{\includegraphics{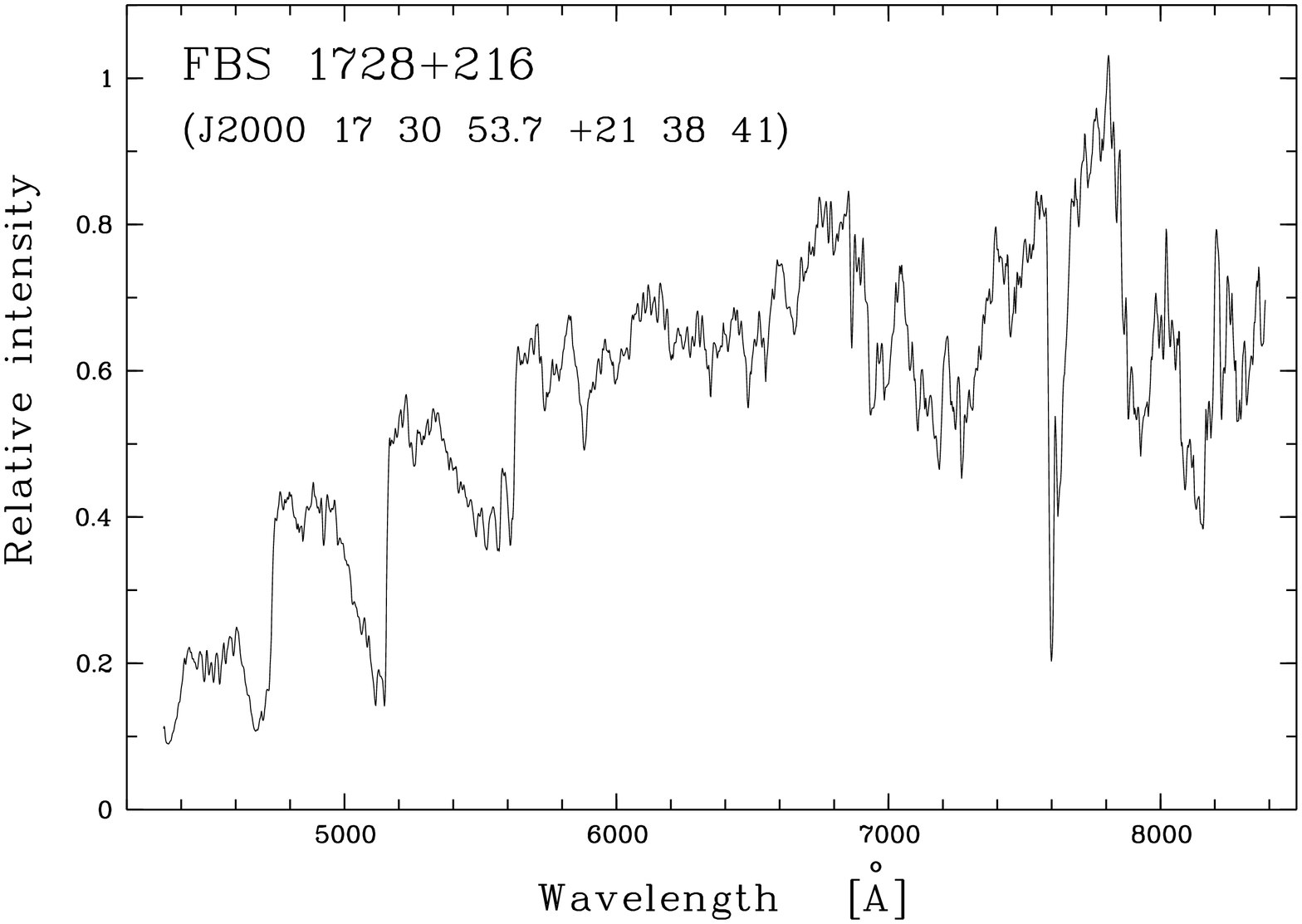}}}
   \resizebox{8.5cm}{!}{\rotatebox{-90}{\includegraphics{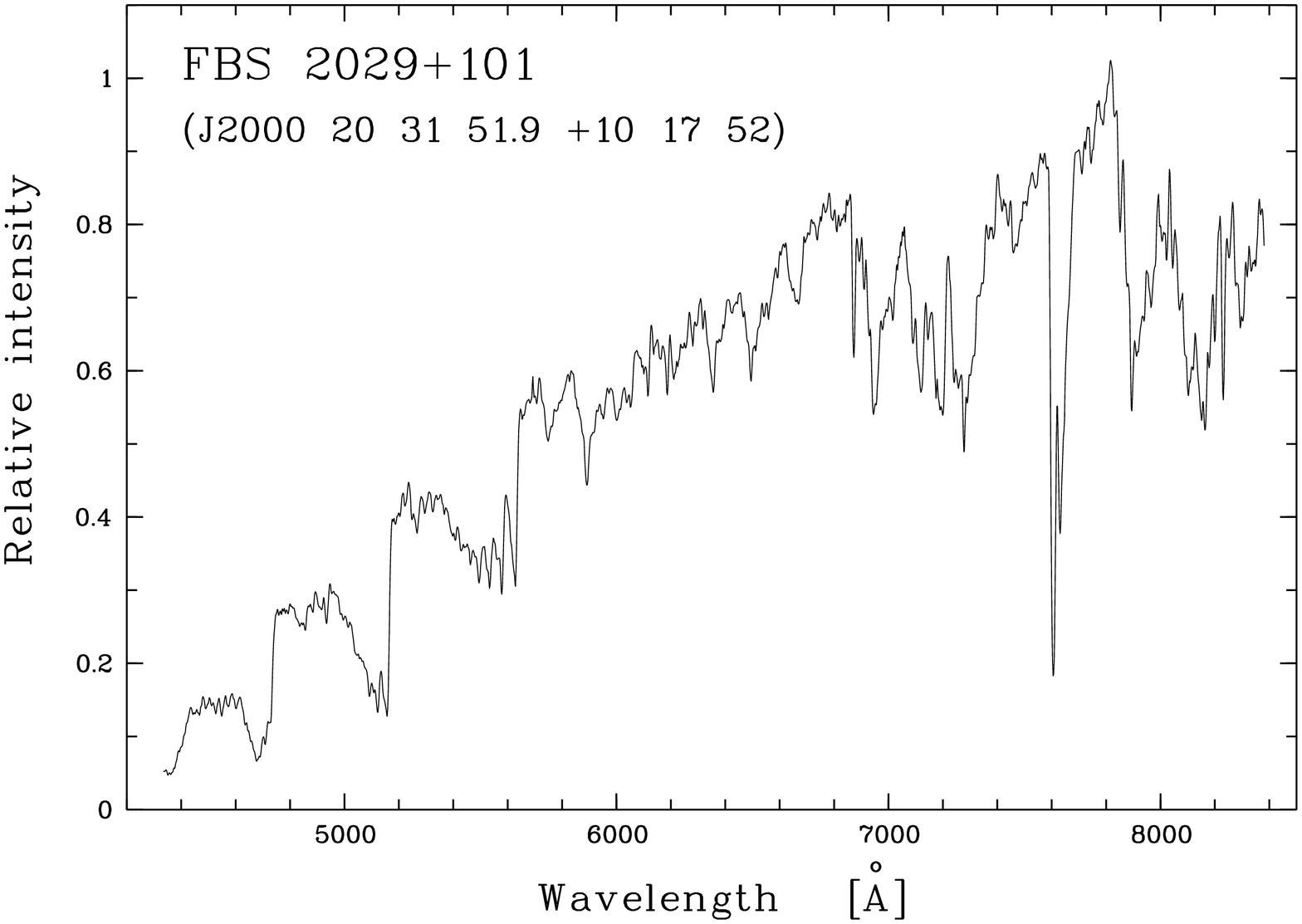}}}
   \resizebox{8.5cm}{!}{\rotatebox{-90}{\includegraphics{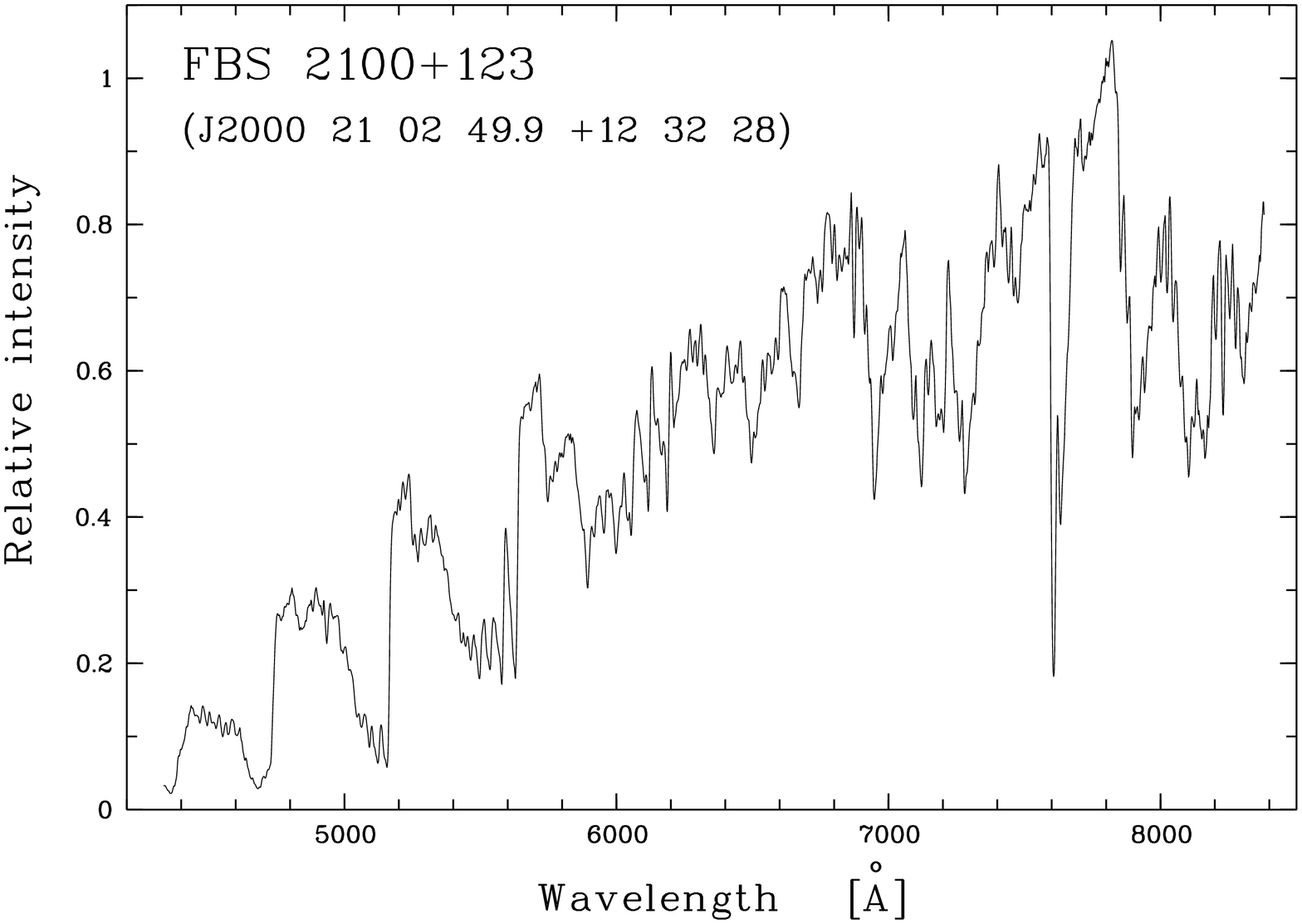}}}
   \resizebox{8.5cm}{!}{\rotatebox{-90}{\includegraphics{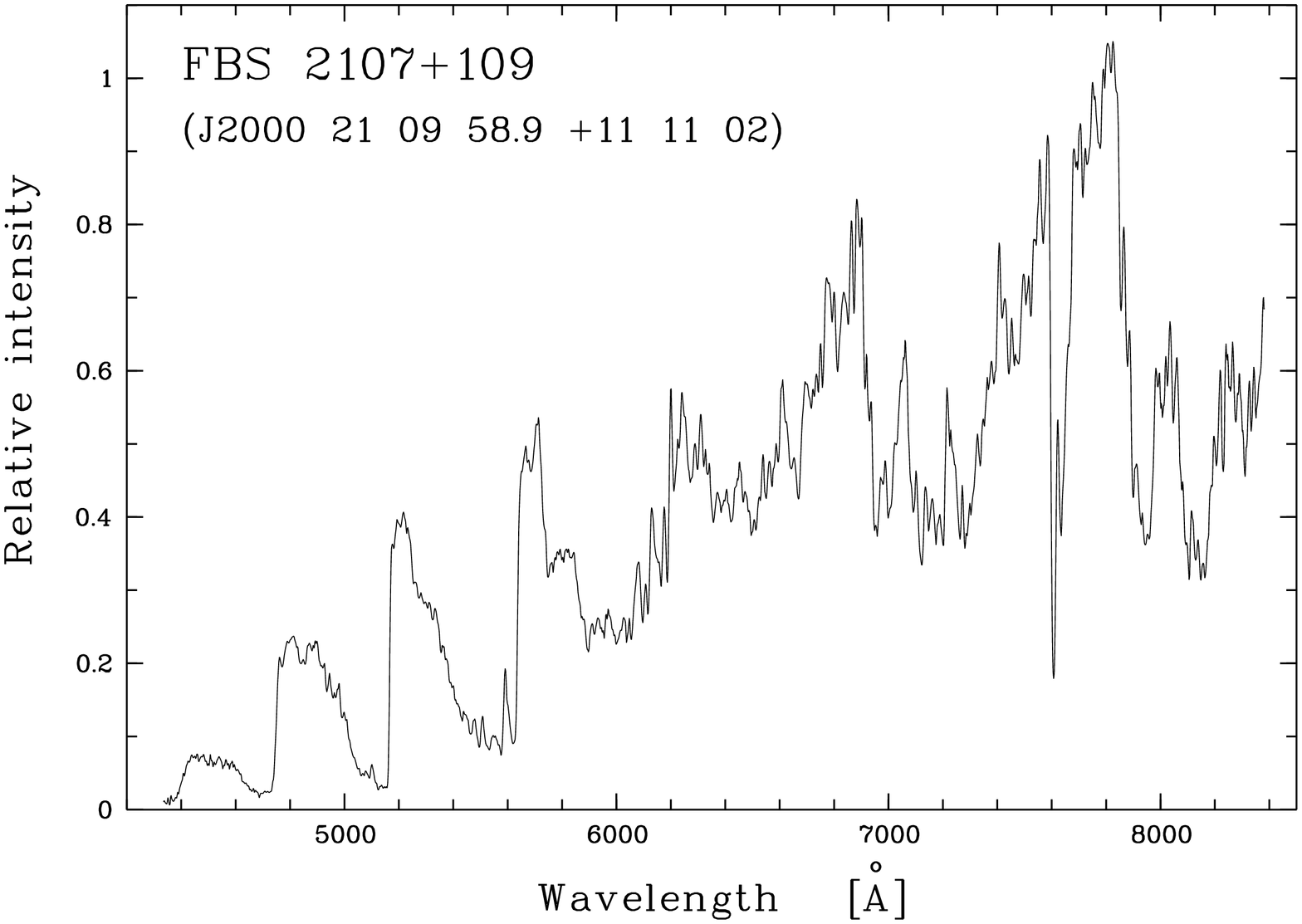}}}
   \resizebox{8.5cm}{!}{\rotatebox{-90}{\includegraphics{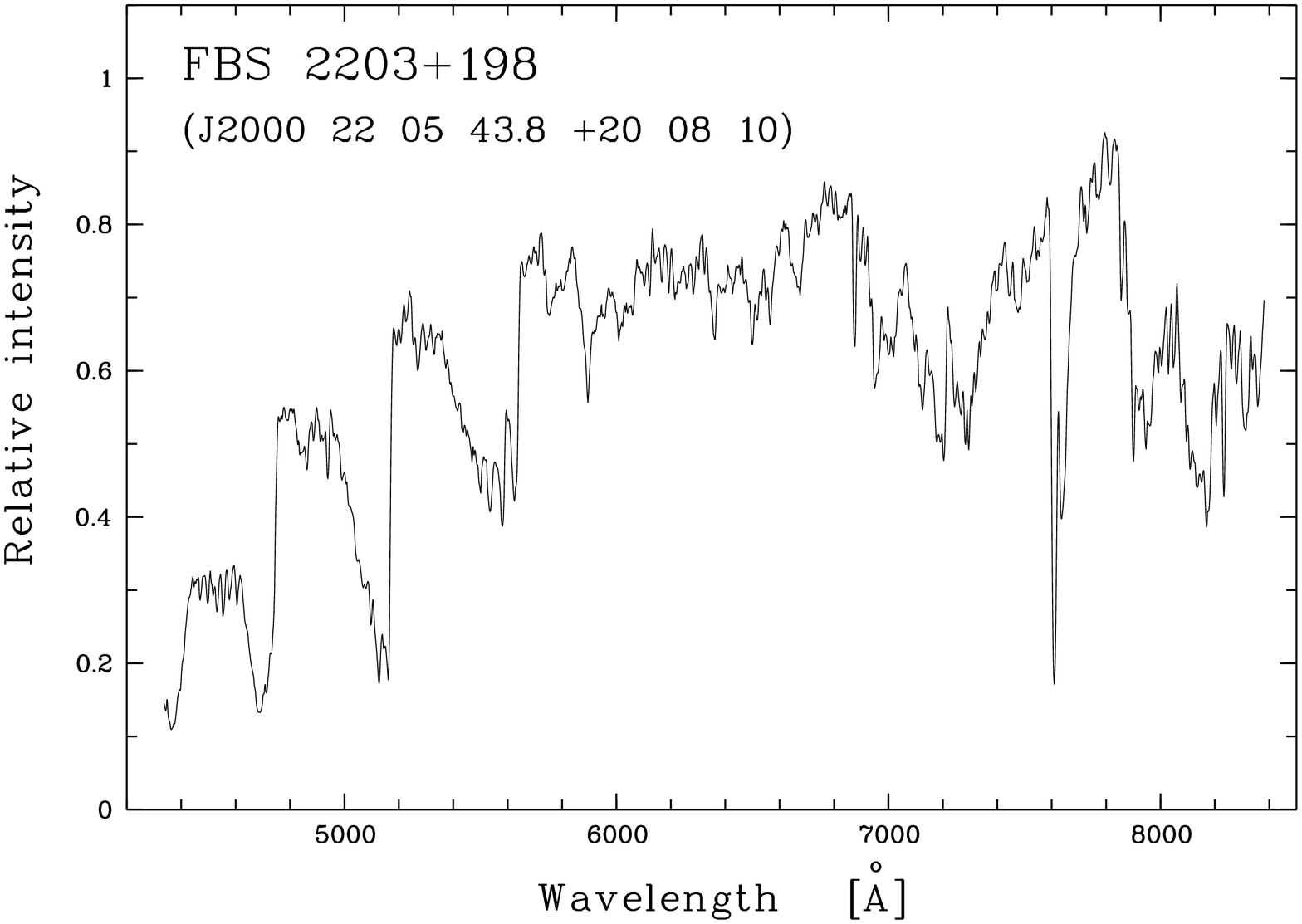}}}
   \resizebox{8.5cm}{!}{\rotatebox{-90}{\includegraphics{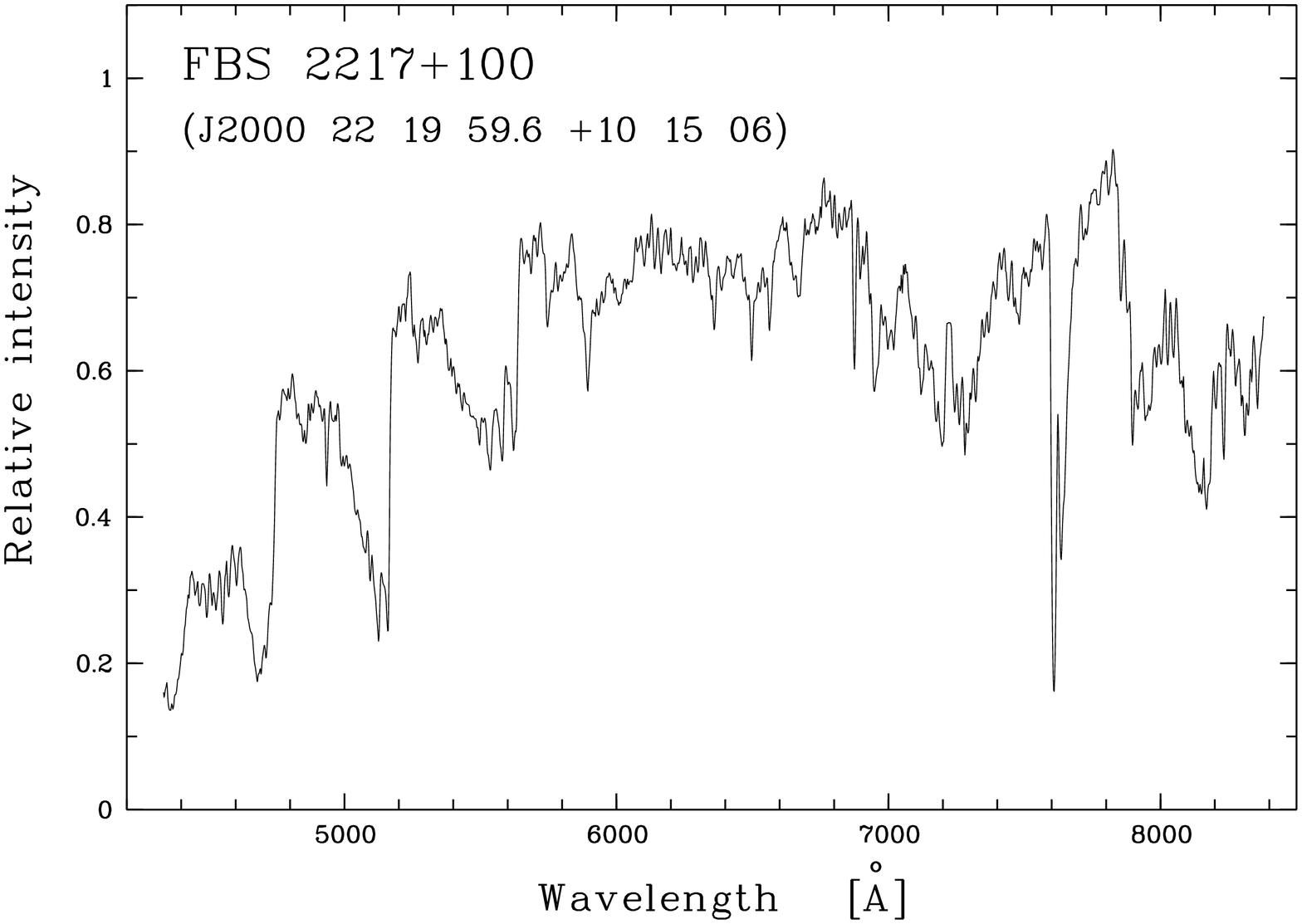}}}
   \end{figure*}


\begin{thebibliography}{}

\bibitem[2001]{alksnis01}
 Alksnis, A., Balklavs, A., Dzervitis, U., et al. 2001, 
 Baltic Astron., 10, 1 (CDS catalogue III/227)

\bibitem[2007]{an07}
An, D., Terndrup, D.M., \& Pinsonneault, M.H. 2007, ApJ, 671, 1640

\bibitem[1989]{beichman88}
Beichman, C., Neugebauer, G., Habing, H.J., \& the Joint IRAS SWG. 1988,
The IRAS catalogue of Point Sources, version 2.0, NASA RP-1190, CDS catalogue II/125

\bibitem[2002]{bergeat02}
 Bergeat, J., Knapik, A., \& Rutily, B. 2002, A\&A, 385,94
 
\bibitem[1991]{bothun91}
Bothun G.L., Elias J.H., MacAlpine, G., et al. 1991, AJ, 101, 2220

\bibitem[2006]{buchanan06}
Buchanan, C.L., Kastner, J.H., Forrest, W.J., et al. 2006, AJ, 132, 1890

\bibitem[1987]{claussen87}
 Claussen, M.J., Kleinmann, S.G., Joyce, R.R., \& Jura, M. 1987, ApJSS, 65,385
 
\bibitem[2003]{clementini03}
Clementini, G., Gratton, R., Bragaglia, A., et al. 2003, AJ, 125, 1309

\bibitem[2007]{cruz07}
Cruz, K.L., Reid, I.N., Kirkpatrick, J.D., et al. 2007,
AJ,133, 439

\bibitem[2003]{cutri03}
Cutri, R.M., Skrutskie, M.F., Van Dyk, S., et al. 2003, The 2MASS All-Sky Catalogue of 
Point Sources, Univ. of Massachusetts \& Infrared Processing \& Analysis Center,
CDS catalogue II/246

\bibitem[1998]{dahlmark98} 
Dahlmark, L. 1998, IBVS 4642

\bibitem[2004]{downes04}
Downes, R.A., Margon, B., Anderson, S.F., Harris, H.C., et al. 2004,
AJ, 127, 2838

\bibitem[1990]{epchtein90}
Epchtein, N., Le Bertre, T., L\'{e}pine, J.R.D. 1990, A\&A, 227, 82

\bibitem[1989]{feast89}
Feast, M.W., Glass, I.S., Whitelock, P.A., Catchpole, R.M. 1989, 
MNRAS, 241, 375

\bibitem[2001]{gigoyan01}
Gigoyan, K., Mauron, N., Azzopardi, M., Muratorio, G., 
\& Abrahamyan, H.V. 2001,
A\&A, 371, 560

\bibitem[1992]{groe92}
 Groenewegen, M.A.T., de Jong, T., van der Bliek, N.S., Slijkhuis, S.,
 \& Willems, F.J. 1992, A\&A, 253, 150

\bibitem[1997]{groe97}
 Groenewegen, M.A.T., Oudmaijer, R.D., \& Ludwig, H.G. 1997, MNRAS, 292, 686

\bibitem[2001]{ibata01}
Ibata, R., Lewis, G.F., Irwin, M., Totten, E.,\& Quinn, T. 2001,
 ApJ, 551, 294

\bibitem[1986]{jura86}
Jura, M. 1986, ApJ, 303, 327

\bibitem[1987]{jura87}
Jura, M. 1987, ApJ, 313,743

\bibitem[2007]{kastner07}
Kastner, J.H., Thorndike, S.L., Romanczyk, P.A., et al. 2007, astro-ph/0703584v2

\bibitem[2000]{kazarovets00}
Kazarovets, E.V., Samus, N.N., \& Durlevich, O.V. 2000, IBVS 4870 

\bibitem[2001]{kontizas01}
Kontizas, E., Dapergolas, A., Morgan, D.H., \& Kontizas, M. 2001,
 A\&A, 369, 932
 
\bibitem[2005]{law05}
Law, D.R., Johnston, K.V., \& Majewski, S.R. 2005, ApJ, 619, 807


\bibitem[1987]{little87}
Little-Marenin, I.R., Ramsay, M.E., Stephenson, C.B., et al. 1987, AJ, 93,663

\bibitem[2003]{macconnell03}
MacConnell, D.J., 2003, PASP 115, 351

\bibitem[2003]{majewski03}
 Majewski, S.R., Skrutskie, M.F., Weinberg, M.D., \&\,\,Ostheimer, J.C. 2003,
 ApJ, 599, 1082

\bibitem[2007]{matsuura07}
Matsuura, M., Zijlstra, A.A., Bernard-Salas, J., et al. 2007, MNRAS, 382, 1889

\bibitem[2004]{mauron04}
 Mauron, N., Azzopardi, M., Gigoyan, K., \& Kendall, T.R. 2004, A\&A, 418, 77

\bibitem[2005]{mauron05}
 Mauron, N., Kendall, T.R., \& Gigoyan, K. 2005, A\&A, 438, 867
 
\bibitem[2007]{mauron07a}
 Mauron, N., Gigoyan, K.S., \& Kendall, T.R. 2007a, A\&A, 463, 969
 
\bibitem[2007]{mauron07b}
 Mauron, N., Gigoyan, K.S., \& Kendall, T.R. 2007b, A\&A, 475, 843

\bibitem[1998]{monet98}
Monet, D., Bird, A., Canzian, B., et al. 1998, The USNO-A2.0 Catalogue,
U.S. Naval Observatory Flagstaff Station and Universities Space 
Research Association
(CDS catalogue I/252)

\bibitem[2003]{monet03}
Monet, D.G., Levine, S.E., Canzian, B.,  et al. 2003, AJ, 125, 984
 
\bibitem[1989]{moshir89}
 Moshir, M., Kopan, G., Conrow, T., et al. 1989, The IRAS Faint 
 Source Catalogue, $|b|$\,$>$\,10, Version 2.0, CDS catalogue II/156A

\bibitem[2003]{newberg03}
Newberg, H.J., Yanny, B., Grebel, E.K., et al. 2003, ApJ, 596, L191


\bibitem[2000]{niko00}
Nikolaev, S., \& Weinberg, M.D. 2000, ApJ, 542, 804 (NW00)

\bibitem[2004]{olofsson04}
Olofsson, H. 2004, Circumstellar Envelopes. In AGB stars, ed. 
 H. Habing \& H. Olofsson, Springer Verlag, New-York, p.330

\bibitem[1998]{schlegel98}
Schlegel, D.J., Finkbeiner, D.P., \& Davis M. 1998, ApJ, 500, 525

\bibitem[1998]{ti98}
Totten, E.J., \& Irwin M.J. 1998, MNRAS, 294, 1

\bibitem[2000]{tiw00}
Totten, E.J., Irwin, M.J., \& Whitelock, P.A. 2000, MNRAS, 314, 630


\bibitem[2000]{vandenbergh00}
Van den Bergh, S. 2000,  The Galaxies of the Local Group, University Press, Cambridge

\bibitem[2007]{vanleeuwen07}
van Leeuwen, F., Feast, M.W., Whitelock, P.A., \& Laney, C.D. 2007, MNRAS, 379, 723


\bibitem[1999]{vanloon99}
Van Loon, J.T., Groenewegen, M.A.T., de Koter, A., et al. 1999, A\&A, 351,559

\bibitem[1998]{walknapp98}
Wallerstein, G., \& Knapp, G.R. 1998, ARA\&A, 36, 369

\bibitem[1999]{whitelock99}
Whitelock, P.A., Menzies, J., Irwin, M.J., \& Feast, M.W. 1999, in
The Stellar Content of Local Group Galaxies, Proceedings of the 192nd IAU Symp., 
ed. P.A. Whitelock \& R. Cannon (ASP), 136

\bibitem[2003]{whitelock03}
Whitelock, P.A., Feast, M.W., van Loon, J.T., \& Zijlstra, A.A. 2003, MNRAS, 342, 86

\bibitem[2006]{whitelock06}
Whitelock, P.A., Feast, M.W., Marang, F., \& Groenewegen, M.A.T. 2006, MNRAS, 369, 751

\bibitem[1992]{wood92}
Wood, P.R., Whiteoak, J.B., Hugues, S.M.G., et al. 1992, ApJ, 397, 552

\bibitem[2004]{zijlstra04}
Zijlstra, A. 2004, MNRAS, 348, L23

\end{thebibliography}
\end{document}